%% file: thesis-temp.tex
\documentclass{report}
\usepackage[a4paper,margin=1in,footskip=0.25in]{geometry}

\pdfoutput=1
\usepackage{amsmath}
\usepackage{xcolor}
\usepackage{appendix}
\usepackage{hyperref}
\usepackage{appendix}
\usepackage[utf8]{inputenc}

\usepackage{setspace}
\usepackage{amsmath, amssymb, amsthm, float, graphicx}
\usepackage{bm}
\numberwithin{equation}{section}

\hypersetup{colorlinks=false, linkcolor=blue, citecolor=red}

\def\beq{\begin{eqnarray}}\def\eeq{\end{eqnarray}}
\def\be{\begin{equation}}\def\ee{\end{equation}}

\def\l{\lambda}

\newcommand{\bea}{\begin{eqnarray}}
\newcommand{\eea}{\end{eqnarray}}

\newcommand{\dsl}{\pa \kern-0.5em /}
\newcommand{\pa}{\partial}

\newcommand{\kfo}{{k^F_{0}}}
\newcommand{\hopt}{{t_{hop}}}
\newcommand{\half}{\frac{1}{2}}

\usepackage{setspace}

\begin{document}

\onehalfspacing

\pagenumbering{gobble}

\begin{center}

\null \vspace{1.6cm}

{\LARGE\text{Topology and quantum phases of low dimensional fermionic systems}}

\vskip 2cm

\large{\textit{A thesis submitted for the degree of} \\ \textit{Doctor of Philosophy} \\\textit{in the Faculty of Sciences}}
\\
\vskip 1cm

{\large \text{{\bf Sayonee Ray}}}

\vspace{1cm}
\begin{figure}[h]
\centering
\includegraphics[scale=0.25]{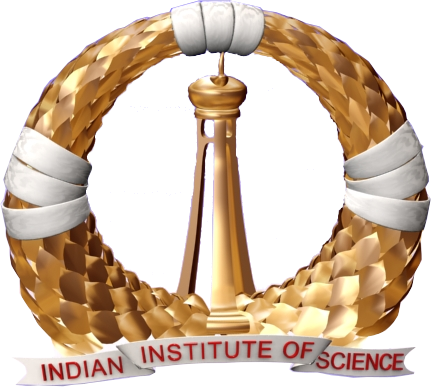}
\end{figure}

\vspace{1cm}
{\Large \textit{Department of Physics\\
Indian Institute of Science \\
\ Bangalore - 560012. India. \\}}

\end{center}



\pagebreak

 \

\vspace{2cm}

\ 

\vskip 2cm

\begin{flushleft}
\textsf{\textbf{\LARGE Declaration}}
\end{flushleft}

\vskip 2cm

{\large I hereby declare that this thesis `Topology and quantum phases of low dimensional fermionic systems" is based on my own research work, that I carried out with my collaborators in the Department of Physics, Indian Institute of Science, during my tenure as a PhD student under the supervision of Prof. Subroto Mukerjee. The work presented here does not feature as the work of someone else in obtaining a degree in this or any other institute. Any other work that may have been quoted or used in writing this thesis has been duly cited and acknowledged.

\vskip 2cm

 \ \\
Date:  8th May, 2017 \hspace{9.5cm} Sayonee Ray

\vskip 1.5cm

 \ \\
Certified by:

\vskip 1.5cm

 \ \\
Prof. Subroto Mukerjee\\
Department of Physics\\
Indian Institute of Science\\
Bangalore: 560012\\
India}

\pagebreak

\begin{center}
\textsf{\textbf{\Large Abstract}}
\end{center}

In this thesis, we study quantum phase transitions and topological phases in low dimensional fermionic systems.
In the first part, we study quantum phase transitions and the nature of currents in one-dimensional systems, using field theoretic techniques like bosonization and renormalization group. This involves the study of currents in Luttinger liquids, and the fate of a persistent current in a 1D system. In the second part of the thesis, we study the different types of Majorana edge modes in a 1D p-wave topological superconductor. Further we extend our analysis to the effect of an additional s-wave pairing and a Zeeman field on the topological properties, and present a detailed phase diagram and symmetry classification for each of the cases. In the third part, we concentrate on the topological phases in two-dimensional systems.  More specifically, we study the experimental realization of SU(3) topological phases in optical lattice experiments, which is characterized by the presence of gapless edge modes at the boundaries of the system. We discuss the specific characteristics required by a such a three component Hamiltonian to have a non-zero Chern number, and discuss a schematic lattice model for a possible experimental realization. 

\paragraph*{}

The thesis is divided into three chapters, as discussed below:

\paragraph*{}

In the first chapter, we study the effect of a boost (Fermi sea displaced by a finite momentum) on one dimensional systems of lattice fermions with short-ranged interactions. In the absence of a boost such systems with attractive interactions possess algebraic superconducting order. Motivated by physics in higher dimensions, one might naively expect a boost to weaken and ultimately destroy superconductivity. However, we show that for one dimensional systems the effect of the boost can be to strengthen the algebraic superconducting order by making correlation functions fall off more slowly with distance. This phenomenon can manifest in interesting ways, for example, a boost can produce a Luther-Emery phase in a system with both charge and spin gaps by engendering the destruction of the former.

\paragraph*{}

In the second chapter, we study the type of Majorana modes and the topological phases that can appear in a one-dimensional spinless p-wave superconductor. We have considered two types of p-wave pairing, $\triangle_{\uparrow\uparrow}=\triangle_{\downarrow\downarrow}$ and $\triangle_{\uparrow\uparrow}=-\triangle_{\downarrow\downarrow}$., and show that in both cases two types of Majorana bound states (MBS) with different spatial dependence emerge at the edges: one purely decaying and one damped oscillatory. Even in the presence of a Zeeman term ${\bf B}$, this nature of the MBS persists in each case, where the value of chemical potential $\mu$ and magnetic field ${\bf B}$ decides which type will appear. We present a corresponding phase diagram, indicating the number and type of MBS in the $\mu$-$B$ space. Further, we identify the possible symmetry classes for the two cases (based on the ten-fold classification), and also in the presence of perturbations like a s-wave pairing and various terms involving magnetic field. It is seen that in the presence of a s-wave perturbation, the MBS will now have only one particular nature, the damped oscillating behaviour, unlike that for the unperturbed p-wave case.

\paragraph*{}

In the third chapter, we study SU(3) topological phases in two dimension. It is shown by Barnett {\it et.al} that $N$ copies of the Hofstadter model with $\frac{2\pi}{N}$ Abelian flux per plaquette is equivalent to an $N$-component atom coupled to a homogeneous non-Abelian SU(N) gauge field in a square lattice. 
Such models have non-zero Chern number and for $N=3$, can be written in terms of the SU(3) generators. In our work, we uncover two salient ingredients required to express a general three-component lattice Hamiltonian in a SU(3) format with non-trivial topological invariant. We find that all three components must be coupled via a gauge field, with opposite Bloch phase (in momentum space, if the NN hopping between two components is $\sim -t e^{i k}$, then for the other two components, this should be $\sim -t e^{-i k}$) between any two components, and there must be band inversion between all {\it three} components in a given eigenstate. For spinless particles, we show that such states can be obtained in a tripartite lattice with three inequivalent lattice sites, in which the Bloch phase associated with the nearest neighbor hopping acts as $k$-space gauge field. The second criterion is the hopping amplitude $t$ should have an opposite sign in the diagonal element for one of the two components, which can be introduced via a constant phase $e^{i\pi}$ along the direction of hopping. The third and a more crucial criterion is that there must also be an odd-parity Zeeman-like term (as $k\rightarrow-k$, the term  changes sign), i.e. $\sin(k)\sigma_z$ term, where $\sigma_z$ is the third Pauli matrix defined with any two components of the three component basis. In the presence of a constant vector potential, the kinetic energy of the electron gets modified when the vector potential causes a flux to be enclosed. This can generate the desired odd parity Zeeman term, via a site-selective polarization of the vector potential. This can be achieved in principle by suitable modifications of techniques used in Sisyphus cooling, and with a suitable arrangement of polarizer plates, etc. The topological phase is affirmed by edge state calculation, obeying the bulk-boundary correspondence.

\pagebreak

\

\vskip 1cm

\begin{center}
\textsf{\textbf{\LARGE Acknowledgements}}
\end{center}

First and foremost, I would like to express my sincere gratitude to my supervisor, Prof. Subroto Mukerjee, for his continuous support, patience and motivation during my research and in writing this thesis. His immense enthusiasm in all aspects of physics has motivated me to participate in different projects covering more than one field of study in condensed matter physics. I could not have imagined having a more suitable mentor for my PhD.

I am sincerely grateful to Prof. Vijay B. Shenoy, Dr. Tanmoy Das and Dr. Subhro Bhattacharjee, who provided me an opportunity to work with them in different projects. It was an enriching experience to learn from them and discussing physics with them. 

I would like to thank all our course instructors and all the faculties in the department, for their continuous effort in teaching us the fundamentals of the subject. A special note of thanks to Prof. Diptiman Sen, for his patience and suggestions during all the discussions and tutorials throughout our coursework and research.

I am grateful to my collaborators Sambuddha Sanyal, Ananya Ghatak and Abhiram Soori. I learnt so much in the projects I worked on with them and look forward to even more such experiences.

I am thankful to my labmates: Manjari, Yogeswar, Arijit, Abhaas, Nairita, Kausik, Sujay, Soumi, Gaurav, for all the inspirational discussions we had and for your wonderful company.  My friends and seniors in the department: Arpan, Oindrila, Sneha, Sabiha, Rathul, Sudip, Adhip, Kingshuk, Ranjan, Manisha, Apratim, Subham, Rahool, Debasmita, Sudipta, Kazi, Prakriti and all others, thank you for making this journey remarkable.

Many thanks to my dance teacher, Anjali Raj Urs, who has taught me there is something beautiful in pushing oneself beyond endurance, and my friends in the dance class for the fun we had together.

I would like to thank my friends, Kallol, Aritra, Jayita, Kausik and Dipanwita, for your valuable support, your friendship and all the entertainment that you provided. I am grateful to my college and school friends for being so inspriring. A very special gratitude goes to Chitrak, who has been extremely supportive and encouraging in all my endeavours.

I gratefully acknowledge Council of Scientific \& Industrial Research (CSIR), India and Department of Science \& Technology, Govt.of India, for funding this research work.

Lastly and most important of all, I am thankful to my family: my parents, my sister and her family, and my grandparents, for their unconditional love and their unfailing support. My dad, for the champion you are and will always be; my mom for being my greatest strength; my sister and Mainak Da for being my biggest inspiration, and my grandparents for gifting me the best childhood ever. You have always provided me with encouragement and inspiration whenever I was in need. Without you this journey would have been impossible.

\begin{flushright}
{\it Sayonee Ray}
\end{flushright}

\pagebreak 

\

\vskip 4cm

\begin{flushright}
\large{To my parents..}

\end{flushright}




\tableofcontents

\chapter*{\textsf{{ List of Publications}}}
\addcontentsline{toc}{chapter}{{{ List of Publications}}}

\begin{flushleft}
\textsf{\textbf{\large The thesis is based on the following works}}
\end{flushleft}

\begin{flushleft}
\textsf{\textbf{\large My other ongoing works during PhD, not included in thesis}}
\end{flushleft}

\pagenumbering{arabic}

\pagebreak

\include{mychapter1}

\include{mychapter2}

\include{chapter4_2}

\include{mychapter3}





\end{document}

%% file: mychapter1.tex
\chapter{\LARGE Introduction} \label{chapter1}
{\large

\section{Quantum phases and phase transitions in 1D fermionic systems}

Quantum phase transitions, differ from classical phase transitions, primarily due to fact that these can occur at temperature $T=0$ and are effected by tuning some non-thermal parameter like chemical potential, magnetic field or chemical composition. At $T=0$, all thermal fluctuations are frozen and the phase transition is caused by quantum fluctuations in the ground state of the system. Common examples of quantum phase transitions are Superfluid (SF) to Mott Insulator (MI) transtion, superconductor (SC) to insulator transition, disorder driven localization-delocalization transition, and so on ~\cite{sachdev,altland,coleman}. \hphantom{} \\

Quantum phase transitions can occur due to the spontaneous breaking of symmetries, treated under the Landau theory of phase transitions. The symmetry broken phase is associated with a local order parameter, and fluctuations over this local order parameter gives the collective excitations in the phase. Collective excitations in such phases are massless Nambu-Goldstone modes, for example, gapless excitations in the broken U(1) phase in the XY model are spin waves ~\cite{abrikosov,altland,coleman}. 

However, there exists a mapping from a quantum system in $D$ dimensions at $T=0$ to a classical system in $D+1$ dimensions. Superfluids and superconductors in 1D in the quantum regime (at $T=0$) , has an equivalent classical description in 2D at finite temperatures. Thus, quantum phase transitions in $D$ dimensions can be understood from its classical counterpart in one higher dimension.\hphantom{} \\

A separate class of quantum phase transitions exist which defy this paradigm of symmetry breaking. Topological phase transitions are quantum phase transitions associated with the topological order of the phase, even though the symmetry is preserved, i.e, there is no local order parameter to distinguish between the two phases. Different phases have different topological numbers associated with global quantities like ground state degeneracy and quasi-particle statistics, for example, fractional quantum hall states, quantum spin liquids, etc ~\cite{altland,chaikin,wen}. \hphantom{} \\

\subsection{Mermin-Wagner theorem in $D\le 2$}
The absence of long-range order in 2D systems with continuous symmetry, even at finite temperatures, can be seen from the Debye Waller factor~\cite{chaikin}.  In the 2D XY model, classical spins in a 2D lattice can rotate in the $x$-$y$ plane. The Hamiltonian with nearest neighbour coupling $J$, is given by:
\begin{equation}
\mathcal{H}=-J\sum_{<ij>}\cos{(\theta_i-\theta_j)},
\end{equation}
where, $\theta_i$ is the angle of the spin at site $i$. The Debye Waller factor is given by $e^{-2W}$~\cite{chaikin}, where,
 \begin{equation}
W = \frac{T \Lambda^{D-2}}{2 \rho_s (D-2)},
\end{equation} 
($\Lambda$ is the wave number cut-off, $T$ is the temperature, $\rho_s$ is the thermodynamic stiffness and $D$ is the spatial dimension) hence showing that as $D\rightarrow2$, $W\rightarrow\infty$.
Thus, there is no long-range order in 2D at finite temperatures. This result in a more general setting is the Mermin Wagner theorem.

However, there is a phase transition at a finite temperature $T_c$ in such 2D systems, which are mediated by topological defects like the vortices. Below $T_c$, there exists a type of  quasi-long range order with power law correlations, associated with bound pairs of vortices and anti-vortices . Above $T_c$, the vortices proliferate, leading to a `disordered phase' with exponentially decaying correlation functions ~\cite{chaikin,girvin,reppy}. This transition at $T_c$ is the Kosterlitz-Thouless (KT) transition.

The universality class of the transition from the SF and SC phase to the normal phase is the XY universality class, and hence quantum superfluids and superconductors in 1D are quasi-long-range ordered (QLRO), even at $T=0$. Stronger interactions in 1D quantum systems, can drive the system to a Mott or a charge density wave (CDW) phase, which occurs through a KT transition ~\cite{giamarchi_1,gnt_1,schulz_1}, as we discuss below.

\subsection{Bosonization and correlations in 1D}

\begin{figure}[h]
\centering 
\includegraphics[width=9cm, height=5.2cm]{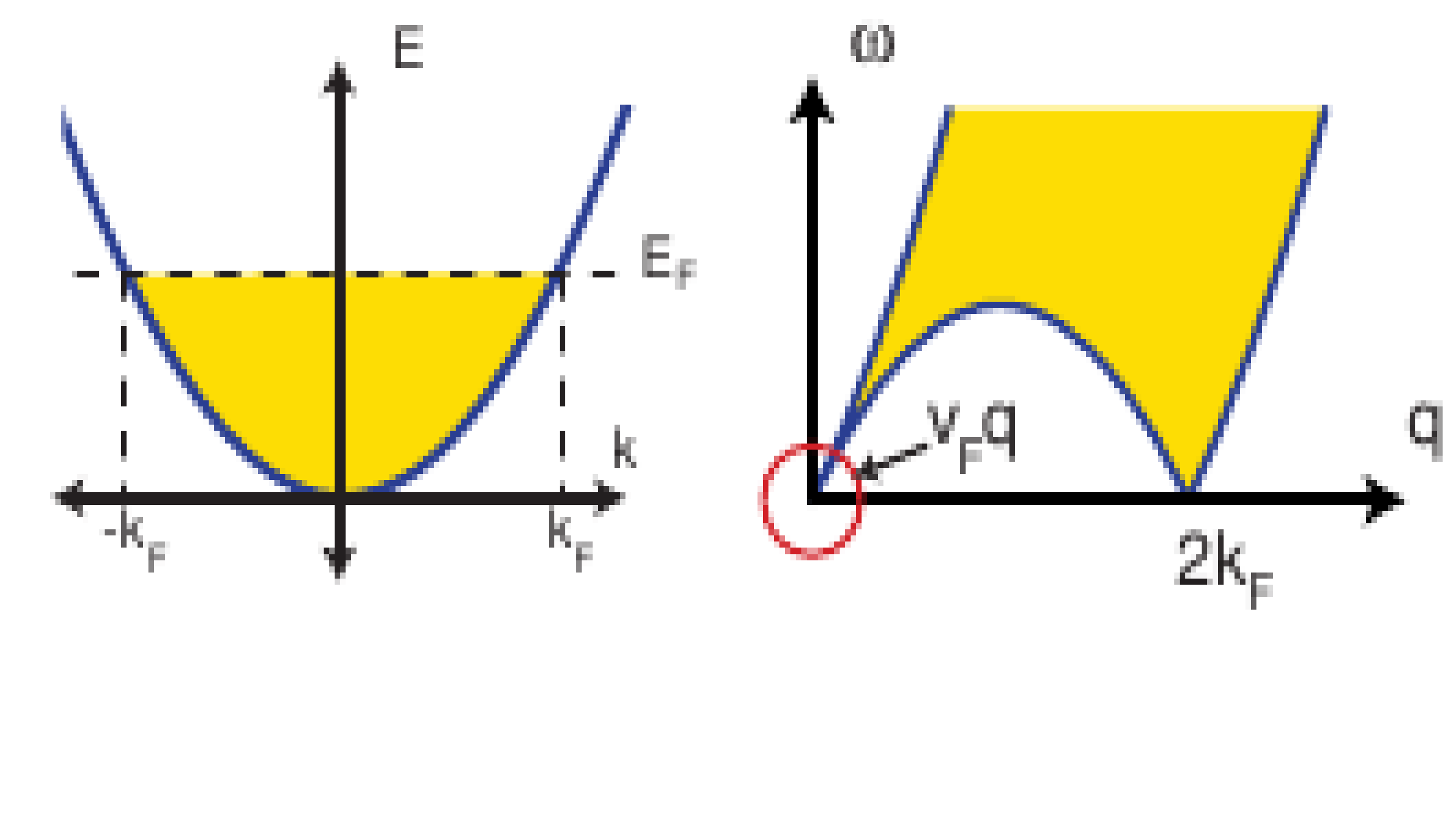}
\caption{(Left) Dispersion of non-interacting electrons in 1D. (Right) Particle-hole excitation spectrum in 1D.
\small{(Ref: "Lectures on Bosonization", C.L. Kane.)}}
\label{excitation1D}
\end{figure}

As can be seen from Fig.~\ref{excitation1D}, low-energy particle-hole excitations in 1D are sharp well-defined excitations with a well defined momentum and energy. The dispersion of any excitation $E(q)$ above the Fermi level is proportional to the momentum of the excitation,
\begin{align} \nonumber
E(q) &= v_F (k+q)-v_F k ,\ \nonumber \\
&= v_F q .\
\end{align}
Such low energy processes can only occur near the Fermi points, and the entire Hilbert space can be written in terms of these collective (bosonic) excitations. Depending on the sign of the velocity near the Fermi points $k_F$ and $-k_F$, the fermions can be categorized into right movers ($v>0$) and left movers ($v<0$).
\begin{equation} \label{ham1_1}
\mathcal{H} = \sum_{k,r=R,L} (rk-k_F) c^{\dagger}_{k,r} c_{k,r} ,\
\end{equation}
where, $r= +1$ and $-1$ for right mover (RM) and left mover (LM) respectively. 
Any single particle operator can be written as:
\begin{eqnarray}\nonumber
\psi(x)&=& \frac{1}{\sqrt{\Omega}} \sum_k c_k e^{i k x} \nonumber\\
&=&\frac{1}{\sqrt{\Omega}} \sum_{k\sim -k_F} c_k e^{i k x}
+ \frac{1}{\sqrt{\Omega}} \sum_{k\sim k_F} c_k e^{i k x} \nonumber\\
&=& \psi_L+\psi_R,
\end{eqnarray}
and, the density operator:
\begin{equation}
\rho(x)=\psi^{\dagger}_R(x)\psi_R(x)+\psi^{\dagger}_L(x)\psi_L(x)
+\psi^{\dagger}_R(x)\psi_L(x)+\psi^{\dagger}_L(x)\psi_R(x)
\end{equation}

Eq.~\ref{ham1_1} can be written in terms of these density operators, which in turn, can be expressed in terms of the bosonic fields $\phi$ and its conjugate field $\theta$ (commutation relation between $\phi$ and $\nabla\theta$ is, $ \big[ \phi(x), \nabla\theta(x') \big] = i \pi \delta(x'-x)$). An interacting spinless Hamiltonian in 1D in the continuum can thus be written in the Gaussian form (note that on a lattice umklapp processes can introduce additional terms),
\begin{equation}
H = \frac{1}{2 \pi} \int dx \big[ u K (\nabla\theta(x))^2+ \frac{u}{K} (\nabla\phi(x))^2 \big]
\end{equation}
where, \begin{eqnarray} \label{phi_theta} \nonumber
\phi(x) &=& - (N_R+N_L) \frac{\pi x}{L} - \frac{i \pi}{L} \sum_{q \neq 0} \frac{1}{q} e^{- \alpha q/2 -i q x} (\rho^{\dagger}_R(q)+\rho^{\dagger}_L(q)),\ \nonumber \\
\theta(x) &=&  (N_R-N_L) \frac{\pi x}{L} + \frac{i \pi}{L} \sum_{q \neq 0} \frac{1}{q} e^{- \alpha q/2 -i q x} (\rho^{\dagger}_R(q)-\rho^{\dagger}_L(q)) ,\
\end{eqnarray}
and $u$ is the Fermi velocity and $K$ is the Luttinger parameter~\cite{giamarchi_1,gnt_1,schulz_1,dsen}. $\nabla\phi$ gives the density fluctuations and $\nabla\theta$ is the current operator in 1D. The details are elaborated in chapter 2.

\begin{figure}
\centering
\includegraphics[width=0.70\linewidth]{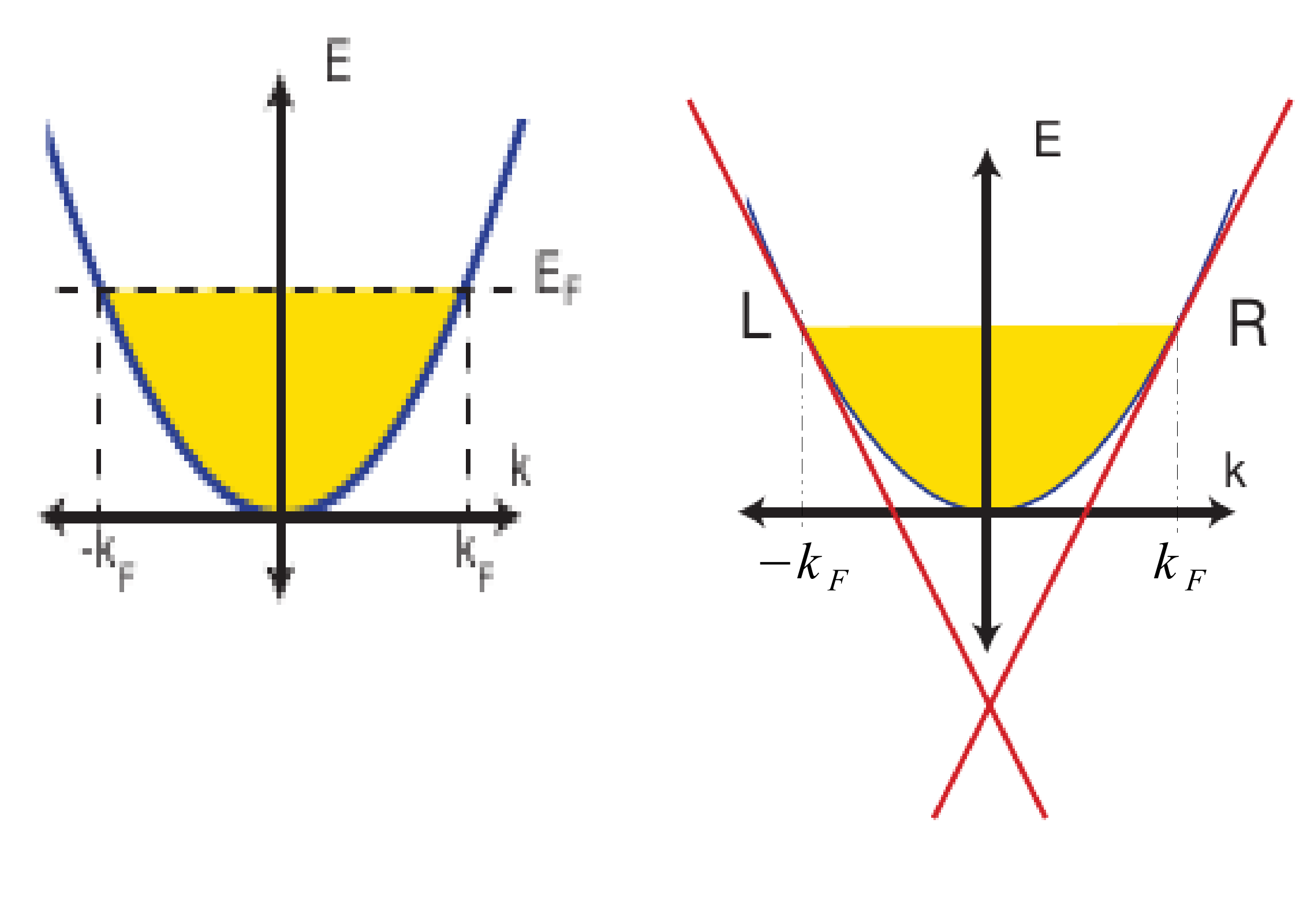}
\caption{Left panel: Dispersion of non-interacting electrons in 1D.  Right panel: The linearized bands at the Fermi points $k_F$ and $-k_F$ of the Luttinger model, extending to $-\infty$ \small[{Ref.: "Lectures on bosonization", C.L. Kane}]}.
\label{fermisea_1D}
\end{figure}

The Luttinger parameter $K$ has the information about the interaction processes in the system:
\begin{eqnarray} \label{u_K} \nonumber
u &=& v_F \bigg[ \bigg(1+ \frac{g_4}{2 \pi v_F}\bigg)^2-\bigg(\frac{g_2}{2 \pi v_F}\bigg)^2 \bigg]^{1/2}, \nonumber \\
K &=& \bigg[ \frac{1+\frac{g_4}{2 \pi v_F}-\frac{g_2}{2 \pi v_F}}{1+\frac{g_4}{2 \pi v_F}+\frac{g_2}{2 \pi v_F}} \bigg]^{1/2},
\end{eqnarray}
where, $g_4$ is the scattering process between the same species of fermions on the same side of the Fermi sea (between fermions of same chirality), $g_2$ is the scattering between species on opposite sides of the Fermi sea (between fermions of opposite chirality), as shown in Fig.~\ref{interaction_1} . $K=1$ is the non-interacting point, $K>1$ being the attractive regime and $K<1$ being repulsive.

\begin{figure}
\centering
\includegraphics[width=0.75\linewidth]{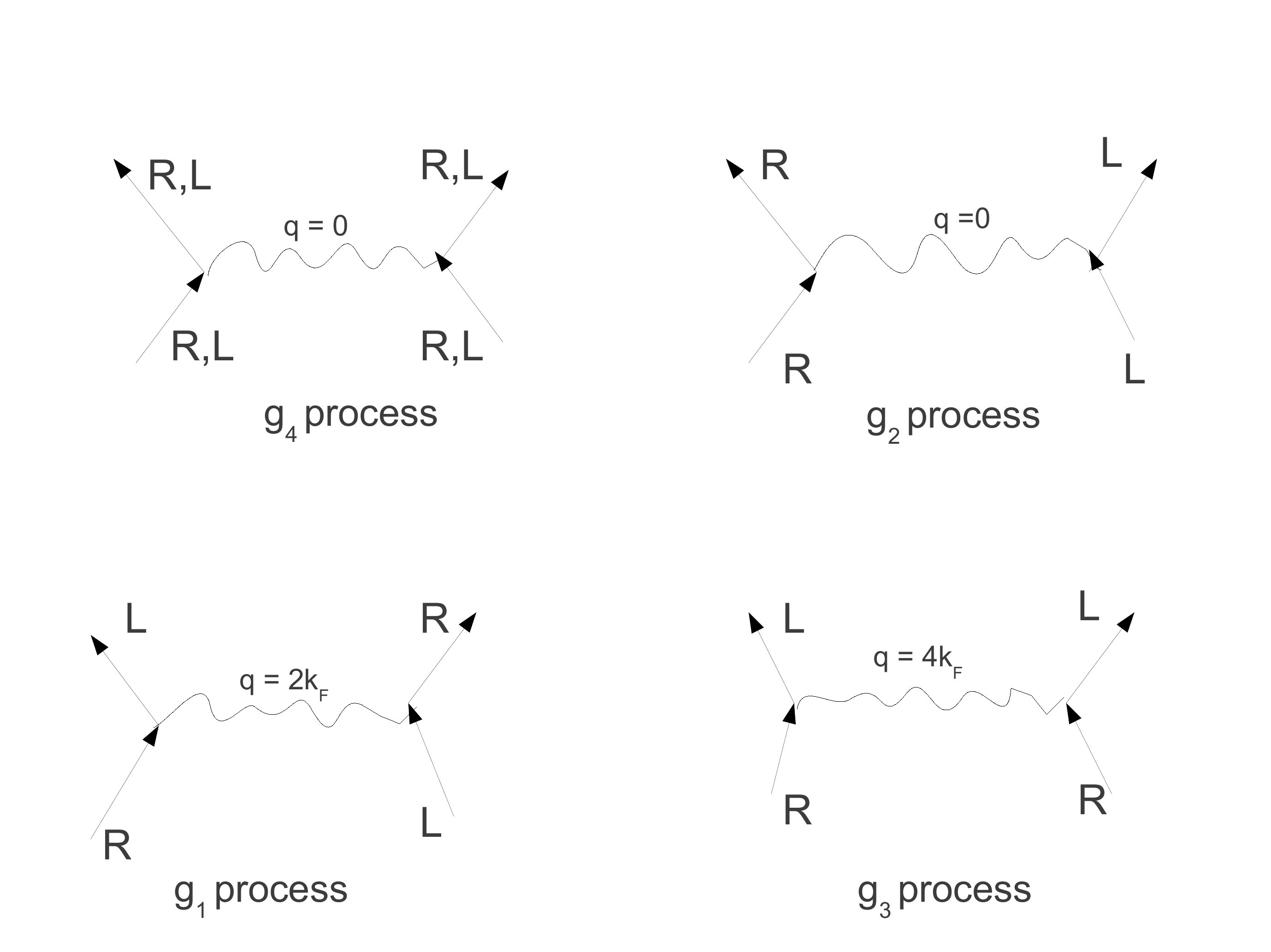}
\caption{Relevant scattering processes in Luttinger liquids. Figure 1: (Clockwise from top left) The $g_4$ process is where the scattering occurs on the same side of the Fermi sea. Figure 2: In the $g_2$ process, the right (left) mover scatters with a left (right) mover but remains a right (left) mover. $g_4$ and $g_2$ processes are relevant in both the charge and the spin sector of the system with the net momentum exchange being $q \sim 0$. Figure 3: Umklapp process in the charge sector is operative only in the presence of a lattice at half-filling. Here ($g_3$ process) two right (left) movers scatter to become two left (right) movers, with the net momentum exchange being $q\sim4k_F$. Figure 4: Umklapp in the spin sector is relevant even in the absence of a lattice. The $g_1$ process is where a right (left) mover (say, with up-spin) scatters with a left (right) mover (with down spin), and becomes a spin-flipped right (left) mover (with down spin). This is a spin exchange process, with the momentum exchange being $q\sim2k_F$}.
\label{interaction_1}
\end{figure}

Note that this description is different from the Fermi liquid paradigm, where instead of single-particle excitations, low-energy  elementary  excitations  are  particle-hole  pairs, which are bosons and can propagate coherently ~\cite{giamarchi_1,gnt_1,schulz_1,dsen,haldane_1}. The Luttinger liquid is the relevant description of such quantum fluids in one dimension. Such systems have a special property of always being `nested', since in 1D the Fermi surface (FS) is just two Fermi points, which can always be connected by a nesting vector $Q=2k_F$.
For values of $k$ near the Fermi points the energy satisfies, $\xi(k+Q)=-\xi(k)$, where the susceptibility for both the particle-particle and particle-hole channel, diverges and the perturbation theory in the interaction becomes singular. 
In higher dimensions, in general, $\chi$ itself does not diverge but its derivatives become singular. Thus, nesting is a rule for 1D systems. This divergence in the susceptibility occurs simulataneously in both the channels, particle-hole and particle-particle, causing competing SC and CDW orders (for spinless fermions) with power law correlations. This causes a quasi-long range order (QLRO) in both these channels. However, for strong enough umklapp interactions, the Luttinger liquid can be destabilized through a KT transition to a gapped CDW phase. 

For fermions with spin, umklapp is operative even in the continuum. The bosonic fields for each spin species are: ~\cite{giamarchi_1,gnt_1,schulz_1}:
\begin{eqnarray} \nonumber
\phi_\rho &=& ( \phi_{\uparrow} + \phi_{\downarrow} ) \nonumber \\
\phi_\sigma &=& ( \phi_{\uparrow} - \phi_{\downarrow} )
\end{eqnarray}
The $\phi_\rho,\theta_\rho$ and $\phi_\sigma,\theta_\sigma$ have the usual commutation relations. The single particle operator in terms of these fields:
\begin{equation}
\psi_r (x) = U_{r, \sigma} \lim_{\alpha \rightarrow 0} \frac{1}{2 \pi \alpha} e^{i r k_F x} e^{-\frac{i}{\sqrt{2}}[(r\phi_\rho(x)-\theta_\rho (x))+\sigma (r\phi_\sigma(x)-\theta_\sigma (x))] } 
\end{equation}
The Hamiltonian can be separated into two independent sectors~\cite{giamarchi_1}:
\begin{equation}
H = H_\rho + H_\sigma
\end{equation}
The charge part of the Hamiltonian is purely quadratic (in the absence of a lattice), only the spin part has an additional term. Together with the quadratic part, the total Hamiltonian is written as ~\cite{giamarchi_1}:
\begin{equation}
H = H_0 + \int dx\ \frac{2 g_{1\perp}}{(2 \pi \alpha)^2} \cos{( 2\sqrt{2} \phi_\sigma (x))},
\end{equation}
where, the Luttinger parameters in $H_0$ are given by:
\begin{eqnarray} \nonumber  
u_\nu &=& v_F \bigg[\bigg(1+\frac{y_{4 \nu}}{2}\bigg)^2-\bigg(\frac{y_\nu}{2}\bigg)^2\bigg]^{1/2}, \nonumber\\
K_\nu &=& \bigg[ \frac{1+\frac{y_{4\nu}}{2}+\frac{y_\nu}{2}}{1+\frac{y_{4\nu}}{2}-\frac{y_\nu}{2}} \bigg]^{1/2}, \nonumber\\
g_\nu &=& g_{1 \parallel}-g_{2 \parallel} \mp g_{2 \perp}, \nonumber\\
g_{4 \nu} &=& g_{4\parallel}\pm g_{4\perp}, \nonumber\\
y_\nu &=& g_\nu/ \pi v_F ,
\end{eqnarray}
where $\nu=\rho, \sigma$ and the upper sign refers to $\rho$ and lower sign to $\sigma$.
Note that the fermionic form of the umklapp term is 
$
g_{1\perp} \sum_{\sigma}\psi^{\dagger}_{L,\sigma}\psi^{\dagger}_{R,-\sigma}\psi_{L,-\sigma}\psi_{R,\sigma},\
$
as will be seen in Sec.~\ref{model_umklapp}).
The entire Hilbert space can be expressed in terms of the product of charge and spin excitations, and ensures the absence of any single particle excitations in the system.

In the presence of spin, now the SC pairing can be both triplet and singlet. Additionally quasi-long-ranged spin density wave (SDW) order is also possible in the system along with CDW. Umklapp can produce a gapped SDW order in the continuum giving a Luther Emery phase~\cite{giamarchi_1,gnt_1}.

\subsection{Landau critical velocity in one dimension}

Let us refer to a gedanken experiment of a fluid, initially in equilibrium, moving with a velocity $v$ in a pipe shaped container. A normal fluid would eventually thermalize and come to rest due to interactions with the walls of the container. However, for a SF, the SF fraction $\frac{\rho_s}{\rho_0}$ will continue to flow indefinitely with velocity $v$. This is because the SF velocity is topologically constrained by the quantization of the condensate phase (It is only beyond a certain velocity above which the phase slips occur, when interaction with the walls becomes relevant and quantized vortices emerge in the fluid)~\cite{leggett}. Observation of the SF response in 2D has been done by torsional oscillator (TO) experiments in $^4 He$ films, where a change in the resonance frequency of the TO has been observed, despite the lack of long range order in 2D at finite temperatures~\cite{reppy}. Thus, superfluidity in 2D, lacking long range phase coherence, can still be understood in terms of the helicity modulus, which is a static property~\cite{fisher}. However, a 1D system lacks a condensate and there is no obvious separation between the SF and the normal component of the fluid. The helicity modulus vanishes in the thermodynamic limit for all temperatures. Thus, at finite temperatures, the problem of defining superfluidity in 1D is a dynamical one. 

The first chapter of the thesis is based on current and anomalous transport in one dimensional systems (like the t-V model) at zero temperature, where currents of both chirality are present. We study the effect of a boost (Fermi sea displaced by a finite momentum) on one dimensional systems of lattice fermions with short-ranged interactions. In the absence of a boost such systems with attractive interactions possess algebraic superconducting order. Motivated by physics in higher dimensions, one might naively expect a boost to weaken and ultimately destroy superconductivity. However, we show that for one dimensional systems the effect of the boost can be to strengthen the algebraic superconducting order by making correlation functions fall off more slowly with distance. This phenomenon can manifest in interesting ways, for example, a boost can produce a Luther-Emery phase in a system with both charge and spin gaps by engendering the destruction of the former.

An electrical current set up in a superconductor continues to flow even in the absence of a driving electric field~\cite{santos_1,bagwell_1,tinkham_1}. Such a persistent current is equivalent to an imbalance in the number of carriers moving along and opposite to the direction of the current, i.e. a boost. In a 1D system, a boost can be realized with different chemical potentials for left and right movers. The magnitude of the boost or the current cannot be arbitrarily large and there is a critical value above which the superconducting state is destroyed. This phenomenon is analogous to the destruction of a superfluid when its flow velocity is larger than the critical velocity. The critical value of the boost can be calculated from the Bogoliubov- de Gennes equations for a superfluid~\cite{baym_1} and superconductor~\cite{wei_1,zagoskin_1}. However, there is a  difference between this phenomenon in SF and in SC. The quasi-particle excitations which makes the ground state unstable in a SF are the collective bosonic modes with linear dispersion and the velocity at which this occurs is the critical velocity.  However, for a SC, the critical velocity can be either the velocity at which the bosonic modes arise, or the Landau critical velocity at which the gap goes to zero and single particle excitations emerge, whichever is lower. 

A natural question is about the fate of superconductors, which do not have long range order (and hence order parameter equal to zero) upon the application of a boost. The most common example of such a system is a one dimensional system of fermions with attractive interactions~\cite{giamarchi_1,gnt_1,schulz_1,haldane_1}. Such one dimensional superconductors have recently come to the fore as they possess interesting topological properties such as the existence of Majorana edge modes under appropriate conditions~\cite{kitaev,nayak,kraus}. Experiments to detect these modes typically involve driving a current through the superconductor \cite{yazdani,shach} and hence it is germane to ask how large the critical current in these systems can be. Moreover, such systems have also been realized in cold atomic gases where it has been possible to make the system left-right asymmetric thereby producing a boost~\cite{zwierlein_1,esslinger_1,wang_1}. 

The critical velocity  $v_c$ at which gapless excitations emerge in the superconducting (SC) ground state, has been calculated in mean-field theory (Bogulibov De-Gennes theory for a 1D p-wave superconductor), and is found to be less than the Landau critical velocity $v_L$, precisely about $\frac{v_L}{\sqrt{2}}$~\cite{wei_1}. 
The above transition of the SC state to the normal state, due to the presence of a flow, is found to be a Clogston-Chandrasekhar type discontinuous  transition~\cite{cc1,cc2}. This occurs due to the competition between the condensation energy and the flow energy associated with the presence of a current in the SC. The equivalence between the exchange field in the Clogston-Chandrasekhar transition, and the flow energy in the boosted superfluid, can be understood from the following analogy. In the presence of the exchange field, the up and down spins in a Cooper pair are separated by energy $\mu_b B$. In the case of the boosted SF, the pairing of spinless electrons is between states $k+\frac{q}{2}$ and $-k+\frac{q}{2}$, which are separated by energy $\frac{2 k_F q}{m}$ near $k_F$. The exchange field in the superconductor is equivalent to the flow energy $k_F v$, where, $v=\frac{2 q}{m}$. In the former case, the normal state can be polarised, thus, reducing the energy by an amount proportional to $B^2$. In the latter case, the stationary normal state has an energy lower than the flowing normal state by an amount proportional to $q^2$. 

At $T=0$, the maximum equilibrium critical current occurs at a velocity $v_L/\sqrt{2}$. At this velocity, the SC state first becomes unstable, and eventually gives way to the normal state, via the pre-emptive transition at $v k_F= \triangle_0/\sqrt{2}$, where $\triangle_0$ is the gap at zero boost at $T=0$.
Contrary to this, the critical velocity in higher dimensions $v_c$ is found to be equal to the Landau critical velocity $v_L$~\cite{baym_1}. 
A similar calculation for a clean one dimensional superconductor incorporating the effects of quantum fluctuations has not been performed so far. However, it has been shown that phase slips induced by the contact of the superconductor with the walls of a container or the presence of statically irrelevant perturbations can dynamically destroy superconductivity at finite frequency and temperature in one dimension~\cite{oshikawa_1}. 

The momentum response function has been calculated in the presence of phase slips at finite temperatures, to study the response of the SF against the moving container~\cite{oshikawa_1},  
\begin{eqnarray}
\chi_{\mu \nu}({\bf r},t) &=& -i \hbar ^{-1} \vartheta(t) \langle\left[\pi_\mu({\bf r},t), \pi_\nu({\bf 0},0)\right]\rangle, \
(\mu,\nu=x,y,z)
\end{eqnarray}
where, 
\begin{eqnarray}
\pi({\bf r})=\frac{\hbar}{2i} \left[\Psi^{\dagger}({\bf r}) \nabla\Psi({\bf r})-\nabla\Psi^{\dagger}({\bf r}) \Psi({\bf r})\right].
\end{eqnarray}
In the absence of phase slips, $\chi(\omega,T)=\rm{lim}_{q\rightarrow0} \chi(q,\omega,T)=0$, implying that there is no normal component and the entire system behaves as a perfect SF at any $\omega$,$T$. In the presence of phase slips, the real part of $\chi(\omega,T)$ is related to the normal component $\rho_n$ and the imaginary part gives the dissipation. It was observed that as the probe frequency is decreased, the superfluid onset temperature decreases. In the limit when $\omega \rightarrow 0^+$, $\chi(\omega\rightarrow0,T)$ is non-zero. The dynamical SF response is thus observable even at very low probing frequencies $\sim 2$kHz, due to constrained dynamics in 1D, which leads to anomalously long lifetime of currents~\cite{taniguchi}.

In the second chapter, we answer the question of how a boost affects the quasi-long-ranged superconducting state in one dimension at zero temperature. Our main result is that the boost can have the counter-intuitive effect of {\em strengthening} the superconductivity (in a sense that we explain later) as opposed to weakening it like in higher dimensions. The boost eventually destroys superconductivity at a critical value but does so discontinuously when one of the Fermi points of the system is boosted to zero momentum. We demonstrate that a similar effect exists even for systems with quasi-long-ranged charge density wave order, i.e, the order is strengthened upon the application of a boost. We also show that for the boost to have any non-trivial effect, the underlying system has to have broken Galilean invariance, which is naturally realized in lattice systems. This has the additional effect of producing interesting phases at commensurate filling upon the application of a boost when umklapp is operative. These observations point to the possibility of new types of phase transitions that can be achieved by boosting the system. For example, we show that a system with a charge and spin gap can be boosted into a Luther-Emery phase \cite{le1_1,le2_1} by closing the charge gap.


\section{Majorana bound states in 1D superconductors}

\subsection{Majorana fermions and their origin}
Majorana fermions arise from the Dirac equation where all the $\Gamma$ matrices are imaginary, but still satisfying the Clifford algebra~\cite{pbpal}. Majorana solutions in the Dirac equation can be obtained from:
\begin{equation}
\left[i \tilde{\gamma}^\mu \delta_\mu -m\right]\Psi_{Maj}=0,
\end{equation}
where, $\tilde{\gamma}^\mu$ are imaginary, $\Psi_{Maj}=\Psi^{\dagger}_{Maj}$ with $m=0, E=0$.
Dirac fermion can be written in terms of two Majorana fermions:
\begin{eqnarray} \nonumber
\Psi_{Dir} &=& \frac{1}{2}(\Psi_{Maj,1}+i\Psi_{Maj,2}) \nonumber\\
\Psi^\dagger_{Dir} &=& \frac{1}{2}(\Psi_{Maj,1}-i\Psi_{Maj,2})
\end{eqnarray}
Majorana particles, being their own  anti-particles, can appear in superconductors in the form of Bogulibov quasi-particles. A quasiparticle excitation at energy $E$ above the BCS ground state (spinless), can be written as, 
\begin{align*}
|\psi\rangle =& \gamma^{\dagger}_E|BCS\rangle \\
=& (\alpha c^{\dagger}_{E,\sigma} + \beta c_{-E,\sigma'})|BCS\rangle  .
\end{align*}
In the above equation, $\gamma^\dagger_E=\gamma_E$ is only satisfied when $\sigma=\sigma'$, giving $\beta=\alpha^*$ and $E=0$ (for a symmetric spectrum). This is automatically true for p-wave SC. For s-wave, $\sigma=-\sigma'$ and hence, $\gamma^\dagger_E \neq\gamma_E$. Hence, it can be seen that one needs p-wave superconductivity to arrive at zero energy Majorana edge states. 

Majorana ``quasiparticles" are theoretically predicted in a number of condensed matter systems like interface of topological insulators with BCS superconductors~\cite{fu-kane}, quantum spin systems~\cite{kitaev_2D} and fractional quantum hall liquids ($\nu=5/2$)~\cite{read}. Several experiments have reported evidence of zero-energy states in nanowire-based systems~\cite{furdyna,deng}, where the appearence of zero energy tunneling resonance is regarded as evidence for the existence of the Majorana zero mode, Fig.~\ref{zerobiaspeak}. However, several other phenomena can give rise to zero-bias peaks, like, an ordinary localized state, the Kondo effect or Andreev states~\cite{potter}. Thus additional experimental signatures unique to Majorana zero modes are more desirable.

\begin{figure}[h]
\centering 
\includegraphics[width=15cm, height=10cm]{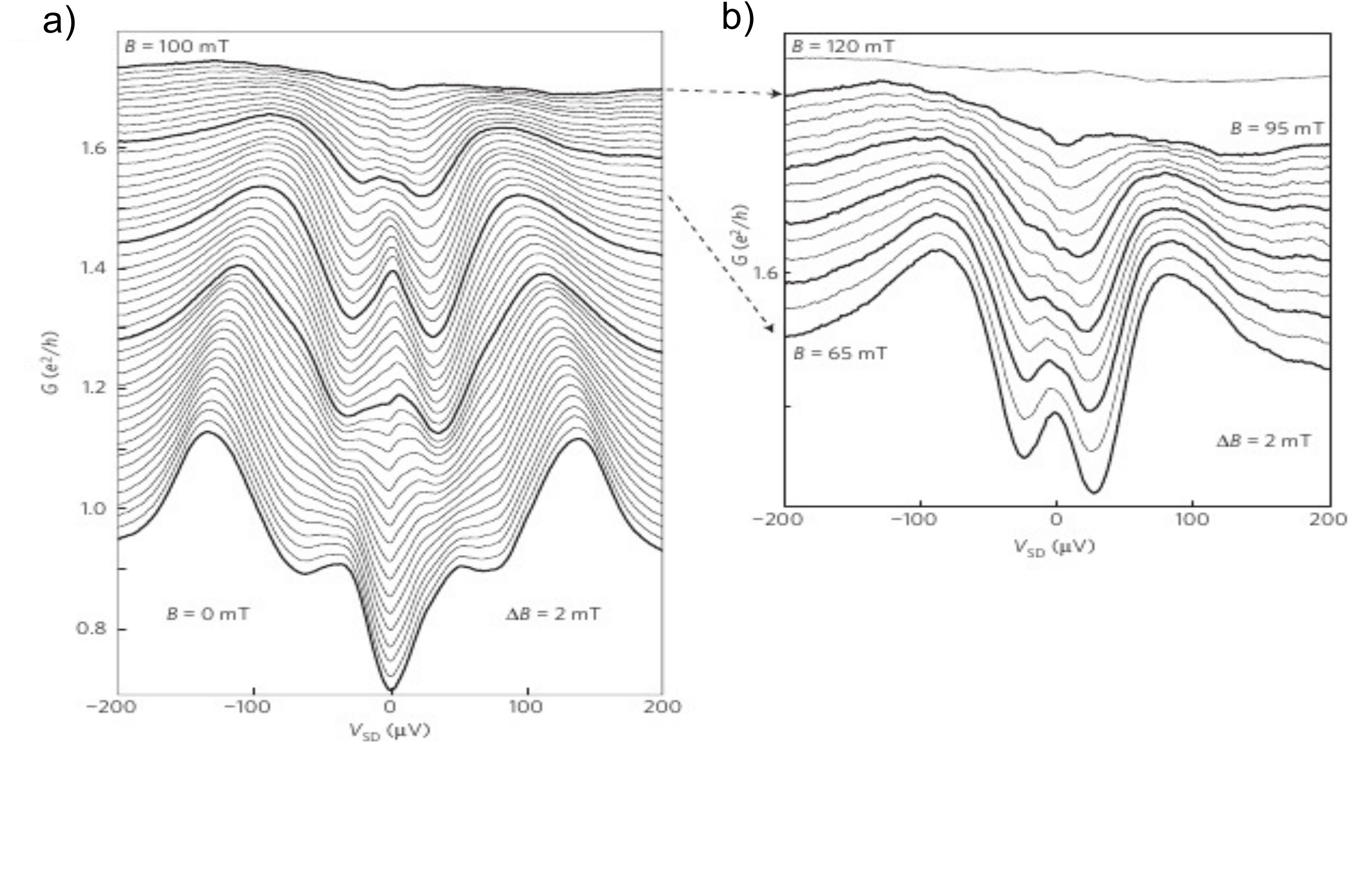} 
\caption{Low bias conductance as a function of applied magnetic field parallel to the wire axis in an induced 1D topological superconductor. A zero-bias peak is seen for the range of $\mu$ which lies within the topological gap, supporting the presence of Majorana fermions at the edges. \small{(Ref: A. Das {\it et.al}, Nat. Phys. {\bf 8}, 887 (2012))}}
\label{zerobiaspeak}
\end{figure}

\subsection{Spinless Kitaev model and Majorana}
One dimensional Kitaev chains (p-wave superconductor) are also  known to host Majorana modes at the edges ~\cite{kitaev,nayak,sau,tang,alicea}. Majorana edge modes can be calculated from the spinless p-wave Hamiltonian:
\begin{equation}\label{1Dpwave}
\mathcal{H} = \sum_k(\xi_k-\mu)c^{\dagger}_k c_k +\left[\triangle(k)c^{\dagger}_k c^{\dagger}_{-k}+\triangle^{*}(k)c_k c_{-k}\right]
\end{equation}
where, $\xi_k$ is the non-interacting electron dispersion, $\mu$ is the chemical potential and $\triangle$ is the p-wave order parameter.
For p-wave we need, $\triangle(k)=-\triangle(-k)$. Bogulibov transformation of Eq.~\ref{1Dpwave} gives,
\begin{equation} \label{bdgham}
\mathcal{H}=\sum_k \Psi^{\dagger}_k H_{BDG} \Psi_k,
\end{equation}
where, the basis is $\Psi_k=\left[c_k\ \ c^{\dagger}_{-k}\right]^{T}$, and, $H_{BDG} = \begin{pmatrix}
\epsilon(k) & \triangle(k) \\
\triangle^{*}(k) & -\epsilon(-k) \\
\end{pmatrix}$,
with, $ \epsilon(k)=-\epsilon(-k) = \frac{1}{2}(\xi_k-\mu)$.

Diagonalizing the BDG Hamiltonian gives the eigenvalues of the quasi-particles, $E(k)=\pm\sqrt{(\epsilon(k))^2+|\triangle(k)|^2}$. Introducing a lattice, and a phase dependent p-wave order parameter, $\triangle e^{i\phi}$, the dispersion now becomes, $E(k)=\pm\sqrt{(-t\cos{(k)}-\mu)^2+ |\triangle_0|^2 \sin^2{(k)}}$. It can be seen that a gap closing occurs at $|\mu|=\pm t$, separating the topologically trivial and non-trivial phase:

\begin{eqnarray} \nonumber
|\mu | &>& t\ \ \ \rm{Gapped}\ (E_+ - E_->0):\ \rm{{\bf trivial}} \nonumber\\
|\mu | &=& \pm t\ \ \ \rm{Gapless\ modes}\ (E=0): k=\pm\pi\ or\
k=0 \nonumber\\
|\mu | &<& t\ \ \ \rm{Gapped}\  \rm{{\bf topological}}\ \ (\triangle \neq 0)
\end{eqnarray}

Majorana states can be explicitly seen in the Kitaev model by mapping the 1D p-wave chain on to a chain of majorana modes by the following transformation,
\begin{eqnarray} \nonumber
c_x &=& \frac{e^{-i \phi}}{2}\left(\gamma_{B,x}+i\gamma_{A,x}\right) \nonumber\\
c^\dagger_x &=& \frac{e^{i \phi}}{2}\left(\gamma_{B,x}-i\gamma_{A,x}\right)
\end{eqnarray}
where, $c$'s are the operators for the Dirac fermions and $\gamma$'s for the Majorana fermions. Each Dirac fermion is composed of two types of Majoranas, $A$ and $B$. 
Replacing $c_x$ and $c^\dagger_x$ in the 1D Kitaev lattice model,
\begin{equation}
H_{Maj} = -\frac{\mu}{2}\sum_x^N (1+i\gamma_{B,x}\gamma_{A,x})
-\frac{i}{4}\sum_x^{N-1} (\triangle+t)\gamma_{B,x}\gamma_{A,x+1}+(\triangle-t)\gamma_{A,x}\gamma_{B,x+1}
\end{equation}
In the topological limit with $\mu=0, \triangle=t\neq0$,
\begin{equation}
H_{Maj}=-\frac{i}{2}\sum_x^{N-1} t\gamma_{B,x}\gamma_{A,x+1},
\end{equation}
indicating that the pairing between the Majoranas are such that two Majorana modes, one at each edge, is left unpaired. Each of these Majoranas are half of the Dirac fermion appearing in the Kitaev chain, Fig.\ref{kitaevchain}

\begin{figure}[h]
\centering 
\begin{tabular}{c} 
\includegraphics[width=7cm, height=5.2cm]{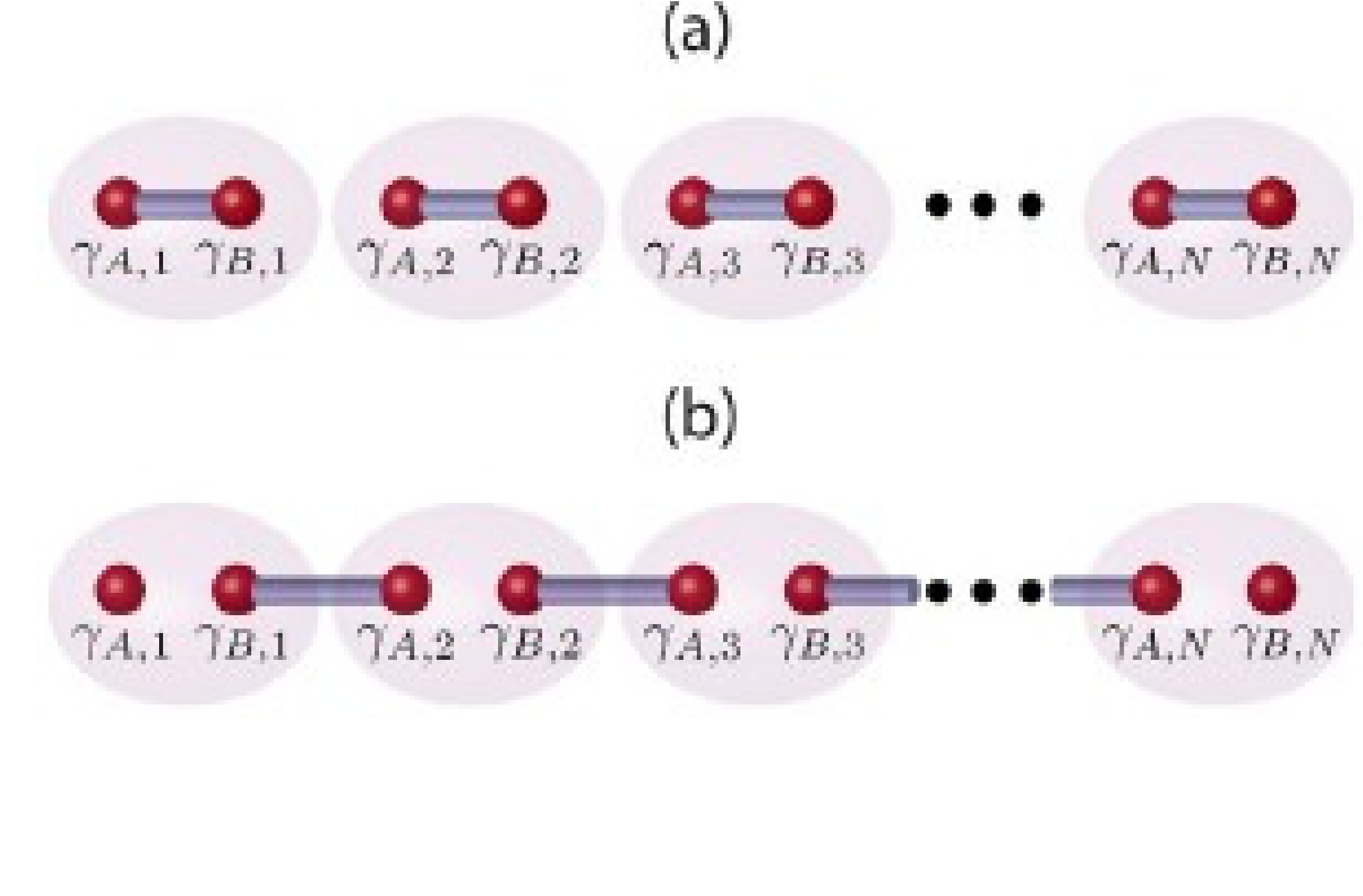} 
\includegraphics[width=8cm, height=5.2cm]{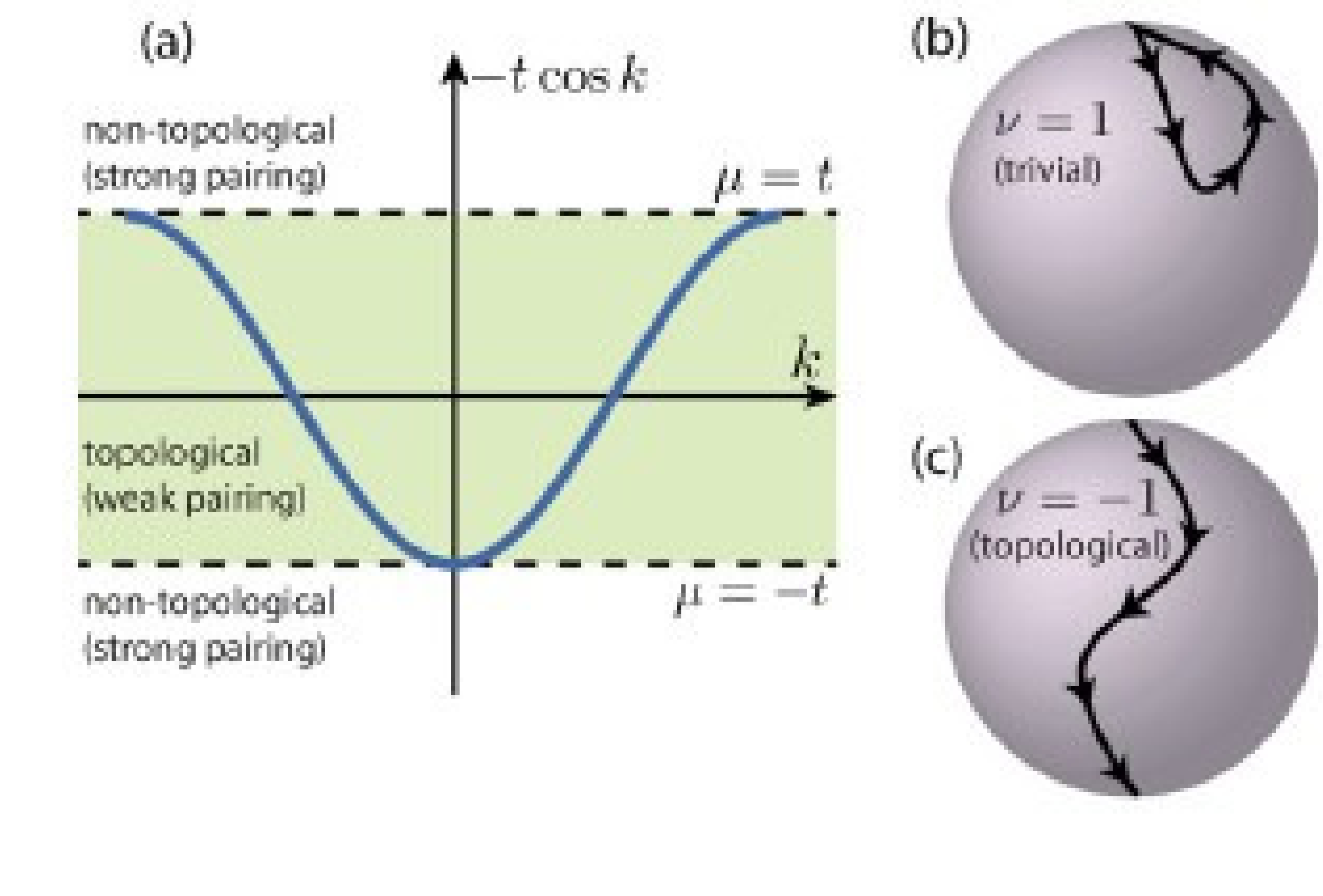}  
\end{tabular}
\caption{Left panel: Schematic diagram indicating the trivial gapped, with $\mu\neq0, \triangle=t=0$, and topological gapped, with $\mu=0, \triangle=t\neq0$, phases. Topological phase has two unpaired Majorana modes on the edges, one on each side. Right panel: K.E dispersion for a 1D spinless Kitaev chain is illustrated, where the p-wave pairing opens a bulk gap except at the chemical potential values $\mu=\pm t$. The system forms a non-topological strong pairing phase for $|\mu|>t$, and a topological weak-pairing phase for $|\mu|<t$. 
For a SU(2) Hamiltonian, $\mathcal{H}(k)=\mathbf{h}(k).\hat{\sigma}$ (where, $\hat{\sigma}$ are Pauli matrices), the topological invariant $\nu$ can be visualized by considering the trajectory that $\hat{h}(k)$ sweeps on the unit sphere as $k$ varies from $0$ to $\pi$; (b) and (c) in the right panel
illustrate the two types of allowed trajectories.
\small{[Ref: J. Alicea, Rep. Prog. Phys. 75, 076501 (2012)]}}
\label{kitaevchain}
\end{figure}

\subsection{p-wave superconductivity with spin}

The general form of a spin-full BdG Hamiltonian with p-wave pairing, in the absence of a magnetic field, is given by:
\begin{eqnarray} \label{pwave_BdG}
H_{BdG} &=& 
\begin{pmatrix}
\frac{\epsilon_{k,\uparrow}}{2} & \triangle_{\uparrow \downarrow}(k) & 0 & \triangle_{\uparrow \uparrow}(k)\\
\triangle^*_{\uparrow \downarrow}(k) & -\frac{\epsilon_{-k,\downarrow}}{2} & \triangle^*_{\downarrow \downarrow}(k) & 0\\
0 & \triangle_{\downarrow \downarrow}(k) & \frac{\epsilon_{k,\downarrow}}{2} & \triangle_{\downarrow \uparrow}(k)\\
\triangle^*_{\uparrow \uparrow}(k) & 0 & \triangle_{\downarrow \uparrow}(k) & -\frac{\epsilon_{-k,\uparrow}}{2}\\
\end{pmatrix}
\end{eqnarray}
in the basis, $\begin{pmatrix}
c^{\dagger}_{k,\uparrow} & c_{-k,\downarrow} & c^{\dagger}_{k,\downarrow} & c_{-k,\uparrow}
\end{pmatrix}$ . 

In the presence of two different species of electrons (say, spin) labelled by $\alpha$ and $\beta$, triplet pairing between them should follow:
$
\triangle_{\alpha \beta}({\bf k}) = -\triangle_{\alpha \beta}(-{\bf k})$, and,
$\triangle_{\alpha \beta}({\bf k}) = \triangle_{\beta \alpha}({\bf k}).
$

With the above properties, some of the possible pairings are: 
\begin{eqnarray} \label{3_pairing} \nonumber
\triangle_{\uparrow \uparrow}({\bf k})&=&\triangle_{\downarrow \downarrow}({\bf k}) \nonumber\\
\triangle_{\uparrow \uparrow}({\bf k})&=&-\triangle_{\downarrow \downarrow}({\bf k})\nonumber\\
\triangle_{\uparrow \downarrow}({\bf k})&=&\triangle_{\downarrow \uparrow}({\bf k})
\end{eqnarray}

The special cases of Hamiltonians with these possible pairings are:
\begin{eqnarray} \label{3_ham} \nonumber
\mathcal{H}_1 &=& \left(\frac{p^2}{2m}-\mu\right)\tau_z + \triangle_0 p\sigma_0 \tau_x \nonumber\\
\mathcal{H}_2 &=& \left(\frac{p^2}{2m}-\mu\right)\tau_z + \triangle_0 p\sigma_z \tau_x \nonumber\\
\mathcal{H}_3 &=& \left(\frac{p^2}{2m}-\mu\right)\tau_z + \triangle_0 p\sigma_x\tau_x
\end{eqnarray}
where, $\triangle_0$ gives the magnitude of the p-wave pairing, $\sigma$ is the Pauli matrix for spin and $\tau$ for particle-hole sector. In the first case, the pairing is between same species with same sign for $\uparrow$ spin and $\downarrow$ spin~\cite{flensberg}, whereas for the second case, they have opposite signs. In the third case, the pairing is of the form $\langle \hat{c}_\sigma({\bf k}) \hat{c}_{-\sigma}(-{\bf k})\rangle$~\cite{KSG}. 

\begin{figure}[h]
\centering
\includegraphics[width=8cm,height=8cm]{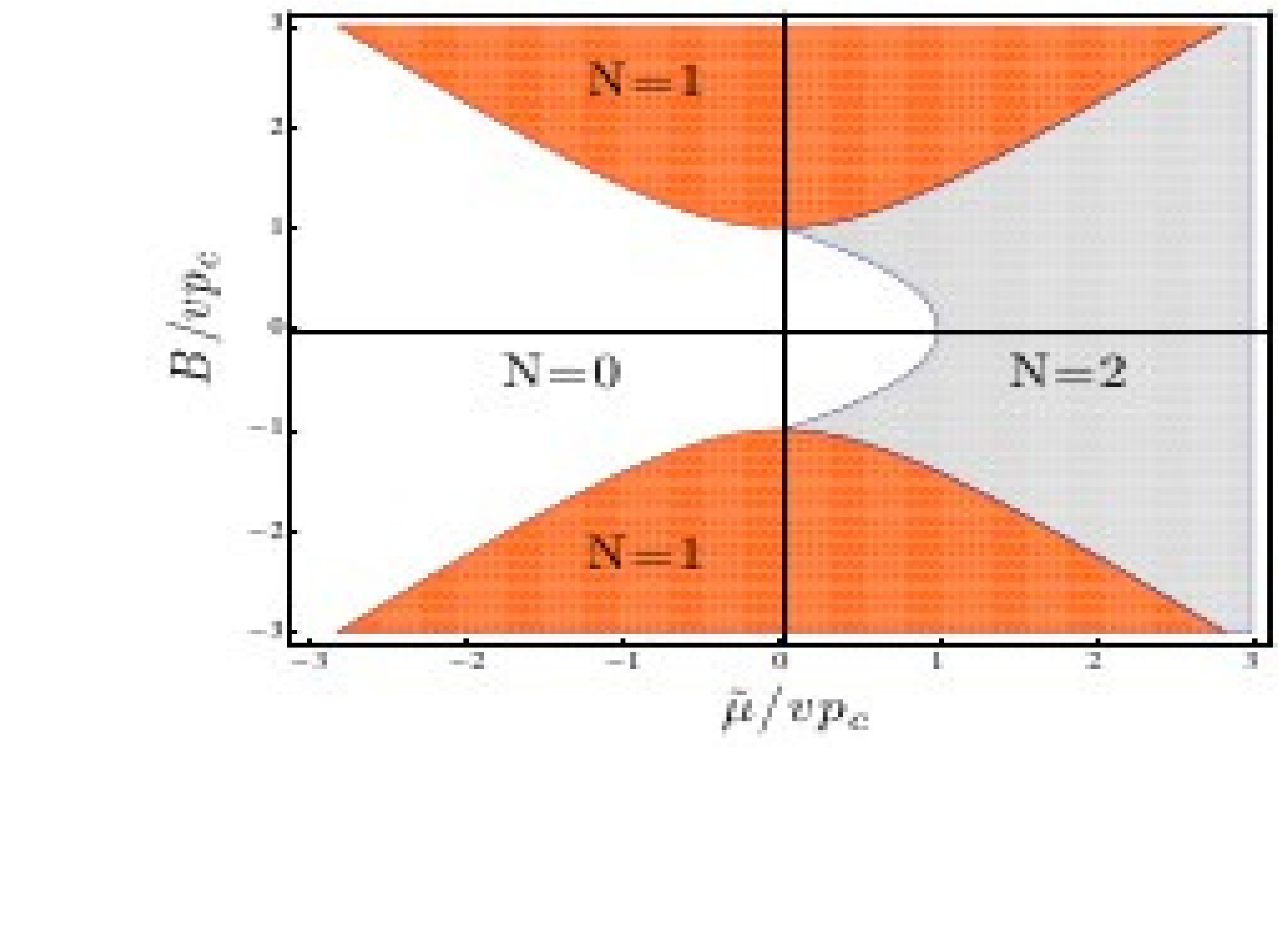}
\caption{Topological phase diagram (chemical potential $\mu$ versus magnetic field $B$) of the low energy model of a TR symmetric two-channel quantum wire, proximity coupled to a conventional s-wave SC, in the presence of a magnetic field, orthogonal to the $z$-axis. Orange regions have single localized MBS whereas the gray region has MBS doublets. \small{( Ref: E. Gaidamauskas {\it et.al} Phys. Rev. Lett. {\bf 112}, 126402 (2014).)}}
\label{phase_MBS}
\end{figure}

The first kind of Hamiltonian is realizable in the low energy sector of a time reversal (TR) symmetric two-channel quantum wire, proximity coupled to a conventional s-wave SC, and is found to support a Kramer's pair of Majorana bound states (MBS) in the topological phase~\cite{flensberg}. Transition between different topological phases, hosting two, one or zero number of MBS, can be effected by applying a magnetic field perpendicular to the spin-orbit direction, as shown in Fig.~\ref{phase_MBS}. It was shown by Tewari {\it et.al}~\cite{tewari1} that MBS in a TR symmetric 1D p-wave chain are topologically robust to perturbations which are TR symmetry breaking, like the magnetic field. It was identified that with perturbations such systems belong to the BDI symmetry class, whereas the TR symmetric Hamiltonian can be characterised as both BDI or DIII. Majorana Kramers pairs can still persist in the absence of TR symmetry if the chiral symmetry is preserved~\cite{tewari2}. This is because broken chirality may not be associated with broken TR symmetry and in such cases MBS survive due to the emergence of the $Z_2$ topological invariant in the system. It was also shown that for s-wave SC with spin-orbit coupling, which has MBS at the edges, only those perturbations which break TR and chiral symmetry simultaneously will split the MBS into finite energies.

In chapter 3, we study the type of Majorana zero modes present at the edges of 1D p-wave superconductors, having different kinds of spinless p-wave pairing ($\triangle_{\uparrow \uparrow}=\triangle_{\downarrow \downarrow}$ and $\triangle_{\uparrow \uparrow}=-\triangle_{\downarrow \downarrow}$). We study the effect of a Zeeman magnetic field on such spinless p-wave SC and the transition from one topological phase to another in the presence of such Zeeman terms. In addition, we analyse the effect of a s-wave term in each of the above cases and present a detailed symmetry classification of the different topological phases attained with combinations of the s-wave and Zeeman terms. In performing the classification, we have redefined the time reversal (TR) and the particle hole (p-h) operators in a general sense, such that the TR operator is an anti-unitary operator which does not mix the particle-hole sector and the p-h operator is an anti-unitary operator which mixes the p-h sector. For cases, where more than one anti-unitary operator satisfies the symmetry condition, the Hamiltonian needs to be block diagonalized into an irreducible form and an effective operator needs to be identified in each block to get the correct symmetry operation~\cite{tenfold}.
We show examples where in spite of TR symmetry being present, the breaking of chiral symmetry in each such block splits the MBS so they have finite energies.  For the spinless p-wave case without perturbations, we observe the presence of two types of MBS: a purely decaying MBS and a damped oscillating MBS, depending on the values of magnetic field $B$ and chemical potential $\mu$, and present phase diagrams for the same. Even in the presence of a Zeeman term ${\bf B}$, this nature of the MBS persists in each case, where the value of chemical potential $\mu$ and magnetic field ${\bf B}$ decides which type will appear.  Further, we identify the possible symmetry classes for the two cases (based on the ten-fold classification)~\cite{tenfold}, and also in the presence of perturbations like an s-wave pairing and various terms involving magnetic field. It is seen that in the presence of an s-wave perturbation, the MBS will now have only one particular nature, the damped oscillating behaviour, unlike that for the unperturbed p-wave case.

\section{SU(3) topological insulators}

\subsection{Quantum Hall effect and Chern number}
The quantum hall effect ushered in a new era in condensed matter physics in the 1980's. It is the remarkable phenomenon of quantization of hall conductance in 2D systems under strong magnetic fields and very low temperatures. The electrons in the bulk form circular localized orbit, rendering the bulk insulating, while the edge remains conducting~\cite{klitzing}. Such phases of matter are associated with certain topological characteristics, which are robust to any perturbations that does not close gap in the band structure of the system. These systems are characterized by certain topological invariants called Chern numbers, which is also related to the number of edge modes in the system ~\cite{TIreviewCK}.

On a lattice, the Hall conductance is given by the Thouless-Kohmoto-Nightingale-den Nijs (TKNN) formula and is called the Chern number~\cite{haldane2_1,TKNN}. 
In the presence of a lattice, the Bloch state $|u(k)\rangle$ is invariant under transformations of the form, $|u(k)\rangle \rightarrow e^{i \phi(k)} |u(k)\rangle $, which is similar to an electromagnetic gauge transformation in momentum space. The Berry connection given by,
\begin{equation}
\mathcal{A}({\bf k}) = -i\langle u(k)|\nabla_k|u(k)\rangle
\end{equation}
is analogous to the electromagnetic vector potential, and under the above transformation of the Bloch state, $\mathcal{A}\rightarrow\mathcal{A}+\nabla_{\bf k}\phi({\bf k})$. Even though $\mathcal{A}$ is not gauge invariant, the integral of $\mathcal{A}$ over a closed loop (analog of the magnetic flux) is.
The integral of the Berry curvature,  $\mathcal{F}=\nabla\times\mathcal{A}$, over a closed path in the Brillouin zone gives the Chern number, which is an integer,
\begin{equation}
n=\frac{1}{2\pi}\int_{BZ} d^2{\bf k} \mathcal{F},
\end{equation}
and characterizes the topological properties of the system. This can be physically visualized in the following way. For a SU(2) Hamiltonian, $\mathcal{H}(k)=\mathbf{h}(k).\hat{\sigma}$ (where, $\hat{\sigma}$ are Pauli matrices), the Chern number $\nu$ can be visualized as the number of times $\hat{h}(k)$ wraps around a unit sphere as $\vec{k}$ goes around the toroidal Brillouin zone once (The reciprocal lattice vector in the Brillouin zone, $\vec{k}=(k_x,k_y)$ lives in a torus, with $k_x=k_x+\frac{2\pi}{a_x}, k_y=k_y+\frac{2\pi}{a_y}$). 
 
\subsection{Haldane model and Chern insulator}

\begin{figure}[h]
\centering 
\includegraphics[width=8cm, height=5.5cm]{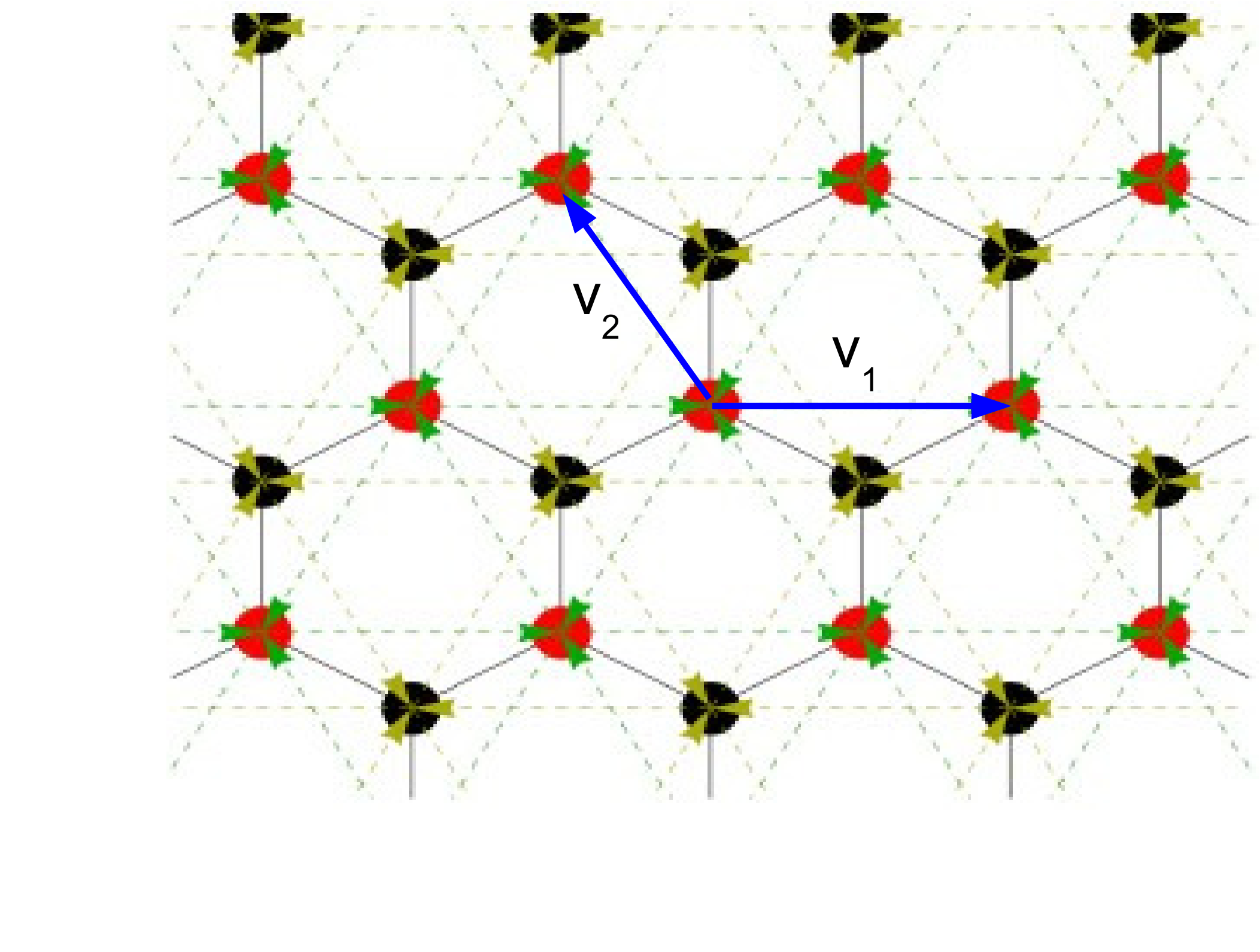}
\caption{Haldane model: A honeycomb structure with two sublattices A (red) and B (black) with nearest neighbour (NN) and next nearest neighbour hopping (NNN). $\vec{v}_1$ and $\vec{v}_2$ denotes the two lattice vectors.}
\label{haldane_lattice}
\end{figure}

The Haldane model~\cite{haldane2_1} is a lattice model, with nearest neighbour (NN) and next nearest neighbour hopping (NNN) on a honeycomb lattice with two sub-lattices Fig.~\ref{haldane_lattice}, exhibiting non-trivial topological properties in the absence of a magnetic field. The complex nature of the NNN hopping breaks TR symmetry, whereas the asymmetry of the on-site potential of the two sub-lattices breaks inversion symmetry. The lattice vectors are: $\vec{v}_1=\{\sqrt{3}a,0\}$ and $\vec{v}_2=\{-\frac{\sqrt{3}}{2}a,\frac{3}{2}a\}$, where, $a$ is the NN distance. The Hamiltonian of this model is:
\begin{eqnarray} \label{haldane_eq}
H &=& H_{NN}+H_{NNN} \nonumber\\
&=& \begin{pmatrix}
a^{\dagger}_k & b^{\dagger}_k
\end{pmatrix}
\begin{pmatrix}
H_{11}(k) & H_{12}(k)\\
H_{21}(k) & H_{22}(k)
\end{pmatrix}
\begin{pmatrix}
a_k\\
b_k\\
\end{pmatrix}
\end{eqnarray}
and, 
\begin{align*}
H_{11}(k) &=\triangle_0/2-2t'\sum_i\cos{(\vec{k}.\vec{v}_i-\phi)}, & H_{12}(k) &= -t\sum_i c^{-i\vec{k}.\vec{e}_i}\\
H_{22}(k)&=-\triangle_0/2-2t'\sum_i\cos{(\vec{k}.\vec{v}_i+\phi)}, & H_{12}(k) &= -t\sum_i c^{i\vec{k}.\vec{e}_i}
\end{align*}
where, $\triangle_0$ is the difference in the on-site potential of A and B atoms, and $\vec{e}_i$ is the lattice vector between NN. Eq.~\ref{haldane_eq}, can be written as,
\begin{equation}
H(\vec{k})=a(\vec{k})\mathcal{I}+d(\vec{k}).\vec{\sigma}
\end{equation}
where, $\vec{\sigma}$ are the Pauli matrices, $a(\vec{k})=\frac{(H_{11}+H_{22})}{2}$, and, $d(\vec{k})=\frac{(H_{11}-H_{22})}{2}\hat{\sigma}_z + \frac{(H_{12}+H_{21})}{2}\hat{\sigma}_x +\frac{i(H_{12}-H_{21})}{2}\hat{\sigma}_y$. The Berry curvature $\mathcal{F}(\vec{k})$ can then be expressed as:
\begin{equation}
\mathcal{F}(\vec{k})=\pm \frac{1}{2\pi}\hat{d}(\vec{k})\left[\partial_{k_x} {\bf \hat{d}}(\vec{k}) \times \partial_{k_y} {\bf \hat{d}}(\vec{k})\right],
\end{equation}
which, on integrating over a closed path in the Brillouin zone gives a non-zero Chern number.

At the Dirac points, $K$ and $K'$, the gap opens either due to breaking of TR symmetry or inversion symmetry. If the band dispersion is given by $\epsilon_\pm(k)$, then, 
the quantities, $\epsilon_+(K)-\epsilon_-(K)$ and $\epsilon_+(K')-\epsilon_-(K')$, having the same sign indicates trivial insulator and opposite sign indicates Chern insulator, with the Chern number being $\pm1$.


\subsection{Hofstadter model}
The Hofstadter model ~\cite{hofstadter} describes yet another system which exhibits the robust quantization of the Hall conductivity. The Hamiltonian of free electrons hopping on a square lattice in the presence of a uniform magnetic field (with vector potential $\mathbf{A} $) is:
\begin{equation} \label{hofstadter1}
H = -t \sum_{<r,r'>} c^{\dagger}_{r'} e^{-i \int {\bf A}.{\bf dl}} c_r +h.c
\end{equation}
With the Landau gauge, the Eq.~\ref{hofstadter1} becomes,
\begin{equation}
H = -t \sum_{<r,r'>} c^{\dagger}_{r'+x} c_r + e^{i \Phi x}c^{\dagger}_{r'+y} c_r + h.c
\end{equation}
where, $\Phi$ is the flux quanta per unit cell. When $\Phi$ is not a rational number, the energy $E$ vs. $\Phi$ shows a fractal structure as shown in Fig.~\ref{hofstadter}. For fixed rational flux, $\Phi=\frac{2\pi p}{q}$, a gapped spectrum with $q$ bands is observed. 

Similar to the Haldane model, the Chern number shows the topologically non-trivial character of the model. However, the calculation in this case is more complicated than the earlier case~\cite{galitski}. $N$ coupled copies of the Hofstadter model with $\frac{2\pi}{N}$ Abelian flux per plaquette, has non-zero Chern number for each $N$. This non-uniform gauge field, can be written in terms of $\hat{S}_z$ (where, $\hat{S}_z = \rm{diag}(s,s-1,...-s)$ with $2 s+1=N$), with only diagonal elements. Under a gauge transformation, this $N$ component Hofstadter model in the presence of a non-uniform Abelian gauge field, can be converted to an $N$ component Hofstadter model with a uniform non-Abelian gauge field. The lowest value of $N$ where a non-zero Chern number appears, in the presence of a uniform non-Abelian gauge field, is at $N=3$. It was shown by Barnett {\it et.al}~\cite{galitski} that such a $N$-component Hofstadter model with $2\pi/N$ Abelian flux is equivalent to a SU(N) model trapped in a uniform SU(N) gauge field, where the topological character of the former for $N>2$ is inherited by the latter system.  

Bloch Hamiltonian on a square lattice with nearest neighbour hopping in the presence of a homogeneous SU(N) gauge field is:
\begin{equation} \label{bloch_su3}
H({\bf k}) = -2t\left[\cos{(k_x-\hat{A}_x)}+\cos{(k_y-\hat{A}_y)}\right]
\end{equation}
where, $t$ is the hopping and $\hat{A}_{x,y}$ are constant $N\times N$ Hermitian matrices.
For SU(2) systems, the cosines in Eq.~\ref{bloch_su3} can be expanded and the Hamiltonian can be rewritten as:
\begin{equation} \label{su2_ham}
H({\bf k})= a({\bf k})+{\bf b(k)}.\hat{\sigma}
\end{equation} 
The Berry curvature can be obtained from the projection operators $\mathcal{P_\pm({\bf k})}$, corresponding to the two eigenstates, with $\mathcal{P_\pm({\bf k})}=\frac{1}{2}\left[1\pm
{\bf b}(k).\hat{\sigma}/|{\bf b(k)}|\right]$,
\begin{equation} \label{berry_su2}
\Omega_\pm ({\bf k})=\pm \frac{{\bf b(k)}}{2|{\bf b(k)}|^3}.\left[\partial_{k_x} {\bf b(k)} \times \partial_{k_y} {\bf b(k)}\right]
\end{equation}

\begin{figure}[h]
\centering 
\includegraphics[width=12cm, height=7.5cm]{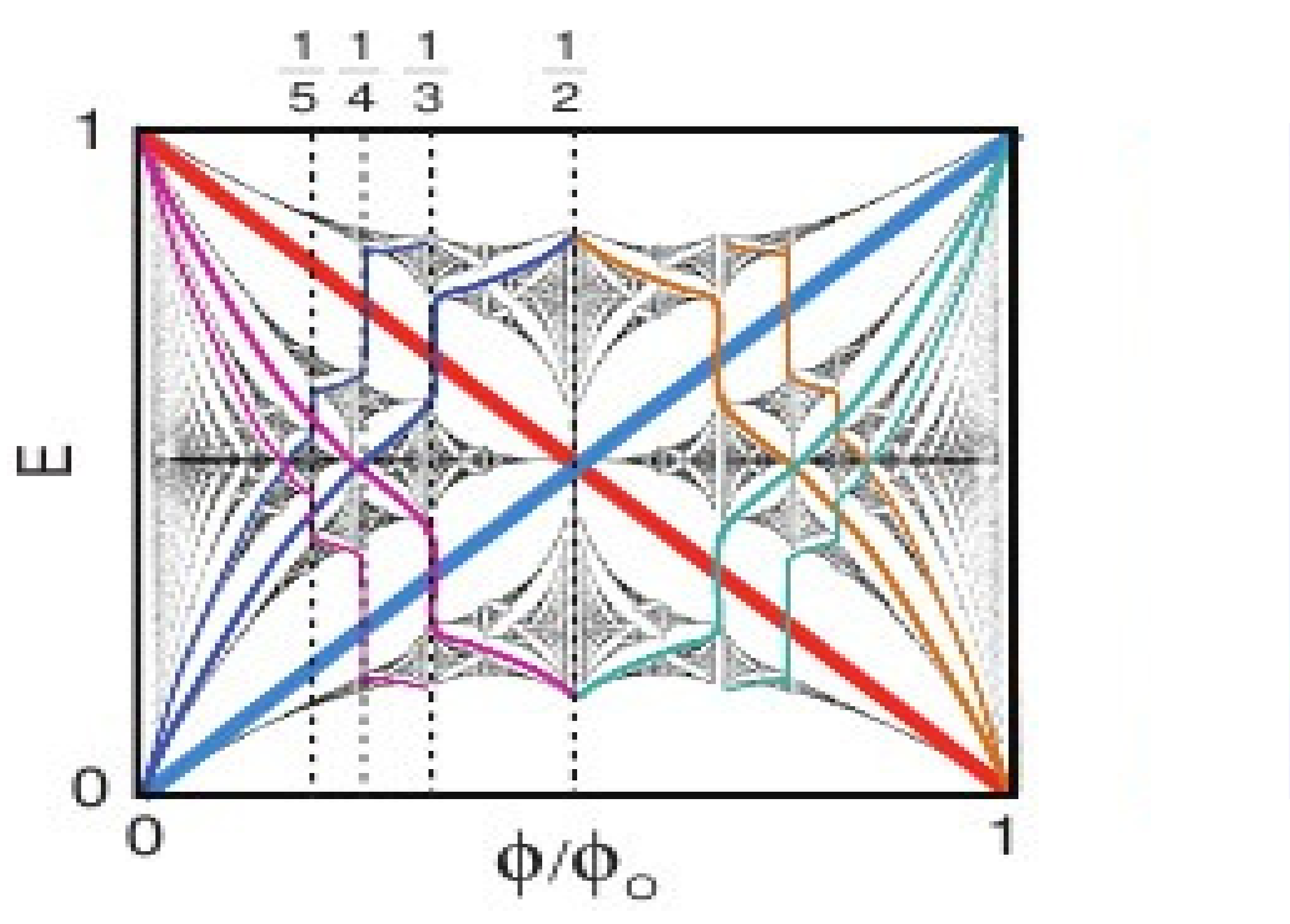}
\caption{Hofstdater butterfly: Energy $E$ vs. $\Phi$ for free electrons in a square lattice, in the presence of magnetic flux $\Phi$ per unit cell. At rational fluxes $\Phi=1/2,1/3,1/4,1/5...$, the spectrum is gapped, for irrational values of flux, it shows a fractal structure. Ref: {\small \url{
www.physics.iisc.ernet.in/~qcmjc/talk_slides/QCMJC.2013.08.22_Yinghai.pdf}}}
\label{hofstadter}
\end{figure}

In SU(2) systems, arbitrary gauge fields in the Hamiltonian can be expressed as linear combinations of the Pauli matrices~\cite{galitski}, such that $\hat{A}_{x,y}=u_{x,y}+\mathbf{v}_{x,y}.\hat{\mathbf{\sigma}}$, which enter the tight-binding lattice Hamiltonian in the following form:
\begin{eqnarray}
H &=& -t\sum_{i} \left(\Psi^{\dagger}_ie^{-i \hat{A}_x}\Psi_{i+\hat{x}}+\Psi^{\dagger}_ie^{-i \hat{A}_y}\Psi_{i+\hat{x}}+h.c\right) \nonumber\\
&=& \sum_k \Psi^{\dagger}_{\bf k} \mathcal{\hat{H}}({\bf k})\Psi_k
\end{eqnarray}

On expanding the exponents to obtain Eq.~\ref{su2_ham}, it can be seen that $\partial_{k_x}\mathbf{b}({\bf k}) \times \partial_{k_y}\mathbf{b}({\bf k}) \propto {\bf v}_x \times {\bf v}_y$, thus rendering the Berry curvature to go to zero for SU(2) systems. This can also be seen physically, where, for $N=2$, the flux per each plaquette is $\pi$, which is itself TR invariant (flux $\phi$ has to be equal to $\phi+2\pi$).
Hence, $N>2$ needs to be considered for topologically non-trivial results, of which the simplest is $N=3$. 

The SU(3) Hamiltonians referred to in the work~\cite{galitski} and in our present work, are $3\times3$ Hamiltonians which can be expressed in terms of the Gell-mann matrices $\hat{\lambda}$ and the identity, 
\begin{equation} \label{su3_ham}
H({\bf k})= a({\bf k})\mathcal{I}+{\bf b(k)}.\hat{\lambda},
\end{equation}
and are not invariant under SU(3) transformations in general. The ${\bf b(k)}$ vector lies on the surface of the eight dimensional sphere of the generators. The term ${\bf b(k)}.\hat{\lambda}$ has an appearance of a spin-orbit coupling and can be used to create the same for spin-full systems.

A SU(N) spin-orbit coupling can be generated in ultra-cold atomic systems, by coupling the internal degrees of freedom of the trapped atoms (N component atoms) to a non-Abelian SU(N) gauge field. More specifically, it was shown that a SU(3) spin-orbit coupling gives rise to a qualitatively different phenomena, with a non-trivial topological state arising in a square lattice, even in the presence of a homogeneous SU(3) gauge field,~\cite{galitski}. This is different from the SU(2) case, where homogeneous gauge fields give topologically trivial phases. 

The experimental realization of such synthetic spin orbit coupling (SOC) have come to the fore with the advent of cold atom physics. Such systems are relevant due to their tunability and wide range of applications in engineering and simulating condensed matter systems~\cite{mazza,bloch}. Furthermore, SOC provides a common origin to a variety of quantum Hall and topological classes of materials.\cite{TIreviewTD,TIreviewCK,TIreviewSCZ} In all these examples, the spinor of the SU(2) representation comprises of two chiral species originating from the momentum locking with two sublattices, or two orbitals, or spin-1/2 particles. As a chiral object forms an orbit, surrounding an external magnetic field, or momentum space Berry curvature, it produces a non-zero flux, which is quantified by the winding number or Chern number (topological invariants). 

\subsection{SU(3) topological insulator} 
Encouraged by the tremendous success of material realization and engineering of the SU(2) based topological systems~\cite{TIreviewTD,Ando,EngineeringTI}, in chapter 4 we study a SU(3) topological insulator (where atoms with three internal degrees of freedom are trapped in a SU(3) gauge field) and illustrate an experimental scheme to realize an equivalent lattice model.  Prediction of a SU(3) topological insulator (TI) in condensed matter systems is rather limited,\cite{galitski,jpsj} and few physical systems or optical lattices are realized to date with such properties. An explicit topology-engineering scheme in a three band model has been discussed in a recent work~\cite{jpsj},  where the generation of an arbitrary Chern number is based on the equivalence of the topological number of a given band on the monopole charge, and can be extended to higher Chern number derivatives. The feasibility of the material realization of the corresponding models has not been discussed in the existing literature.

We start with a generalized form of the SU(3) Hamiltonian Eq.~\ref{su3_ham}, where ${\bf b}({\bf k})$ is the eight component vector, which in our case represents electron hopping in a lattice. 
The product of two Gell-mann matrices can be written as $\hat{\lambda}_a\hat{\lambda}_b=\frac{2}{3} \delta_{ab}+d_{abc}\hat{\lambda}_c+if_{abc}\hat{\lambda}_c$, where $d_{abc}$ and $f_{abc}$ are symmetric and anti-symmetric structure constants of SU(3) respectively,~\cite{galitski,mukunda}. These structure constants define three bilinear operations for the eight-component vectors, which are, $ \mathbf{u\cdot v}=u_av_a$, $(\mathbf{u}\times\mathbf{v})_a=f_{abc}u_bv_c$ and the star product $(\mathbf{u}*\mathbf{v})_a=\sqrt{3}d_{abc}u_bv_c$, where $\mathbf{u}$ and $\mathbf{v}$ are two arbitrary vectors. It can be seen here that for SU(2) systems, only the first two operations exist, with $f_{abc}$ being the usual Levi-civita symbol. 

The eigenstate projection operator for the SU(3) case can be written in terms of the Gell-mann matrices:
\begin{equation} \label{proj_su3}
\hat{P}_{{\bf k},n}=\frac{1}{3}\left(1+\sqrt{3}\mathbf{n}_{{\bf k},n}.\mathbf{\hat{\lambda}} \right),
\end{equation}
where, $\mathbf{n}_{{\bf k},n}$ lies on the surface of the eight-dimensional sphere of the $\mathbf{\hat{\lambda}}$ vectors, and $Tr\hat{P}_{{\bf k},n}=1$. The condition $(\hat{P}_{{\bf k}n})^2=\hat{P}_{{\bf k}n}$ gives two constraints on $\mathbf{n}_{{\bf k},n}$. They are $\mathbf{n}_{{\bf k},n}\cdot\mathbf{n}_{{\bf k},n}=1$ and $\mathbf{n}_{{\bf k},n}*\mathbf{n}_{{\bf k},n}=\mathbf{n}_{{\bf k},n}$.
To express $\mathbf{n}_{{\bf k},n}$ in terms of the $\mathbf{b}({\bf k})$'s, the relation $\left[\hat{P}_{{\bf k},n},\mathcal{\hat{H}}({\bf k})\right]=0$ can be used, (true for projection operators for the eigenstates) which leads to the condition, $\mathbf{b}({\bf k})\times \mathbf{n}_{{\bf k},n}=0$. This relation along with the above conditions on $\mathbf{n}_{{\bf k},n}$, give the resulting expression for the Berry curvature which can be obtained from the projection operator Eq.~\ref{proj_su3},
\begin{align} \label{su3_berry}
\Omega_n({\bf k}) &= -\frac{4}{3^{3/2}}f^3_{1{\bf k}n} \bigg(f_{2{\bf k}n}^2 \partial_{k_x} {\bf b} \times \partial_{k_y} {\bf b} + f_{2{\bf k}n} \partial_{k_x} {\bf b} \times \partial_{k_y} ({\bf b}*{\bf b}) 
+ f_{2{\bf k}n}  \partial_{k_x} ({\bf b}*{\bf b}) \times \partial_{k_y}{\bf b} \nonumber \\
&+ \partial_{k_x} ({\bf b}*{\bf b}) \times \partial_{k_y}({\bf b}*{\bf b}) \bigg). \bigg(f_{2{\bf k}n} {\bf b}+({\bf b}*{\bf b})\bigg)
\end{align}
where, $f_{1{\bf k}n}=\frac{1}{|b(k)|^2\left(4 \cos^2{(\theta_k+\frac{2\pi}{3} n)}-1\right)}$, $f_{2{\bf k}n}=2|b(k)|\cos{\big(\theta_k+\frac{2\pi}{3}n\big)}$, and, $\theta_k=\frac{1}{3}\arccos{\left[\frac{b(k).b(k)*b(k)}{|b(k)|^3} \right]}$ ($n$ runs from one to three). For SU(2) systems, the above equation reduces to the familiar form with only the cross product term in Eq.~\ref{berry_su2}, with $f_{1{\bf k}n}$ and $f_{2{\bf k}n}$ defined accordingly.

The flux of the Berry curvature $\Omega({\bf k}^*)$ ($k$-points where it attains peaks is denoted by $k^*$) through the first Brillouin zone for each band is quantized, and is quantified by the associated Chern number. Therefore, the Chern number essentially dictates the number of orbit centers, while its sign corresponds to the direction of associated phase winding.
For banded system, total chern number has to vanish with or without a magnetic field. Without magnetic field, the Chern number for each band is zero. In the presence of TR symmetry, the bands need to appear in Kramer's pairs, with chern numbers $C$ and $-C$. In such cases, one could have a Quantum Spin Hall effect~\cite{TIreviewCK,TIreviewSCZ}.

The above-mentioned features can also be understood simply from the corresponding band topology. The center of orbits at ${\bf k}^*$-points are those discrete points where band degeneracy occurs. Therefore, the number of vortices dictate the number of band inversions in both $k_x$ and $k_y$ directions. In SU(2) topological systems, the odd number of band inversions (at the TR invariant momenta, if this symmetry is present) between two basis components gives a finite Chern number or $Z_2$ invariant. For SU(3) systems, the band inversion must happen between all three bands, or at least, the band with the highest Chern number ($C_1$) must undergo inversion with {\it both} the other two bands. This is an important distinction of the SU(3) framework. Another requirement of the SU(3) topological state is that here not only a gauge field is required to be present in the off-diagonal term of the Hamiltonian, but an odd parity Zeeman-like term, i.e. $\sin{k}\sigma_z$ term must also be present between any two basis. Such an odd parity Zeeman-like term does not arise naturally from nearest neighbour hopping in a lattice.

In chapter 4, we study a minimal model realization of a SU(3) topological insulator, using atoms in a square lattice coupled to a homogeneous SU(3) gauge field. Unlike SU(2) systems, where non-trivial topology arises only with inhomogeneous gauge fields (where hopping amplitude becomes position dependent), SU(3) systems have finite Chern numbers even with homogeneous gauge fields. (We re-emphasize that the SU(3) system referred to here does not necessarily have SU(3) symmetry. The name only suggests that the Hamiltonian can be expressed as linear combinations of the Gell-mann matrices and the identity.) We study the specific criteria and the particular terms necessary in the Hamiltonian to get a non-zero Chern number. Another focal point of our work is to construct a lattice model which inherits these features -- suitable gauge fields and odd-parity Zeeman term -- and thus intrinsically perform as a topological material. We present a model Hamiltonian of a spinless SU(3) topological system in a tripartite lattice with three inequivalent sub-lattices. Each sublattice is sitting in distinct 1D chains, and they are coupled by nearest-neighbor (NN) quantum tunneling. This would naturally give an uncompensated Bloch phase (in momentum space, NN hopping between two components is $\sim t e^{i k}$, which when not associated with a $t e^{-i k}$, will give a $\cos$ term and a $\sin$ term, the latter of which is required in the off-diagonal element)  for all three inter-site hoppings as: $e^{i{\bf k}.{\bf r}_{12}}$, $e^{i{\bf k}.{\bf r}_{23}}$, and, $e^{-i{\bf k}.{\bf r}_{13}}$, where ${\bf r}_{ij}$ are distances between $i$ and $j$ sublattice belonging to nearest neighbour unit cells. 
Finally, for the generation of the odd parity `onsite' matrix-element, we propose to utilize the interaction of an electron with a constant vector potential with spatially dependent polarization, such that it encloses a non-zero flux. We show with a modified tight-binding model in a tripartite lattice that the interaction term $\propto {\bf k}\cdot{\bf A}$, where ${\bf A}$ is the vector potential, naturally leads to a $\pm\sin{(\beta {\bf k}\cdot\hat{\epsilon})}$ term (where $\hat{\epsilon}$ is the unit vector along the polarization direction, and $\beta$ is a tunable parameter). Tuning the direction of the polarization parallel or orthogonal to ${\bf k}$ (or with other methods as discussed in chapter 4), one can selectively generate, reverse, and destroy this term in different sublattices. This is a simplistic picture for visualizing the odd-parity Zeeman term, by using opposite polarization in two sub-lattice sites, while it is destroyed in the third site. This generates the SU(3) topological system, which we characterize by the detailed analysis of the Berry curvature, Chern number, and edge states. 



\cleardoublepage

}


%% file: mychapter2.tex
\chapter{{\LARGE Bosonization and boosted superfluids in one-dimension}\\
\vspace{0.5cm}
\small{Ref: S. Ray, S. Mukerjee, and, V. B. Shenoy, Annals of Physics, {\bf 384}, 71-84 (2017)}} \label{chapter2}
{\large

\section{Summary of main results}
We study the effect of a boost (Fermi sea displaced by a finite momentum) on one dimensional systems of lattice fermions with short-ranged interactions. In the absence of a boost such systems with attractive interactions possess algebraic superconducting order. Motivated by physics in higher dimensions, one might naively expect a boost to weaken and ultimately destroy superconductivity. However, we show that for one dimensional systems the effect of the boost can be to strengthen the algebraic superconducting order by making correlation functions fall off more slowly with distance. This phenomenon can manifest in interesting ways, for example, a boost can produce a Luther-Emery phase in a system with both charge and spin gaps by engendering the destruction of the former.

\section{Outline}
In spite of the failure of perturbation theory in one dimension, one can have an exact treatment of interactions by using a low energy effective field theory called bosonization~\cite{giamarchi,gnt,schulz}. All excitations can be described in terms of certain bosonic fields, and even interacting systems with forward scattering terms can yield quadratic Hamiltonian (Properties like nesting, make the treatment of interactions in one-dimension much simpler, effectively leaving out interactions with higher order momentum exchange ($> 2 k_F$)). Since in one dimension the Fermi surface is reduced to Fermi points at $+k_F$ and $-k_F$, all low energy excitations are restricted to momentum exchange $k \sim 0$ or $k\sim 2 k_F$.

The chapter is arranged as follows. In Sec.~\ref{spinless_model}, we discuss the technique of bosonization in a 1D spinless fermionic system (including interactions like forward scattering) and elaborate on the nature of correlations. In Sec.~\ref{model_umklapp}, we extend our discussion to back-scattering interaction terms, in the presence of a lattice or spin-exchange term (umklapp), and review the RG analysis. In Sec.~\ref{current_1D}, we present our work on boosted one dimensional fermionic superfluids on a lattice, and explore the effect on the correlation functions of such 1D systems. In Sec.~\ref{conclude}, we conclude by summarizing our results.

\section{Spinless model} \label{spinless_model}
\begin{figure}[h]
\centering
\includegraphics[width=0.60\linewidth]{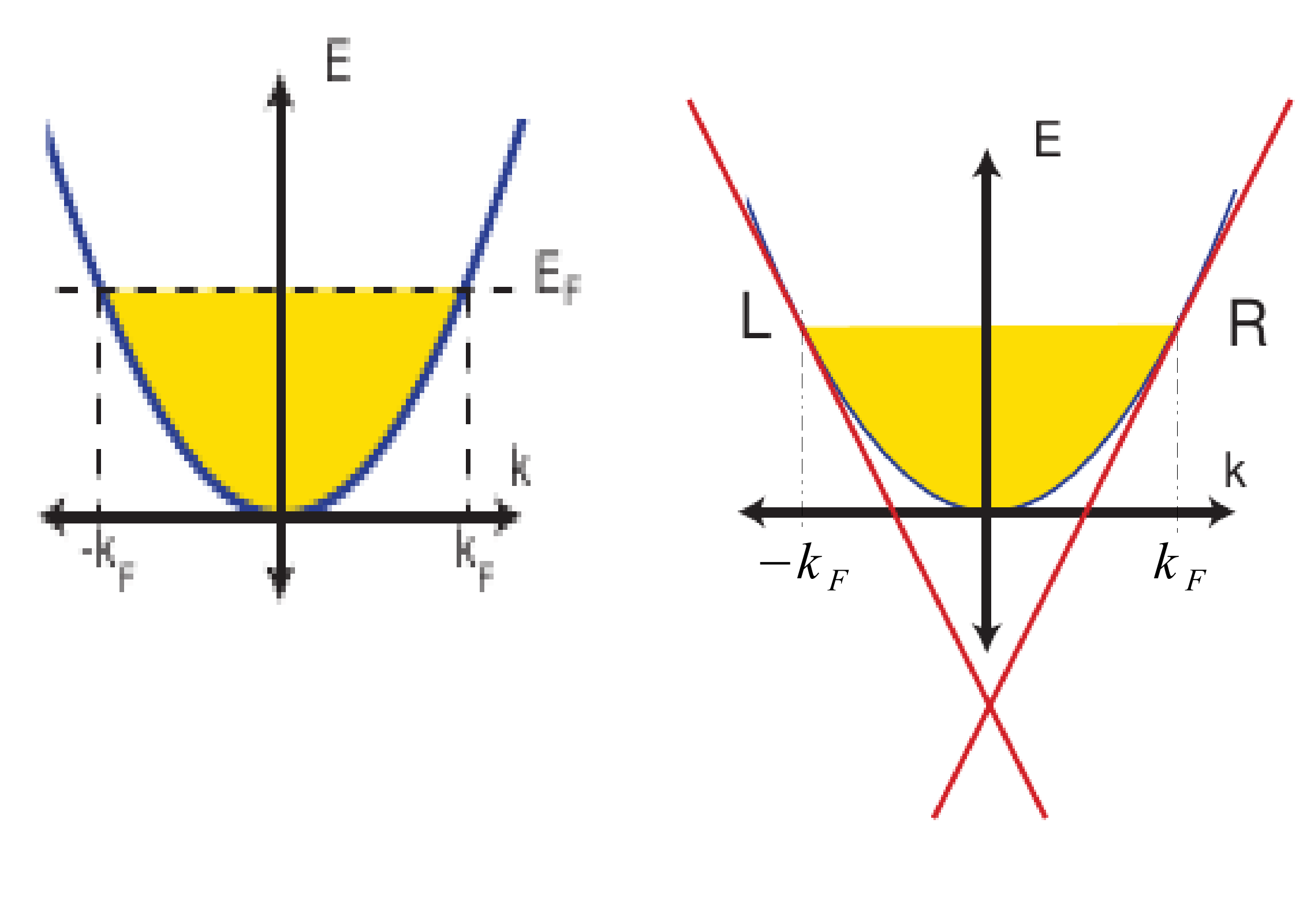}
\caption{(From left) Fig.1: Dispersion of non-interacting electrons in 1D.  Fig.2: The linearized bands at the Fermi points $k_F$ and $-k_F$ of the Luttinger model, extending to $-\infty$ \small[{Ref.: "Lectures on bosonization", C.L. Kane}]}.
\label{fermisea1D}
\end{figure}

The system can be divided into two species: right movers (RM) and left movers (LM), depending on the positive and negative velocities at $+k_F$ and $-k_F$, respectively. The spectrum is then linearized about these Fermi points, Fig.~\ref{fermisea1D}, and the Hamiltonian of the system becomes ~\cite{giamarchi}:
\begin{equation} \label{ham1}
\mathcal{H} = \sum_{k,r=R,L} (rk-k_F) c^{\dagger}_{k,r} c_{k,r} ,\
\end{equation}
where, $r= +1$ and $-1$ for RM and LM respectively. The low energy excitations are similar to those in a Dirac Hamiltonian and hence when extended to all the states below the chemical potential, the Fermi Sea can be mapped on to the `Dirac Sea', with an infinite number of negative energy states filled. \hphantom{} \\
The particle-hole excitations in such a system are well-defined and depends on only the momentum exchanged ($q$) during the scattering processes:
\begin{align*} 
E_{R,L}(q) &= v_F (k+q)-v_F k ,\ \\
&= v_F q .\
\end{align*}
The Hamiltonian in Eq.~\ref{ham1} can be written in terms of the basis of these particle-hole excitations (since they are well-defined excitations and only collective excitations survive in 1D). Thus, density fluctuations can be chosen as a natural basis for such systems with ~\cite{giamarchi,sen}:
\begin{equation} \label{density}
\rho^{\dagger} (q) = \sum_k c^{\dagger}_{k+q}c_{k} .\
\end{equation} 
Thus, introducing bosonic creation operators for each chirality:
\begin{eqnarray} \label{bbs} \nonumber
b^{\dagger}_{(R,L),q} &=& \frac{1}{n_q} \sum_{-\infty}^{\infty} c^{\dagger}_{(R,L), k+q} c_{(R,L),k} ,\ \nonumber \\
b_{(R,L),q} &=& \frac{1}{n_q} \sum_{-\infty}^{\infty} c^{\dagger}_{(R,L), k-q} c_{(R,L),k} ,\ \nonumber \\
q &=& \frac{2 \pi}{L} n_q ,\ \nonumber \\
\end{eqnarray}
where $L$ is the system size and $n_q=1,2,3..$ . The fermion number operator $\hat{N}_{(R,L)}=\sum_{-\infty}^{\infty} :c^{\dagger}_{(R,L), k} c_{(R,L),k}:$ (where, $: :$ denotes normal ordering) and the bosonic operators follow the following commutation relations:
\begin{eqnarray} \nonumber
\big[\hat{N}_R, b_{R,q}\big] &=& \big[\hat{N}_R, b^{\dagger}_{R,q}\big] = 0 ,\ \nonumber \\
\big[b^{\dagger}_{R,q}, b_{R,q'}\big] &=& \delta_{q,q'} ,\ \nonumber \\
\big[b_{R,q}, b_{R,q'}\big] &=& 0 .\
\end{eqnarray}
Therefore, the vacuum or the Fermi Sea of the system given by :
\begin{eqnarray} \nonumber
c_k |0\rangle &=& 0, \  \ \text{$\forall\ \ k>0$},\ \nonumber \\
c^{\dagger}_k |0\rangle &=& 0, \  \ \text{$\forall\ \ k<0$} ,\
\end{eqnarray}
is also the one defined by the $b$ and $b^{\dagger}$ operators. The bosonic operators in terms of the density operators in \ref{density} and \ref{bbs} , can be written as :
\begin{eqnarray} \nonumber
b^{\dagger}_q &=& \bigg(\frac{2 \pi}{L |q|}\bigg)^{\frac{1}{2}}\sum_ r Y(rq) \rho^{\dagger}_r (q) ,\ \nonumber \\
b_q &=& \bigg(\frac{2 \pi}{L |q|}\bigg)^{\frac{1}{2}}\sum_ r Y(rq) \rho^{\dagger}_r (-q),\
\end{eqnarray}
where $Y(rq)$ is a step function. The bosonic operators are only defined for $q \neq 0$.
Single particle fermionic fields $\psi_r$ for each chirality can be obtained from the commutation relations with the density operators (or the bosonic operators):
\begin{eqnarray} \label{rhopsicom} \nonumber
\big[ \rho^{\dagger}_r (q), \psi_r(x) \big] &=& \frac{1}{\sqrt{L}} \sum_ {k,k'} e^{i k' x} \big[ c^{\dagger}_{r, k+q} c_{r,k}, c_{r,k'} \big] \nonumber \\
&=& e^{-i q x} \psi_r (x) .\
\end{eqnarray}
If we operate the LHS and the RHS of Eq. ~\ref{rhopsicom}, we have:
\begin{equation}
\rho^{\dagger}_r (q) \psi_r(x) |0\rangle = e^{-i q x} \psi_r (x) |0 \rangle .\
\end{equation}
Thus, $\psi_r (x)|0\rangle$ is an eigenstate of $\rho^{\dagger}_r (q)$ for every value of $q$, and hence is a coherent state,
\begin{equation} \label{psi}
\psi_r (x) = U_r e^{\sum_q e^{i q x} \rho^{\dagger}_r(-q)(\frac{2 \pi r}{q L})} ,\
\end{equation}
where, $U_r$ are the Klein factors. \footnote{To reproduce all possible states of the original fermionic Fock space, it is necessary to introduce two additional operators (one for each species) that change the total number of fermions. These operators in the field theory language, are the Klein factors, which commute with the boson operators ~\cite{giamarchi}.}
Similarly, the commutator with the Hamiltonian can be calculated, using Eq.~\ref{ham1} and Eq.~\ref{density}, and is:
\begin{equation}
\big[ \rho_ q, H \big] = v_F q \rho_q , \ \ \text{$\forall\ \ q>0$\ \ and $\forall\ \ q<0$}
\end{equation}
This allows us to write the Hamiltonian in terms of the density operators $\rho$ or the bosonic field operators $b$ and $b^{\dagger}$:
\begin{equation} \label{ham_b}
H \simeq \sum_ {q \neq 0} v_F |q| b^{\dagger}_q b_q .\
\end{equation}
Introducing the fields $\phi$ and $\theta$ ~\cite{giamarchi}:
\begin{eqnarray} \label{phi_theta} \nonumber
\phi(x) &=& - (N_R+N_L) \frac{\pi x}{L} - \frac{i \pi}{L} \sum_{q \neq 0} \frac{1}{q} e^{- \alpha q/2 -i q x} (\rho^{\dagger}_R(q)+\rho^{\dagger}_L(q)),\ \nonumber \\
\theta(x) &=&  (N_R-N_L) \frac{\pi x}{L} + \frac{i \pi}{L} \sum_{q \neq 0} \frac{1}{q} e^{- \alpha q/2 -i q x} (\rho^{\dagger}_R(q)-\rho^{\dagger}_L(q)) ,\
\end{eqnarray}
Eq. ~\ref{psi} and Eq. ~\ref{ham_b} can be written in terms of the bosonic fields $\phi$ and $\theta$:
\begin{eqnarray} \label{psi_phi_theta} \nonumber
\psi_r (x) &=& U_r \lim_{\alpha \rightarrow 0} \frac{1}{2 \pi \alpha} e^{i r(k_F-\pi/L)x} e^{-i(r\phi(x)-\theta (x))} \nonumber \\
H &=& \sum_{q\neq0} v_F |q|b^{\dagger}_q b_q + \frac{\pi v_F}{L} \sum_r N_r^2
\end{eqnarray}
The term with $N_r$ is the $q=0$ contribution. $\alpha$ is an arbitrary  cut-off. $\phi$ and $\theta$ follow the commutation relations:
\begin{eqnarray} \label{phi_theta_comm} \nonumber
\big[ \phi(x), \theta(x') \big] &=& i \frac{\pi}{2} \rm{Sgn}(x'-x) \nonumber \\
\big[ \phi(x), \nabla\theta(x') \big] &=& i \pi \delta(x'-x)
\end{eqnarray}
This shows the conjugate momentum of $\phi(x)$ is $\nabla\theta(x)$, and, in the limit $L\rightarrow\infty$,
\begin{eqnarray} \label{rho_phi_theta} \nonumber
\nabla\phi(x) &=& -\pi \big[ \rho_R(x)+\rho_L(x)] \nonumber\\
\nabla\theta(x) &=& \pi \big[ \rho_R(x)-\rho_L(x)]
\end{eqnarray}
where, $\nabla\phi(x)$ is the $q \sim 0$ part of the density fluctuations at $x=0$ and $\nabla\theta(x)$ is the current operator in 1D. The Hamiltonian Eq.~\ref{ham_b}, written in terms of the new fields is:
\begin{equation} \label{ham_phi}
H = \frac{1}{2 \pi} \int dx\ v_F \big[ (\nabla\theta(x))^2+(\nabla\phi(x))^2 \big]
\end{equation}

\subsubsection{With interactions:} 
\begin{figure}
\centering
\includegraphics[width=0.75\linewidth]{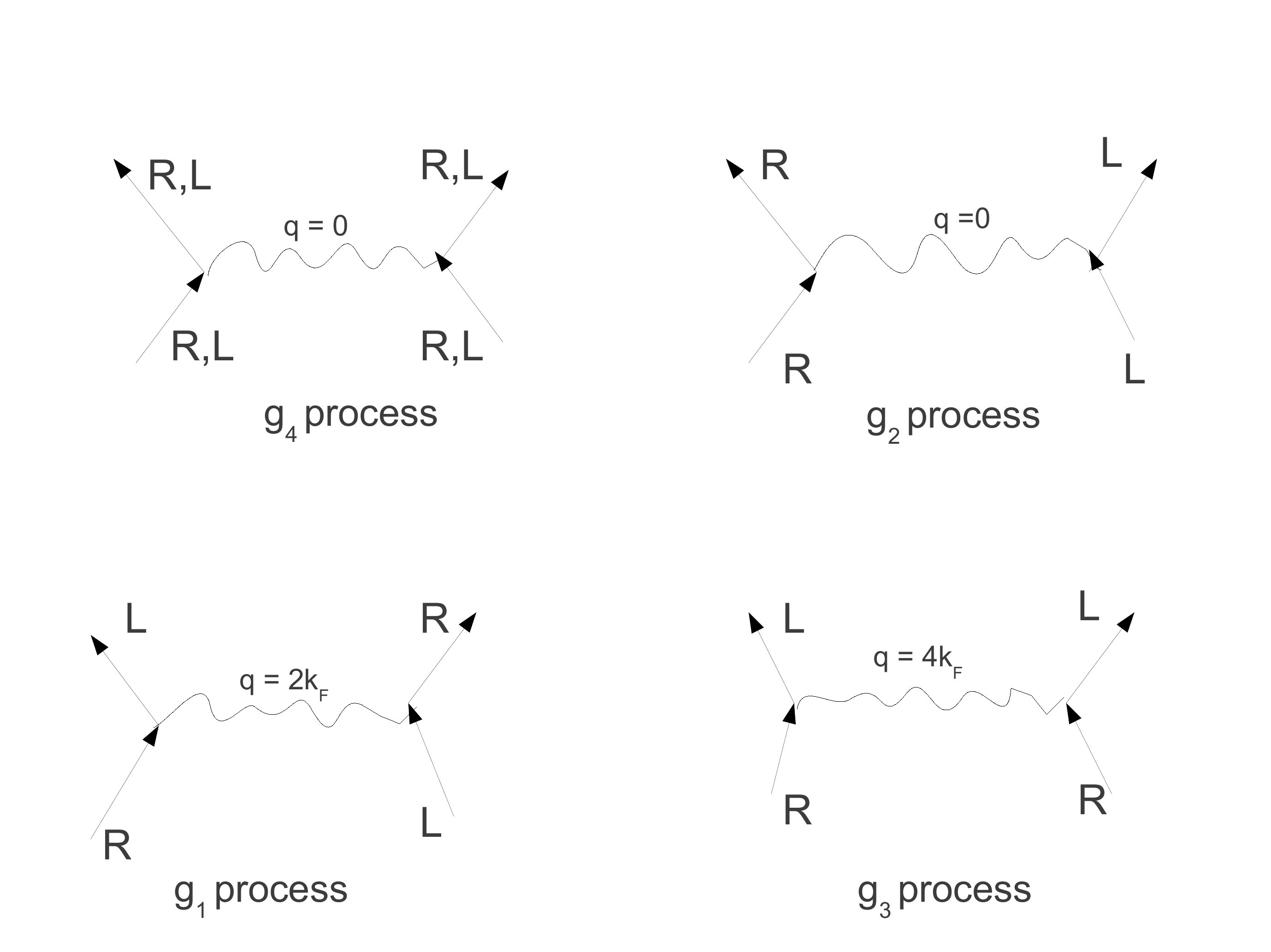}
\caption{Relevant scattering processes in Luttinger liquids. Figure 1: (Clockwise from top left) The $g_4$ process is where the scattering occurs on the same side of the Fermi sea. Figure 2: In the $g_2$ process, the right (left) mover scatters with a left (right) mover but remains a right (left) mover. These processes are relevant in both the charge and the spin sector of the system with the net momentum exchange being $q \sim 0$. Figure 3: Umklapp process in the charge sector is operative only in the presence of a lattice at half-filling. Here ($g_3$ process) two right (left) movers scatter to become two left (right) movers, with the net momentum exchange being $q\sim4k_F$. Figure 4: Umklapp in the spin sector is relevant even in the absence of a lattice. The $g_1$ process is where a right (left) mover (say, with up-spin) scatters with a left (right) mover (with down spin), and becomes a spin-flipped right (left) mover (with down spin). This is a spin exchange process, with the momentum exchange being $q\sim2k_F$.}
\label{interaction_fig}
\end{figure}

Since 1D systems are always nested, the low-energy scattering processes are restricted and only a finite number of them are allowed, Fig.~\ref{interaction_fig}. According to the convention ~\cite{giamarchi}, scattering processes occurring on the same side of the Fermi Sea are the $g_4$ process, where a right (left) mover scatters with a right (left) mover and remains a right (left) mover. $g_2$ process is one where a right (left) mover scatters against a left (right) mover but remains a right (left) mover.  In both of these processes, the exchange of momentum is zero. There are also interactions with higher momentum exchange, the $g_1$ process with $q \approx 2 k_F$ and the $g_3$ process with $q \approx 4 k_F$. $g_1$ process is a backscattering process, where the fermion exchanges sides. For spinless fermions, $g_1$ and $g_2$ processes are identical. Thus, for spinless fermions, we consider only $g_4$ and $g_2$ processes.
\begin{eqnarray} \label{g4_1} \nonumber
\frac{g_4}{2} \psi^{\dagger}_R(x)\psi_R(x)\psi^{\dagger}_R(x)\psi_R(x) &=& \frac{g_4}{2} \rho_R(x)\rho_R(x) \nonumber \\
&=& \frac{g_4}{2} \frac{1}{(2 \pi)^2}(\nabla\phi(x)-\nabla\theta(x))^2
\end{eqnarray}
Similarly for the left movers, \begin{eqnarray} \label{g4_2} \nonumber
\frac{g_4}{2} \psi^{\dagger}_L(x)\psi_L(x)\psi^{\dagger}_L(x)\psi_L(x) &=& \frac{g_4}{2} \rho_L(x)\rho_L(x) \nonumber \\
&=& \frac{g_4}{2} \frac{1}{(2 \pi)^2}(\nabla\phi(x)+\nabla\theta(x))^2 
\end{eqnarray}
The $g_2$ process gives, \begin{eqnarray} \label{g2} \nonumber
g_2 \psi^{\dagger}_R(x)\psi_R(x)\psi^{\dagger}_L(x)\psi_L(x) &=& \frac{g_4}{2} \rho_R(x)\rho_L(x) \nonumber \\
&=& \frac{g_2}{(2 \pi)^2}((\nabla\phi(x))^2-(\nabla\theta(x))^2)
\end{eqnarray}
From Eqs. ~\ref{ham_phi},~\ref{g4_1}, ~\ref{g4_2} and ~\ref{g2}, we see that $g_4$ renormalizes $v_F$, whereas $g_2$ changes the relative weights of $\nabla\phi$ and $\nabla\theta$. Any spinless 1D Hamiltonian with interactions is thus quadratic in terms of the bosonic fields and has the following form ~\cite{giamarchi,gnt,schulz,sen}:
\begin{equation}
H = \frac{1}{2 \pi} \int dx \big[ u K (\nabla\theta(x))^2+ \frac{u}{K} (\nabla\phi(x))^2 \big]
\end{equation}
where, $u$ is the renormalized velocity and K is the Luttinger parameter and is dimensionless,
\begin{eqnarray} \label{u_K} \nonumber
u &=& v_F \bigg[ \bigg(1+ \frac{g_4}{2 \pi v_F}\bigg)^2-\bigg(\frac{g_2}{2 \pi v_F}\bigg)^2 \bigg]^{1/2} \nonumber \\
K &=& \bigg[ \frac{1+\frac{g_4}{2 \pi v_F}-\frac{g_2}{2 \pi v_F}}{1+\frac{g_4}{2 \pi v_F}+\frac{g_2}{2 \pi v_F}} \bigg]^{1/2}
\end{eqnarray}
Thus, $g_4$ modifies the Fermi velocity whereas $g_2$ changes the relative weights of the $\nabla \phi$ and $\nabla\theta$ fields. For attractive interactions, $g_4, g_2<0$ and $K>1$ and vice versa for repulsive interactions.

\subsection{Correlation Functions}
There are two `ordering' (quasi-long-range ordering) tendencies for spinless fermions: one in the particle-hole  channel (charge density wave) and the other in the particle-particle channel (superconducting). The physical properties of these QLRO phases are studied from the behaviour of their correlation functions, as given below.

\subsubsection{Superconducting correlations}
The pairing operator is given by ~\cite{giamarchi}:
\begin{equation}
O_{SC}(x)=\psi^{\dagger}(x)\psi^{\dagger}(x+a) ,\
\end{equation}
where, each single particle fermionic operator can be separated into right and left movers, $\psi^{\dagger}(x)=\psi^{\dagger}_R(x)+\psi^{\dagger}_L(x)$, giving,
\begin{eqnarray} \nonumber
O_{SC}(x) &=& (\psi^{\dagger}_R(x)+\psi^{\dagger}_L(x))(\psi^{\dagger}_R(x+a)+\psi^{\dagger}_L(x+a)) \nonumber \\
&=& \psi^{\dagger}_R(x)\psi^{\dagger}_R(x+a)+\psi^{\dagger}_L(x)\psi^{\dagger}_L(x+a) \nonumber \\
&+& \psi^{\dagger}_R(x)\psi^{\dagger}_L(x+a)+\psi^{\dagger}_L(x)\psi^{\dagger}_R(x+a)
\end{eqnarray}
In the limit $a \rightarrow 0$, the first two terms are suppressed due to Pauli's principle. The dominant contribution is then,
\begin{eqnarray}
O_{SC}(x)
&=&\psi^{\dagger}_R(x)\psi^{\dagger}_L(x+a)+\psi^{\dagger}_L(x)\psi^{\dagger}_R(x+a)
\end{eqnarray}
Using Eq.~\ref{psi_phi_theta}, 
\begin{eqnarray} \label{corr_SC} \nonumber
O(x)O^\dagger(0) &=& \frac{e^{-i 2(\theta(x)-\theta(0))}}{\pi \alpha} \nonumber\\
\langle O(x)O^\dagger(0) \rangle &=& \langle \frac{e^{-i 2(\theta(x)-\theta(0))}}{\pi \alpha} \rangle \nonumber\\
&=& \frac{1}{(\pi \alpha)^2} \bigg( \frac{\alpha}{r}\bigg)^{1/K}
\end{eqnarray}
Superconducting correlations have a power law decay, with a non-universal decay constant depending on the interactions of the system. For attractive interactions ($K>1$), the correlations are stronger and decay further slowly.

\subsubsection{Charge density wave (CDW) correlations}
The density operator can be written in terms of the right and left movers: $ \psi^{\dagger}_L(x)\psi_L(x)+ \psi^{\dagger}_R(x)\psi_R(x)$, which written in terms of the bosonic fields gives~\cite{giamarchi}:
\begin{equation}
\rho(r)= -\frac{1}{\pi}\nabla \phi(r)+\frac{1}{2 \pi \alpha}[e^{2 i k_F x}e^{-i2\phi(r)}+h.c] 
\end{equation}
The density-density correlations is then calculated in a similar fashion to give (for $r \gg\alpha$):
\begin{eqnarray} \label{corr_CDW} \nonumber
\langle \rho(r)\rho(0)\rangle &=& -\frac{1}{\pi^2}\langle\nabla \phi(r)\nabla \phi(0)\rangle ++\frac{1}{(2 \pi \alpha)^2}[e^{2 i k_F x}\langle e^{-i2(\phi(r)-\phi(0))}\rangle +h.c] \nonumber \\
&=& \frac{K}{2 \pi^2} \frac{y_\alpha^2-x^2}{(x^2+y_\alpha^2)^2}+\frac{2}{(2 \pi \alpha)^2} \cos{(2 k_F x)} \bigg(\frac{\alpha}{r}\bigg)^{2K}
\end{eqnarray}
where, $y_\alpha=u \tau + \alpha \rm{Sgn}(\tau)$, is the imaginary time, regularised by the cut-off $\alpha$, and $u$ is the velocity,
The $q\sim0$ part gives Fermi-liquid like behaviour, whereas the second term has the Luttinger liquid behaviour. For repulsive interactions ($K<1$), the CDW correlations decay slower than the superconducting correlations and hence are more dominant. 

Correlations of Luttinger liquids in the massless phase corresponds to the correlation functions of a classical two-dimensional system at criticality. Such correlation functions are invariant by a large class of transformations, like continuous rotation (between $x$ and $\tau$ direction), and scale transformations. In addition, systems at criticality are also invariant under a broader class of transformations called conformal transformation. Transformations that are locally identical to dilatations, rotations and translations fall under this category. These transformations preserve angles between any three points. It can be seen that the space and time coordinates in the above correlations scale similarly, preserving angles, thus establishing the conformal invariance of the system.

\section{Model with umklapp} \label{model_umklapp}
In the presence of a lattice, the wave-vector is defined modulo a reciprocal lattice vector (in 1D, the reciprocal lattice vector is a multiple of $2 \pi/a$, where $a$ is the lattice spacing). The momentum in the scattering processes is now conserved modulo the reciprocal lattice vector. Thus, along with the processes that truly conserve momentum, such as  $k_1+k_2=k_3+k_4$, there are additional processes with momentum exchange $k_1+k_2-k_3-k_4=Q$, where $Q$ is the reciprocal lattice vector. Such interactions are called umklapp processes, Fig.~\ref{interaction_fig}. Since the interaction processes in 1D can only occur across the two Fermi points $+k_F$ and $-k_F$, umklapp scattering with $k_1+k_2-k_3-k_4=2 k_F$ can only occur when $4k_F=2\pi$, i.e, at half-filling. For spinless fermions, umklapp process can only exist for such commensurate fillings.
The presence of spins also brings in similar interaction terms like scattering processes with momentum exchange $q \sim 2 k_F$, due to which we no longer have a quadratic Hamiltonian. We need to do a renormalization group (RG) analysis to find the fixed point for the theory.

\subsection{Model with spin}
Umklapp is operative even in the continuum for fermions with spin, Fig.~\ref{interaction_fig}. We represent the bosonic fields for each spin species and define the total charge and spin degrees of freedom ~\cite{giamarchi,gnt,schulz}:
\begin{eqnarray} \nonumber
\rho &=& ( \rho_{\uparrow} + \rho_{\downarrow} ) \nonumber \\
\sigma &=& ( \rho_{\uparrow} - \rho_{\downarrow} )
\end{eqnarray}
Defining the bosonic fields for the charge and the spin sectors:
\begin{eqnarray} \nonumber
\phi_\rho &=& ( \phi_{\uparrow} + \phi_{\downarrow} ) \nonumber \\
\phi_\sigma &=& ( \phi_{\uparrow} - \phi_{\downarrow} )
\end{eqnarray}
The $\rho$ and $\sigma$ fields commute with each other, whereas the $\phi_\rho,\theta_\rho$ and $\phi_\sigma,\theta_\sigma$ have the usual commutation relations. The single particle operator in terms of these fields:
\begin{equation}
\psi_r (x) = U_{r, \sigma} \lim_{\alpha \rightarrow 0} \frac{1}{2 \pi \alpha} e^{i r k_F x} e^{-\frac{i}{\sqrt{2}}[(r\phi_\rho(x)-\theta_\rho (x))+\sigma (r\phi_\sigma(x)-\theta_\sigma (x))] } 
\end{equation}
The Hamiltonian can be separated into two independent sectors~\cite{giamarchi}:
\begin{equation}
H = H_\rho + H_\sigma
\end{equation}

The umklapp interaction term in the spin sector is typically of the form:
\begin{eqnarray} \nonumber
H_1 &=& \int dx\ g_{1\parallel} \sum_\sigma [ \psi^{\dagger}_{L, \sigma} \psi^{\dagger}_{R, \sigma} \psi_{L, \sigma} \psi_{R, \sigma}] + g_{1\perp} \sum_\sigma [ \psi^{\dagger}_{L, \sigma} \psi^{\dagger}_{R, -\sigma} \psi_{L, -\sigma} \psi_{R, \sigma}] \nonumber\\
&=& \int dx (-g_{1\parallel}) \sum_\sigma [ \psi^{\dagger}_{L, \sigma} \psi_{L, \sigma} \psi^{\dagger}_{R, \sigma} \psi_{R, \sigma}] + g_{1\perp} \sum_\sigma [ \psi^{\dagger}_{L, \sigma} \psi_{R, \sigma} \psi^{\dagger}_{R, -\sigma} \psi_{L, -\sigma} ] \nonumber\\
&=& \int dx (-g_{1\parallel}) \sum_\sigma [ \rho_{R,\sigma} \rho_{L,\sigma}] + \frac{g_{1\perp}}{(2 \pi \alpha)^2} \sum_{\uparrow,\downarrow} [ e^{i (-2 \phi _s (x))}e^{i (2 \phi_{-s} (x))}]
\end{eqnarray}
The $g_{1 \parallel}$ term is effectively similar to $g_2$ term and together they are included in a $g_{2, eff}$ term. The $g_{1\perp}$ is the spin exchange term, and can be bosonized to give:
\begin{equation}
H_{1 \perp} = \int dx\ \frac{2 g_{1\perp}}{(2 \pi \alpha)^2} \cos{( 2\sqrt{2} \phi_\sigma (x))}
\end{equation}
The charge part of the Hamiltonian is purely quadratic (in the absence of a lattice), only the spin part has an additional term. Together with the quadratic part, the total Hamiltonian is written as ~\cite{giamarchi}:
\begin{equation}
H = H_0 + \int dx\ \frac{2 g_{1\perp}}{(2 \pi \alpha)^2} \cos{( 2\sqrt{2} \phi_\sigma (x))},
\end{equation}
where, the Luttinger parameters in the quadratic part is given by:
\begin{eqnarray} \nonumber  
u_\nu &=& v_F \bigg[\bigg(1+\frac{y_{4 \nu}}{2}\bigg)^2-\bigg(\frac{y_\nu}{2}\bigg)^2\bigg]^{1/2}, \nonumber\\
K_\nu &=& \bigg[ \frac{1+\frac{y_{4\nu}}{2}+\frac{y_\nu}{2}}{1+\frac{y_{4\nu}}{2}-\frac{y_\nu}{2}} \bigg]^{1/2}, \nonumber\\
g_\nu &=& g_{1 \parallel}-g_{2 \parallel} \mp g_{2 \perp}, \nonumber\\
g_{4 \nu} &=& g_{4\parallel}\pm g_{4\perp}, \nonumber\\
y_\nu &=& g_\nu/ \pi v_F ,
\end{eqnarray}
where $\nu=\rho, \sigma$ and the upper sign refers to $\rho$ and lower sign to $\sigma$.
The entire Hilbert space can be expressed in terms of the product of charge and spin excitations, and ensures the absence of any single particle excitations in the system.

\subsection{Model with umklapp on a lattice}
Interactions with momentum exchange $q \sim 4k_F$ are contributions from the $g_3$ process in Fig.~\ref{interaction_fig}. Similar kinds of interactions written in terms of $\psi_R$ and $\psi_L$ for such a backscattering term give ~\cite{giamarchi,gnt}:
\begin{equation}
H_3 = \int dx\ \frac{2 g_{3}}{(2 \pi \alpha)^2} \cos{( \sqrt{16} \phi_\rho (x))}
\end{equation}
The charge sector of the Hamiltonian is now no longer quadratic (even though the spin sector might be, depending on whether the spin exchange process is present). The total Hamiltonian in the charge sector is given by:
\begin{equation}
H = H_0 + \int dx\ \frac{2 g_{3}}{(2 \pi \alpha)^2} \cos{( \sqrt{16} \phi_\rho (x))},
\end{equation}


\section{Currents in one-dimensional systems} \label{current_1D}
One dimensional systems with short-ranged interactions can be modelled as 1D Hubbard type models (spinless fermions can be modelled as a t-V type model) or 1D spin chains ~\cite{giamarchi,gnt,schulz,sen}. The Luttinger liquid theory successfully explains all classes of 1D models, both on a lattice and in the continuum. The current carrying states in these models are free bosonic modes with both chirality. The current operator in such systems are given by the difference between the total number of modes with opposite chirality. \hphantom{} \\
Luttinger liquids can be experimentally realized in several systems, ranging from edge states in the fractional quantum Hall effect ~\cite{chang1,chang2,teo} (in this case, a chiral Luttinger liquid is found) to quantum wires in semiconductor heterostructures and single-walled carbon nanotubes (SWNT) ~\cite{bockrath,shi}. The latter can be used to perform electrical transport experiments
using different geometries, for example, in junctions manipulated by mechanical means. 


\subsection{Boosted superfluid in one-dimension}

Currents set up in a superconductor continues to flow even in the absence of a driving electric field~\cite{santos,bagwell,tinkham}. Such currents can be realized with different chemical potentials for the number of carriers moving along and opposite to the direction of the current, i.e, a boost. In 1D, this is equivalent to having different chemical potentials for the right and left movers. Such systems have also been realized in cold atomic gases where it has been possible to make the system left-right asymmetric thereby producing a boost~\cite{zwierlein,esslinger,wang} Such a boost in three dimensions, are known to have a critical value beyond which the superconducting state gets destroyed. This is similar to the phenomenon of destruction of superfluidity when the flow velocity is larger than the critical velocity~\cite{baym,wei,zagoskin}. The critical velocity of a clean one dimensional superconductor has been calculated in mean-field theory and was found to be smaller than the standard Landau critical velocity due to a pre-emptive Clogston-Chandrasekhar-type discontinuous transition~\cite{wei}. A similar calculation for a clean one dimensional superconductor incorporating the effects of quantum fluctuations has not been performed so far. However, it has been shown that phase slips induced by the contact of the superconductor with the walls of a container or the presence of statically irrelevant perturbations can dynamically destroy superconductivity at finite frequency and temperature in one dimension~\cite{oshikawa}.\hphantom{} \\

A natural question is about the fate of superconductors which do not have long-range order upon the application of boost. Most common example being a one dimensional system of fermions with attractive interactions~\cite{giamarchi,gnt,schulz,haldane}. Such systems have order parameter equal to zero and power law correlations. \hphantom{} \\

\subsubsection{Framework}
\begin{figure}[h]
\centering
\includegraphics[width=7cm, height=4.5cm]{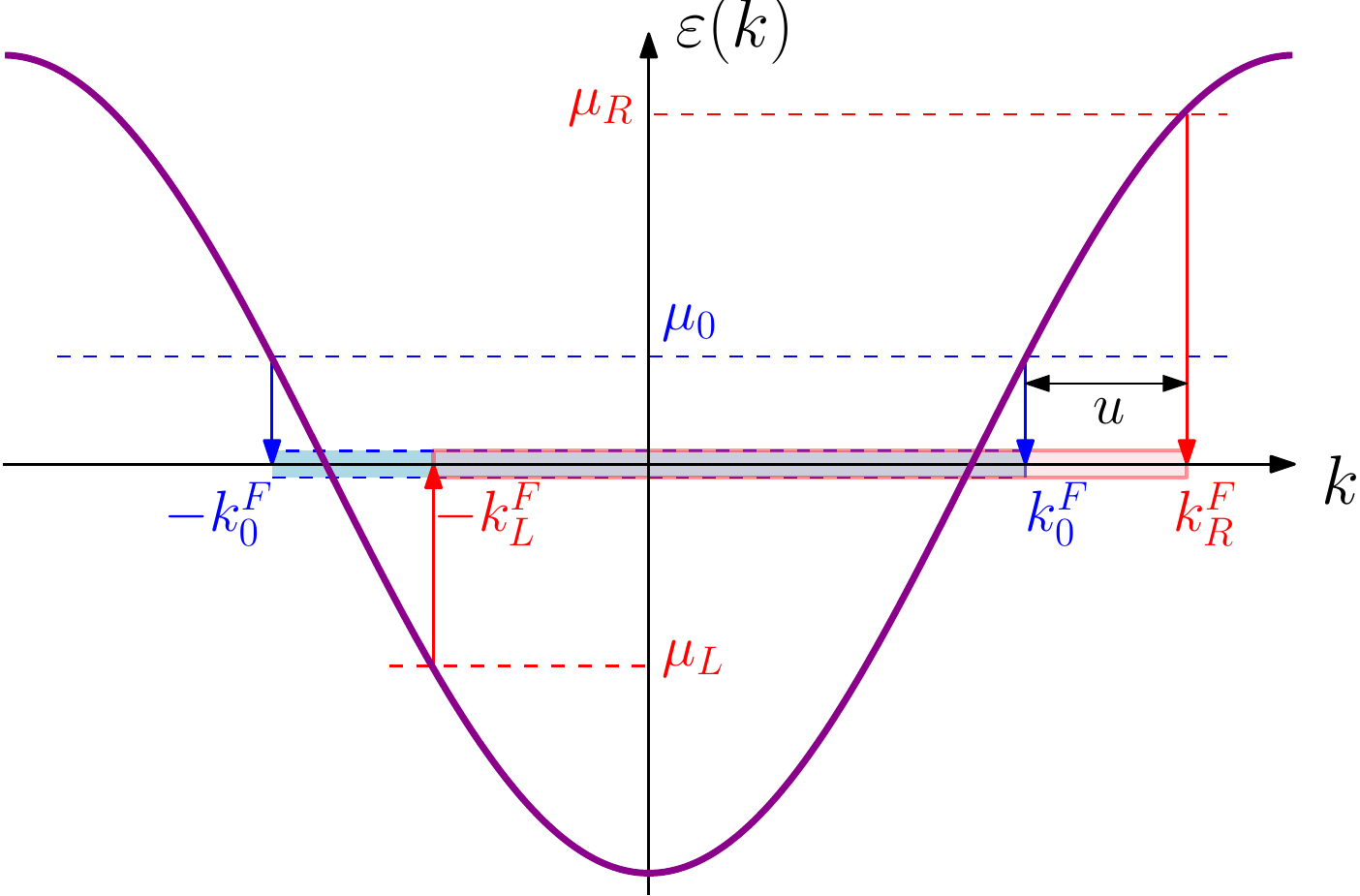} 
\caption{Schematic of the boosted Fermi sea. $\kfo$[$\mu_0$]  unboosted Fermi wavevector [chemical potential]. With boost: Fermi wavevector [chemical potential] of left[right] movers is $k^F_L[\mu_L]$ and $k^F_R[\mu_R]$; see Eq. (\ref{Eq:kFu}). }
\label{fig:scheme}
\end{figure}

One dimensional systems of spinless fermions with a dispersion $\epsilon(k)$ symmetric in $k$ have two Fermi points at $\kfo$ and $-\kfo$ corresponding to right and left movers respectively. $\kfo=\pi n$, where $n$ is the density of fermions. The boost $u$ is defined by a transformation $k \rightarrow k+u,~\forall k$. This destroys the left-right symmetry and the two Fermi points are now at,
\begin{equation} \label{Eq:kFu}
k ^F_s=\kfo +s u, 
\end{equation}
where $s=+1[-1] \equiv R[L]$ for the right[left] movers with Fermi velocity $v^F_s (u)= \left. \frac{d \epsilon}{dk} \right|_{k=k^F_s(u)}.$ Assuming $\epsilon(k)$ to be analytic, 
$v^F_s=v^F(u)+s w(u),$ 
where $v^F(u)$ and $w(u)$ are even and odd functions of $u$ respectively.  

Under the effect of the boost, the right movers and left movers have different effective Fermi velocities:
\begin{equation} \label{RL_vf}
\begin{split}
v^F_R\ &=\ v^F(u)+w(u) \\
v^F_L\ &=\ v^F(u)-w(u).
\end{split}
\end{equation}

The kinetic part of the Hamiltonian is:
\begin{equation}
H_K\ =\ \int \bigg[ v^F_R \rho_R^\dagger(x) \rho_R(x) + v^F_L \rho_L^\dagger(x) \rho_L(x) \bigg] dx,
\end{equation}
where the $\rho_{R, L}$ are the density operators of the right and left movers.
We can introduce the field $\phi$ and conjugate field $\Pi$ ~\cite{giamarchi}, 
\begin{equation} \label{phi-theta}
\begin{split}
\nabla \phi(x)\ &=\ -\pi [ \rho_R(x)+\rho_L(x) ], \\
\Pi(x)\ &=\ \pi [ \rho_R(x)-\rho_L(x) ],
\end{split}
\end{equation}
where $\nabla \phi$ is the $ q \sim 0$ part of the density fluctuation and $\Pi$ represents the current operator.
In terms of these bosonic fields ($[\phi (x_1), \Pi (x_2)]=i \pi \delta (x_1-x_2)$), the kinetic part of the Hamiltonian becomes ~\cite{giamarchi}:
\begin{equation}
H_K\ =\ \frac{1}{\pi} \int \bigg[ v^F(u) \bigg( (\nabla\phi (x))^2 + (\Pi (x))^2 \bigg) -2 w(u) (\nabla\phi (x))(\Pi (x)) \bigg] dx.
\end{equation}
For spinless fermions, the relevant interaction processes are the $g_4$ and $g_2$ scattering processes ~\cite{giamarchi}:
\begin{equation}
\begin{split}
V_{g_4}\ &=\ \frac{g_4}{2} [\rho_R(x) \rho_R(x)+\rho_L(x) \rho_L(x)], \\
V_{g_2}\ &=\ g_2 \rho_R(x) \rho_L(x).
\end{split}
\end{equation}
These interaction processes are not affected by the boost. To see this, we note that they arise from density-density interactions (as also do umklapp terms to be considered later) which preserve the translational symmetry of the lattice. Such interactions are of the form 
\begin{equation}
V \sim \sum_{k_1,k_2,k_3,k_4} V(k_1-k_3) c^\dagger_{k_1} c^\dagger_{k_2}c_{k_3}c_{k_4} \delta(k_1+k_2-k_3-k_4),
\label{Eq:interaction}
\end{equation}
in momentum space,  where $k_1,k_2,k_3$ and $k_4$ are the momenta of the fermions. Since all momenta are fully summed over in the above form and terms in the summation only involve differences of momenta, a boost leaves it unaffected since it adds $u$ to {\em all} momenta.

\subsection{Modified Luttinger parameter and correlation functions}
Using \ref{phi-theta}, the total Hamiltonian with boost is:
\begin{equation} \label{htot}
\mathcal{H}= \frac{v (u)}{\pi}\int_{-L/2}^{L/2} dx \left[K (u) :\left(\Pi(x) \right)^2: + \frac{1}{K(u)} :\left(\nabla \phi(x) \right)^2: -\frac{2 w(u)}{v(u)} :(\nabla\phi (x))(\Pi(x)): \right],
\end{equation}
where \begin{eqnarray}\nonumber
v(u) & = & \sqrt{\left(v^F(u) + \frac{g_4}{2\pi} \right)^2 - \left( \frac{g_2}{2\pi} \right)^2},\\ 
K(u)& = & \sqrt{\frac{v^F(u) - \left(\frac{g_2}{2\pi}-\frac{g_4}{2\pi}\right)}{v^F(u) + \left(\frac{g_2}{2\pi}+\frac{g_4}{2\pi}\right)}}.
\label{Eq:boost_para}
\end{eqnarray}
\vspace{0.5cm}

Eq. (\ref{Eq:boost_para}) shows $v(u)$ and $K(u)$ pick up $u$ dependences only from $v^F(u)$, and $u$ also introduces a coupling between $\Pi$ and $\nabla \phi$ with strength $w(u)$. It can also be seen that Eqs. (~\ref{htot}) and (~\ref{Eq:boost_para}) reduce to their standard forms when $u=0$ ~\cite{giamarchi,gnt,schulz,haldane}. 
When $u=0$, the system possesses conformal invariance since the coupling between the $\Pi$ and $\phi$ fields is absent, which the boost clearly breaks. However, conformal invariance can be restored by introducing new fields:
$\tilde{\phi}(x,t) = \phi[x+w(u)t,t],  \;\; \tilde{\theta}(x,t) = \theta[x+w(u)t,t],$
(where, $\tilde{\Pi} =\nabla \tilde{\theta}$, is the conjugate field of $\tilde{\phi}$) in terms of which the Hamiltonian can again be written in the standard form with Fermi velocity $v(u)$ and Luttinger parameter $K(u)$ :
\begin{equation}
\mathcal{H}= \frac{v(u)}{\pi}\int_{-L/2}^{L/2} dx \left[K(u) :\left(\tilde{\Pi}(x) \right)^2: + \frac{1}{K(u)} :\left(\nabla \tilde{\phi}(x) \right)^2: \right],
\label{Eq:Ham_tilde}
\end{equation}
where:
\begin{eqnarray}\nonumber 
\partial_t \tilde{\phi} & = & \partial_t \phi + w(u)\nabla \phi\\ 
\nabla \tilde {\phi} & = & \nabla \phi,
\label{Eq:new_fields}
\end{eqnarray}
with $\tilde{\Pi}=\partial_t \tilde{\phi}$. Eq. (\ref{Eq:new_fields}) implies 
\begin{equation}
\tilde{\phi}(x,t)=\phi(x+w(u)t,t),
\label{Eq:field_boost}
\end{equation}
Thus, the factor $w(u)$ acts as a velocity in the Galilean transformation of the coordinates $(x,t)$ to obtain the new field $\tilde{\phi}$.

Thus, conformal invariance can be restored by effecting a (non-conformal) Galilean transformation on the space-time co-ordinates. $w(u)$ acts as the velocity in this transformation.
For small $u$ (compared to $\kfo$), 
\begin{equation}
K(u) \approx K(0) + \frac{u^2}{2} \left. \frac{dK(u)}{dv^F(u)} \frac{d^2 v^F(u)}{du^2} \right|_{u=0}.
\label{Eq:small_boost}
\end{equation}
 It can be seen that for a system with Galilean invariance (and hence $\epsilon(k)$ quadratic in $k$), $K(u)=K$. Thus, Galilean invariance needs to be broken for a non-trivial effect of the boost for which we put the fermions on a lattice with nearest neighbour hopping $- \hopt$ and dispersion $\epsilon(k)=-2 \hopt \cos k$. 
Consequently, 
\begin{eqnarray}
v^F(u)  =  v^F(0)\cos u, \;\;\;
w(u)  =  2 \hopt \cos (\kfo) \sin u,
\label{Eq:lattice}
\end{eqnarray} 
where $v^F(0)= 2 \hopt \sin (\kfo)$ and thus
\begin{equation}
K(u) \approx K(0) +\frac{u^2}{4} \left[ \frac{v^F(0)}{v(0)} \left(K(0)^2-1 \right)\right].
\label{Eq:Kwithu}
\end{equation}

Eq. (\ref{Eq:Kwithu}) implies that $K(u) > K(0) [K(u) < K(0)]$ for $K(0) >1 [K(0)<1]$ for small $u$. This is true at larger values of $u$ as well. Further, $K(u)$ depends not just on $K(0)$ and $u$ but also the microscopic parameters $v^F(0)$, $g_2$ and $g_4$ (the latter two through $v(0)$), showing that the way the boost modifies the Luttinger parameter is "not universal". Also, note that the non-interacting point remains unaffected by the boost, i.e, when $K(0)=1$, $K(u)=1$.

\begin{figure}[h]
\centering 
\begin{tabular}{c} 
\includegraphics[width=7cm, height=4.5cm]{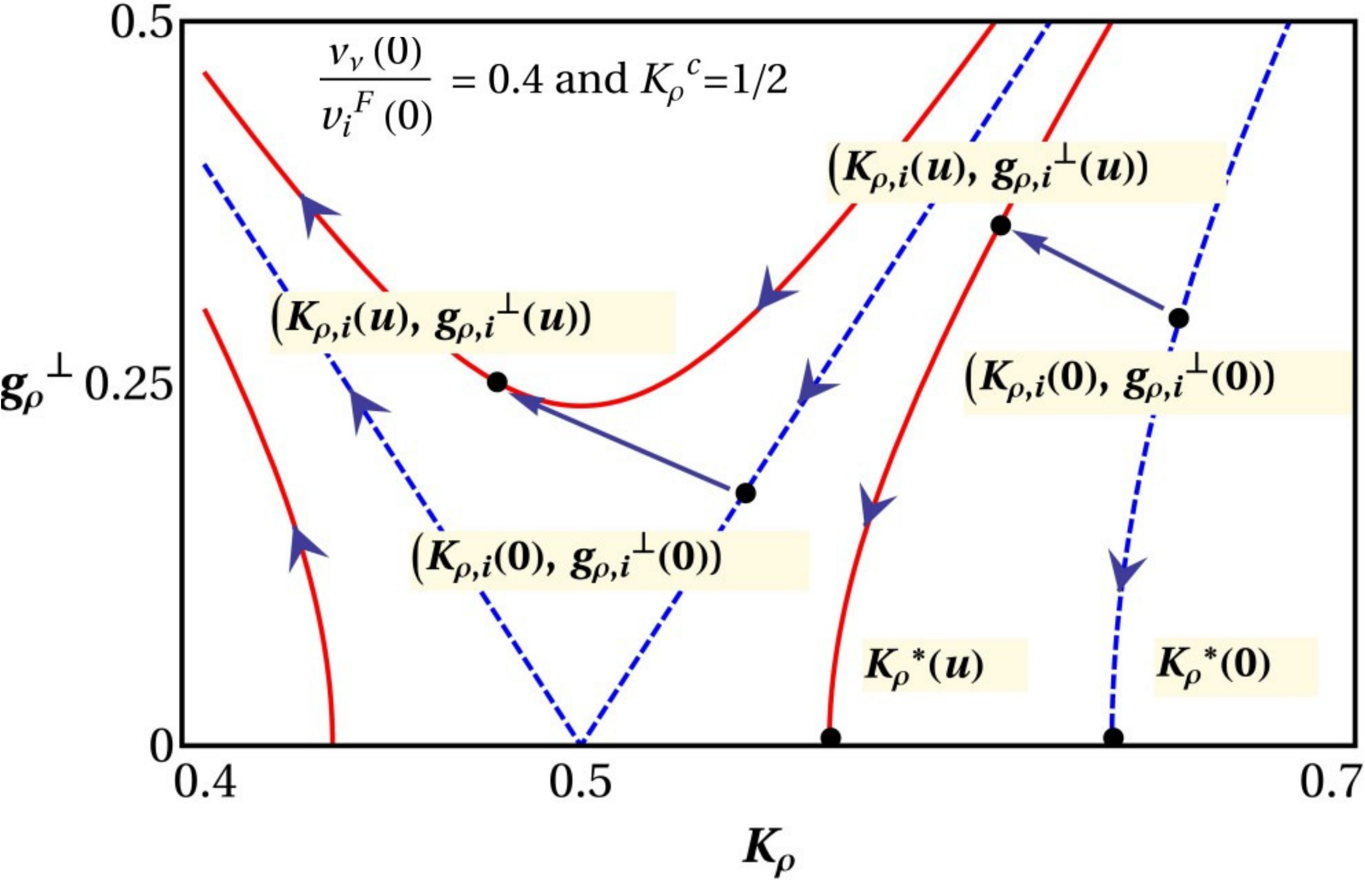} 
\includegraphics[width=7.2cm, height=4.7cm]{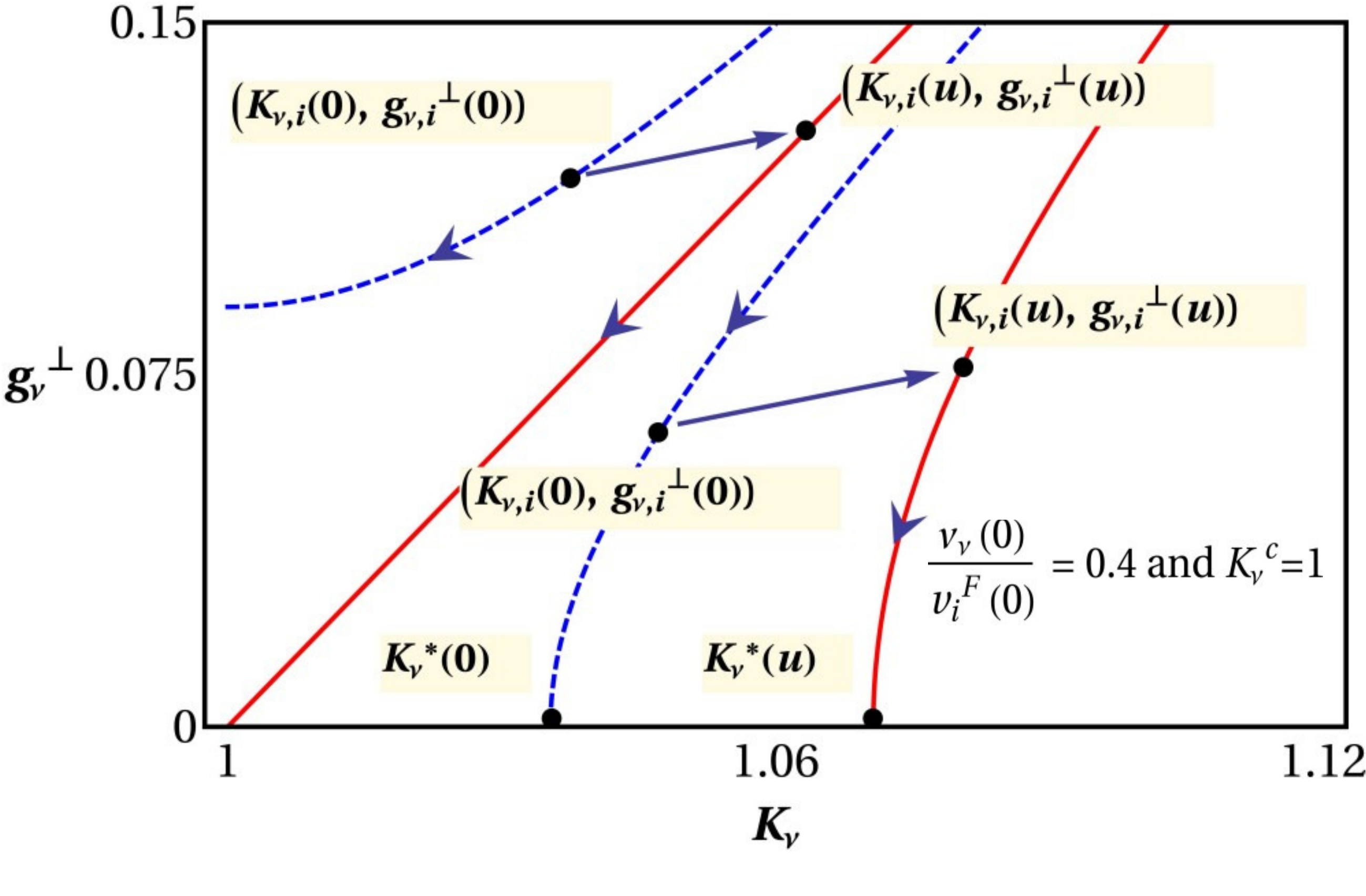}  
\end{tabular}
\caption{Renormalization group flows (see Eq.~(\ref{Eq:RG_flow}). The boost alters the initial values $(K_{\nu,i}(0), g^\perp_{\nu,i}(0))$ to $(K_{\nu,i}(u), g^\perp_{\nu,i}(u))$ as indicated by solid arrows. This changes the flows from dashed blue lines (no boost) to solid red lines (with boost). Top panel: For spinless fermions with umklapp a boost can transform a gapless phase to a gapped phase. Bottom panel: For spin $\half$ fermions, with two decoupled channels ($\nu=\rho,\sigma$), the boost can transform a gapped phase to a gapless one.}
\label{fig:RGflow}
\end{figure}

It is known that $K>1$ and $K<1$ result in dominant quasi-long-range ordered superconducting (SC) and charge density wave (CDW) order respectively ~\cite{giamarchi,gnt,schulz,haldane}. 
This continues to be true even for $u \neq 0$ since the system is described by the standard harmonic Hamiltonian under the transformations $K \rightarrow K(u)$, $v^F \rightarrow v^F(u)$ and $\phi \rightarrow \tilde{\phi}$ but in terms of transformed space-time coordinates. The SC and CDW correlation functions are thus given by \cite{giamarchi,miranda}
\begin{eqnarray}\nonumber 
\langle O_{\rm SC}(x,t) O_{\rm SC}^\dagger (x',t') \rangle & \sim & e^{i 2 u(x-x')} \left( \frac{1}{\ell} \right)^{1/K(u)}, \\
\langle O_{\rm CDW}(x,t) O_{\rm CDW} (x',t') \rangle & \sim & \cos{[2 \kfo(x-x')]} \left( \frac{1}{\ell} \right)^{K(u)},
\label{Eq:6}
\end{eqnarray}
\noindent
where $\ell = \sqrt{\left[x-x' + w(u)(t-t')\right]^2+\left[v^F(u) \right]^2 (t-t')^2}$ $O_{\rm SC[CDW]}$ is the SC[CDW] order parameter. $\langle O_{\rm SC[CDW]}(x,t) \rangle =0$ since there is no long range order. The loss of conformal invariance upon the application of a boost can be clearly seen from the asymmetric way in which the space and time coordinates appear in Eq. (\ref{Eq:6}). The equal time correlation function (with $t=t'$) for the SC[CDW] order decays algebraically with distance with exponent $1/K(u) [K(u)]$. Thus, the order that was dominant in the absence of a boost is strengthened by it while the sub-dominant one is weakened. {\em Hence, a system that is superconducting has its superconductivity strengthened in the presence of a boost.} A CDW system also has its CDW order strengthened similarly. 

Note, however, that this is true only for $u < \kfo$ since beyond that value, there is only one species (either left or right moving) of fermions. Superconductivity is thus discontinuously destroyed at this critical value of the boost. Suppose $\kfo < \pi/2$, one of the Fermi points will move to $k=0$ when a boost $u=\kfo$ is applied. It will no longer be possible to linearize about this Fermi point and so $u=\kfo$ might be a natural limit to the applicability of our treatment. Even though, it once again becomes possible to linearize when the Fermi point moves to $k > 0$, the Fermi vacuum becomes unstable and hence the system is not amenable to our treatment. If both the Fermi points $k^F_L$ and $k^F_R$ are positive, then linearization of the dispersion at $k^F_L$ becomes problematic since the unoccupied states $(k < k^F_L)$ have lower energy rendering the Fermi sea unstable.

\subsection{Correspondence with the microscopic t-V model} \label{tvmodel}
The Hamiltonian for the t-V model at a fixed filling $n$:
\begin{equation}
H= -\hopt \sum_i c_{i}^{\dagger}c_{i+1}- V\sum_i n_i n_{i+1} 
\end{equation}
with the dispersion : $ \varepsilon(k) = -2 \hopt \cos{k}$ . At filling $n$, $\kfo=n \pi$. We need to express $K(u)$ and $v(u)$ in terms of the parameters $\hopt$, $V$ and $n$.
For low energy excitations about $\kfo$, the $c^{\dagger}_i$'s can be written in terms of the $\psi_R$'s and $\psi_L$'s in the following way:
\begin{equation}
c^{\dagger}_i \sim \sqrt{a}\ ( e^{i \kfo x} \psi^{\dagger}_R +  e^{-i \kfo x} \psi^{\dagger}_L)
\end{equation}
 
The interaction is of the form $ -V \sum n_{i}n_{i+1} $, which written in terms of $\psi_R$ and $\psi_L$ gives: 
\begin{eqnarray} \label{interaction} \nonumber
 -V \sum n_{i}n_{i+1} &=& -Va\int dx\ ( e^{i \kfo x} \psi^{\dagger}_R +  e^{-i \kfo x} \psi^{\dagger}_L) ( e^{-i \kfo x} \psi_R +  e^{i \kfo x} \psi_L) \nonumber \\
& &( e^{i \kfo (x+a)} \psi^{\dagger}_R +  e^{-i \kfo (x+a)} \psi^{\dagger}_L)( e^{-i \kfo (x+a)} \psi_R +  e^{i \kfo (x+a)} \psi_L) .
\end{eqnarray}
The low energy interaction processes are then:
\begin{eqnarray} \nonumber
-V \sum n_{i}n_{i+1} &=& -V a \int dx\ (\psi^{\dagger}_R \psi^{\dagger}_R \psi_R \psi_R + \psi^{\dagger}_L \psi^{\dagger}_L \psi_L \psi_L + \psi^{\dagger}_R  \psi_R \psi^{\dagger}_L \psi_L + \psi^{\dagger}_L \psi_L \psi^{\dagger}_R \psi_R ) \nonumber \\
& & + V a \int dx\ (e^{-2 i \kfo a} \psi^{\dagger}_R \psi_L \psi^{\dagger}_L \psi_R + e^{2 i \kfo a} \psi^{\dagger}_L \psi_R \psi^{\dagger}_R \psi_L)
\end{eqnarray} 
$\tilde{g}_4$ and $\tilde{g}_2$ processes are scattering with zero momentum exchange, and hence, their magnitudes are equal. However, $\tilde{g}_1$ processes are scattering with momentum exchange $2 \kfo$, and hence have an additional factor $\cos{2 \kfo}$: 
\begin{eqnarray} \label{g4g1}
\tilde{g}_{4} &=&  \tilde{g}_{2} = -V a, \nonumber\\
\tilde{g}_{1} &=& -2 V a \cos{2 \kfo a}. 
\end{eqnarray}
For spinless fermions in the absence of a boost, the $\tilde{g}_{1}$ and $\tilde{g}_{2}$ processes are effectively the same, hence $ \tilde{g}_{2,eff} $ becomes:
\begin{eqnarray} \label{g2eff}
 \tilde{g}_{2,eff} &=& \tilde{g}_{2} -\tilde{g}_{1} \nonumber\\
           &=& -V a (1 - 2\cos{2 \kfo a}) \ \approx\ -V a \bigg( 4(\kfo a)^{2}-1 \bigg). \nonumber\\
\end{eqnarray}

The t-V model in momentum space is:
\begin{equation}
\mathcal{H}(k)= \sum_{k}(-2t_{hop} \cos{k}-\mu)c^{\dagger}_{k} c_{k}-\frac{V}{N}\sum_{k_1,k_2,q}c^{\dagger}_{k_2+q}c^{\dagger}_{k_1-q}c_{k_2} c_{k_1} .
\end{equation}
The boost is applied by the transformation $k \rightarrow k+u$, for both right movers and left movers. Since the microscopic interactions involve zero momentum exchange, the boost affects only the kinetic term. The Fermi velocity becomes different for the right movers and left movers in the presence of boost (as shown in Eq.~\ref{RL_vf}), with $v^F(u)=v^F(0)\cos(u)$ and $w(u)=2t_{hop}\cos(k^F_0)\sin(u)$ ($v^F(0)$ is the Fermi velocity at zero boost). 
Consequently, the only difference between the Luttinger parameters, $K(u)$ and $v(u)$, in the presence and absence of boost, is due to $v^F(u)$, as shown below. $\tilde{g_4}$ and $\tilde{g}_{2,eff}$ continue to be given by Eqs.~\ref{g4g1} and~\ref{g2eff}, as in the unboosted case.

Consequently, with $a=1$,
\begin{eqnarray}
v^F(u)  =  v^F(0)\cos u, \;\;\;
w(u)  =  2 \hopt \cos (\kfo) \sin u,
\end{eqnarray} 
where $v^F(0)= 2 \hopt \sin (\kfo) =2 \hopt \sin (\pi n) $ and hence,
\begin{eqnarray} \nonumber
v(0) & = & \sqrt{\left(v^F(0) + \frac{\tilde{g}_4}{2\pi} \right)^2 - \left( \frac{\tilde{g}_{2,eff}}{2\pi} \right)^2} \nonumber\\
& =& \sqrt{\left(2\hopt \sin{(\pi n)} - \frac{V}{2 \pi}\right)^2 - \left( \frac{V }{2\pi} (4 \pi^2 n^2-1) \right)^2}, \label{v0} \\
K(0) &=& \sqrt{\frac{v^F(0) - \left(\frac{\tilde{g}_{2,eff}}{2\pi}-\frac{\tilde{g}_4}{2\pi}\right)}{v^F(0) + \left(\frac{\tilde{g}_{2,eff}}{2\pi}+\frac{\tilde{g}_4}{2\pi}\right)}} \nonumber\\
&=&  \sqrt{\frac{ 2\hopt \sin{(\pi n)}+ \frac{V}{\pi}(2 \pi^2 n^2-1)}{2\hopt \sin{(\pi n)} - V (2 \pi n^2)}}. \label{K0}
\end{eqnarray}

In the presence of boost $u$, 
\begin{eqnarray}
v(u) & = & \sqrt{\left(v^F(u) + \frac{\tilde{g}_4}{2\pi} \right)^2 - \left( \frac{\tilde{g}_{2,eff}}{2\pi} \right)^2} \nonumber\\
& =& \sqrt{\left(2\hopt \sin{(\pi n)} \cos{u} - \frac{V}{2 \pi}\right)^2 - \left( \frac{V }{2\pi} (4 \pi^2 n^2-1) \right)^2}, \\
K(u) &=& \sqrt{\frac{v^F(u) - \left(\frac{\tilde{g}_{2,eff}}{2\pi}-\frac{\tilde{g}_4}{2\pi}\right)}{v^F(u) + \left(\frac{\tilde{g}_{2,eff}}{2\pi}+\frac{\tilde{g}_4}{2\pi}\right)}} \nonumber\\
&=&  \sqrt{\frac{ 2\hopt \sin{(\pi n)} \cos{u}+\frac{V}{\pi}(2 \pi^2 n^2-1)}{2\hopt \sin{(\pi n)}\cos{u} - V (2 \pi n^2)}}.
\end{eqnarray}
Also, using Eqs. ~\ref{v0} and ~\ref{K0}, in Eq. ~\ref{Eq:Kwithu}, one can arrive at $K(u)$ for small $u$, in terms of the microscopic parameters $\hopt$, $V$ and $n$.

In the presence of a boost $u$ in Eq. ~\ref{interaction}, the magnitudes of $\tilde{g}$'s remain as they are for the unboosted case, since the scattering processes involve only the difference in the scattering momenta.

\subsection{Pairing Susceptibility}
The pairing susceptibility is given by:
\begin{equation} \label{suscp}
\chi_{\rm{pair}}(q=0,\omega)\ =\ \frac{1}{\Omega} \sum _k \frac{ f(\xi _k) - f(-\xi _{-k}) }{\omega - \xi(k) - \xi(-k) + i\delta},
\end{equation}
where $ f(\xi_k)$ is the Fermi distribution at energy $\xi(k)$, where  $\xi(k) = -2 \hopt \cos{k} - \mu$. $\Omega$ is the volume of the system. 

Linearizing the dispersion about the Fermi points, we get~\cite{giamarchi}:
\begin{eqnarray}\nonumber 
\xi(k)& \simeq & v_R^F (k - k_R^F),\ \ k \sim k_R^F \\
\xi(-k)& \simeq & v_L^F (-k - k_L^F),\ \ k \sim k_L^F,
\label{Eq:xi}
\end{eqnarray}
where: $ v^F_s=v^F(u)+s w(u) $ and $s=1(-1)$ for the right(left) movers, as given in the main text. $k_s^F$ is the Fermi momentum form the right($s=R$) and left($s=L$) movers. 

The pairing susceptibility is largest when the Fermi level is in the middle of the band, which corresponds to half-filling. We thus, calculate it as a function of boost for this value of filling using Eq. (\ref{Eq:xi}). The result is shown in Fig.~\ref{Fig:susceptibility}, which is a plot of the ratio of the susceptibility of the boosted system to that of the unboosted system $\chi(u)/\chi(0)$. It can be seen that the susceptibility increases as a function of the boost consistent with the strengthening of superconducting order. This is true even for values of filling different from half-filling. Note that there is a divergent factor of $\log T$, where $T$ is the temperature that cancels between the numerator and denominator of the quantity $\chi(u)/\chi(0)$.

\begin{figure}[H]
\centering
\includegraphics[width=8cm, height=6cm]{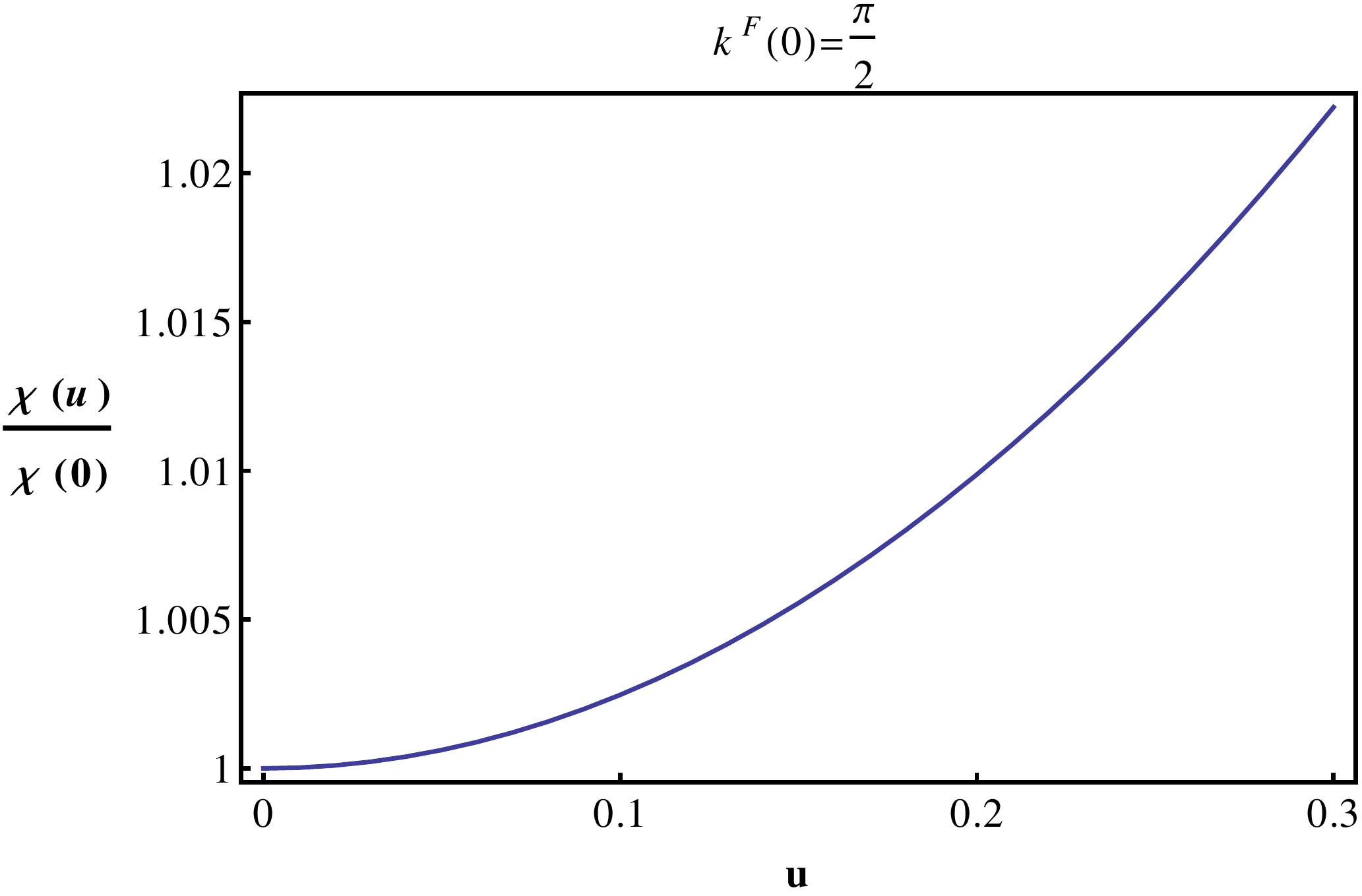} 
\caption{The ratio of the boosted and unboosted pairing susceptibility $\chi(u)/\chi(0)$ at half-filling as a function of $u$. It can be seen that the susceptibility increases with $u$ which is consistent with the strengthening of superconducting order. Note that there is a divergent factor of $\log T$, where $T$ is the temperature that cancels between the numerator and denominator of the quantity $\chi(u)/\chi(0)$.}
\label{Fig:susceptibility}
\end{figure}

\subsection{Effect of spin and umklapp}
Having analyzed the effect of a boost on a system of spinless fermions, we now turn our attention to spin $1/2$ systems. In the absence of a boost, it is known that the charge and spin degrees of freedom can be separated in the low-energy physics, each being described by its own hamiltonian $H_\nu$, fields $\Pi_\nu$ and $\phi_\nu$, Fermi velocity $v_\nu$ and Luttinger parameter $K_\nu$, where $\nu=\rho [\sigma]$ for the charge [spin] sector ~\cite{giamarchi}. 
A point of difference between the sectors is that the spin sector has umklapp even when the underlying system possesses Galilean invariance while the charge sector does not. Umklapp can be relevant in the charge sector only for systems with broken Galilean invariance and commensurate filling. Since we need to break Galilean invariance for the boost to have a non-trivial effect, the low energy physics of the spin and charge sectors is described by the Hamiltonian in Eq. (\ref{Eq:Ham_umklapp}).
 \begin{equation}
 \begin{split}
\mathcal{H_\nu} & = \frac{v_\nu (u)}{\pi}\int_{-L/2}^{L/2} dx \left[K_\nu (u):\left(\tilde{\Pi}_\nu(x) \right)^2: \right. \\& \left. + \frac{1}{K_\nu (u)} :\left(\nabla \tilde{\phi}_\nu (x) \right)^2:  + \frac{g_\nu (u) }{a^2} :\cos(\alpha_\nu  \tilde{\phi}_\nu ): \right],
\end{split}
\label{Eq:Ham_umklapp}
\end{equation}
where $g_\nu (u)$ is a dimensionless parameter and is a function of $u$ [$v_\nu (u) g_\nu (u)$ is the strength of the umklapp term and has the dimensions of energy], $a$ an ultra-violet cutoff and $\alpha_\rho= \sqrt{16 \pi}$ and $\alpha_\sigma= \sqrt{8 \pi}$  for charge and spin respectively. We emphasize again that $g_\rho$ is operative only at commensurate filling although for a system in contact with a container, a similar term may arise with a phase oscillating in space with a minimum wavenumber ~\cite{oshikawa}.
Again, the Hamiltonian reduces to the standard form when $u=0$ ~\cite{giamarchi,gnt,schulz,haldane} . 

\subsection{RG Flow and phase diagram} 
The renormalization group flow equations for the parameters $g_\nu$ and $K_\nu$ at tree level, in terms of two new parameters $h_\nu=2\left( \frac{K_\nu}{K^c_\nu}-1 \right)$ and $g^\perp_\nu=K^c_\nu g_\nu$ with $K^c_\nu = \frac{8 \pi}{\alpha_\nu^2}$ are ~\cite{gnt}:  \begin{equation} \label{Eq:RG_flow}
\frac{d g^\perp_\nu}{d l}\ =\ - h_\nu g^\perp_\nu, \;\;\;\;
  \frac{d h_\nu}{d l}\ =\ - \left( g^\perp_\nu \right)^2.
\end{equation}
The above equations can be integrated to obtain flow lines and for $h_\nu<0$, $g^\perp_\nu$ is a relevant perturbation and opens a gap. For $h_\nu>0$, the flow terminates at $g^\perp_\nu=0$ and $h_\nu=h^*_\nu$ (i.e. a Luttinger liquid results). This has the value $h_{\nu,i}^2 - \left( g^\perp_{\nu,i} \right)^2 = \left(h^*_\nu \right)^2$, where $h_{\nu,i}$ and $g^\perp_{\nu,i}$ are the initial (bare) values of $h_\nu$ and $g^\perp_\nu$.

\begin{figure}[h]
\centerline{\includegraphics[width=6cm, height=5.5cm]{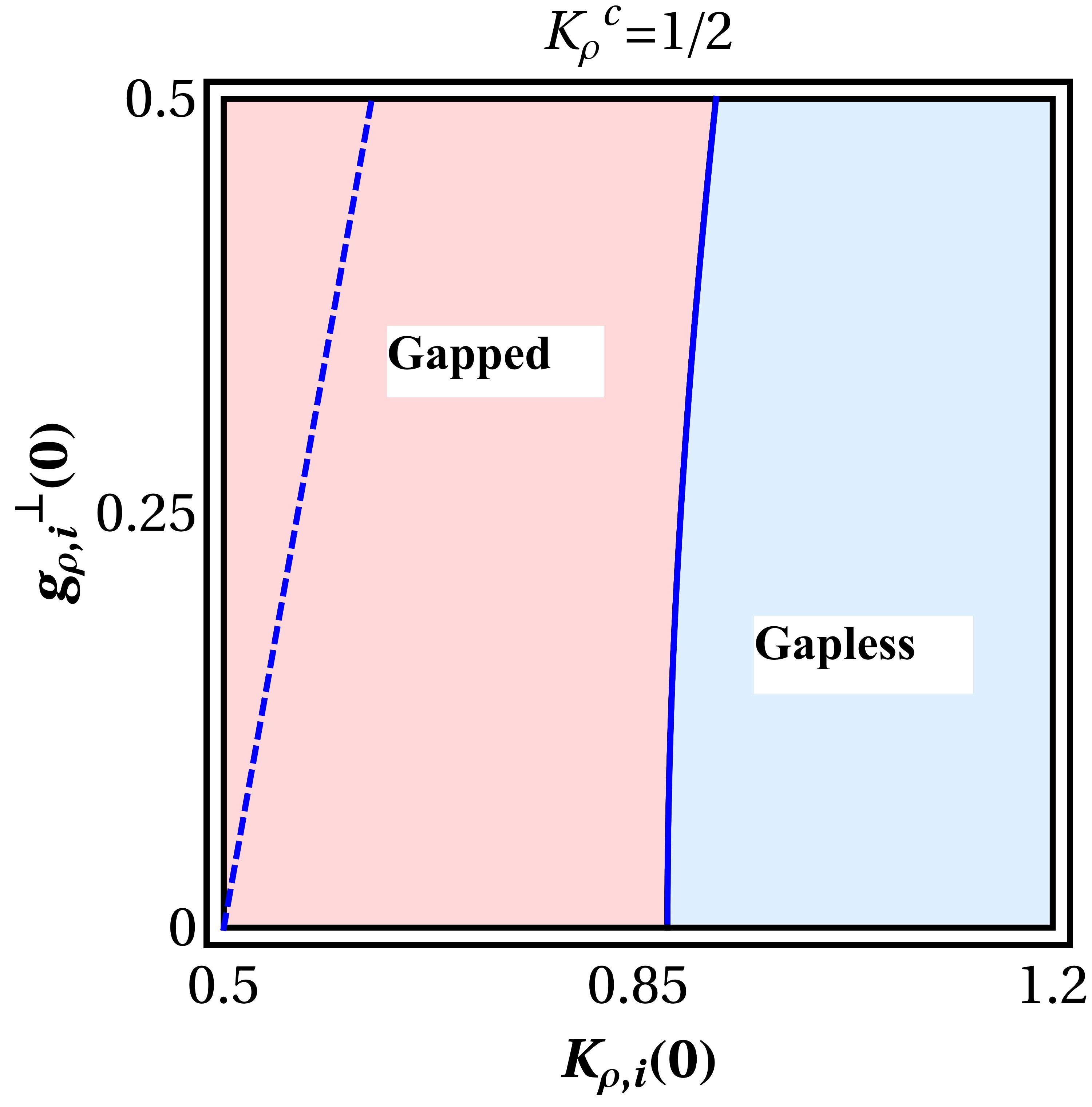}~
\includegraphics[width=6cm, height=5.5cm]{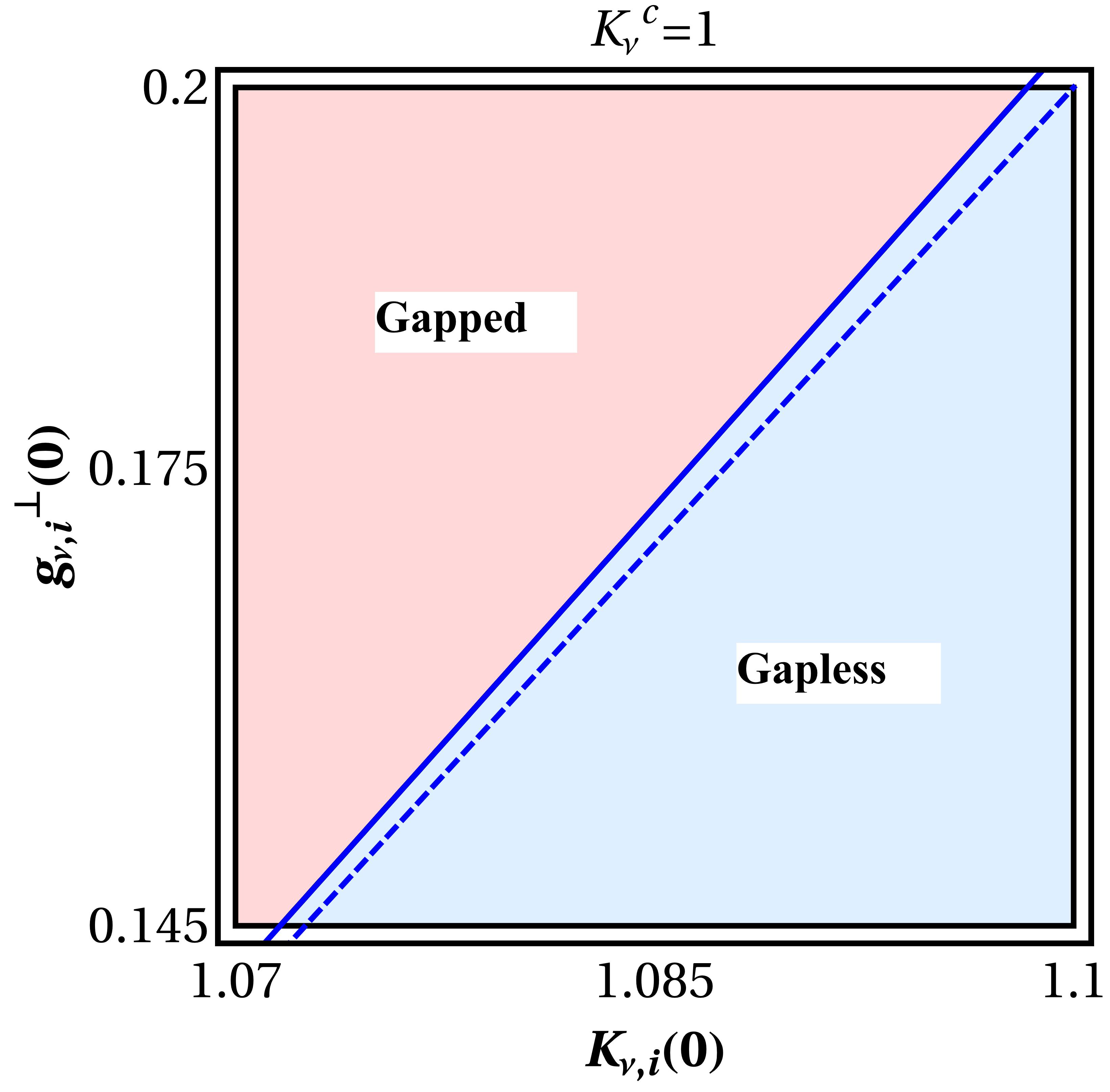}} 
\caption{The phase diagram of spinless (left) and spin $\half$ (right) boosted fermions. The dashed line separates the gapped and gapless phases at zero boost. Suitable boosts can be applied to transform this to the solid line which separates the new gapped and gapless regions as indicated.}
\label{fig:PhaseDiagram}
\end{figure}

Eqs. (~\ref{Eq:RG_flow}) are valid even for $u\neq 0$. 
Suppose the unboosted system starts the flow from $h_{\nu,i}(0)$ and $g^\perp_{\nu,i} (0)$, then, the flow follows:
\begin{equation}
 h_{\nu,i}(0)^2 - g^\perp_{\nu,i} (0)^2\ =\ h^{*}_\nu (0)^{2} .
\end{equation}
We need to see where the flow is headed under the effect of the boost, i.e, we need to find $h^{*}_\nu (u)$ .
Then the modifications to the unboosted equations are obtained by introducing the parameters as functions of $u$:
\begin{equation}
 \begin{split}
  K_{\nu,i} (u)\ &=\ K^c_{\nu} \bigg[1+ \frac{h_{\nu,i}(u)}{2}\bigg] ,\\
  g^{\perp}_{\nu,i}(u)\ &=\ g^\perp_{\nu,i} (0) \bigg[ \frac{v_{\nu,i}(0)}{v_{\nu,i}(u)} \bigg] ,
  \label{Eq:renormkg}
 \end{split}
\end{equation}
where  $v_{\nu,i}(0)$ and $v_{\nu,i}(u)$ are the initial values of the renormalized Fermi velocity in the absence of boost and in the presence of boost respectively.
Note that the interaction parameter $V$ in Eq. (\ref{Eq:interaction}) corresponds to the product $v_\nu g_\nu$ in Eq. (\ref{Eq:Ham_umklapp}). $V$ does not change under the effect of the boost which implies that $g_\nu(u) \sim 1/v_\nu(u)$, which yields the second of Eqs. (~\ref{Eq:renormkg}).
Define:
\begin{align}
 a_{\nu,i}(u) &= v^{F}_{i} (u) + \frac{\tilde{g}_{4 \nu,i}}{2 \pi} - \frac{\tilde{g}_{2 \nu,i}}{2 \pi} ,\\
 b_{\nu,i}(u) &= v^{F}_{i} (u) + \frac{\tilde{g}_{4 \nu,i}}{2 \pi} + \frac{\tilde{g}_{2 \nu,i}}{2 \pi} .
\end{align} 
Then, $K_{\nu,i}(u) = \sqrt{\frac{a_{\nu,i}(u)}{b_{\nu,i}(u)}}$ ,
$ v_{\nu,i}(u) = \sqrt{a_{\nu,i}(u) b_{\nu,i}(u)}$ and
$ v^{F}_{i} (u) = v^F_{i}(0) (1+ f(u))$ , where,
$ f(u) = \cos{u} -1$ for the usual tight-binding model,  and $v^F_{i}(0)$ is the initial (bare) value of the Fermi velocity in the absence of boost.

Solving for $K_{\nu,i}(u)$ in terms of $a_{\nu,i}(u)$ and $b_{\nu,i}(u)$ :
\begin{eqnarray} \nonumber
 K_{\nu,i}(u)\ & = & \ \sqrt{\frac{a_{\nu,i}(u)}{b_{\nu,i}(u)}} \\ \nonumber
 & = & \sqrt{\frac{a_{\nu,i}(0) + v^F_{i}(0) f(u)}{b_{\nu,i}(0) + v^F_{i}(0) f(u)}} \\ \nonumber
 & = & \sqrt{\frac{a_{\nu,i}(0)}{b_{\nu,i}(0)} \bigg(1 + \frac{v^F_{i}(0)}{a_{\nu,i}(0)} f(u)\bigg)\bigg(1 - \frac{v^F_{i}(0)}{b_{\nu,i}(0)} f(u) \bigg)} \\ 
 & = & K_{\nu,i}(0) \bigg[1 + \frac{v^F_{i}(0) f(u)}{2 a_{\nu,i}(0) b_{\nu,i}(0)} (b_{\nu,i}(0) - a_{\nu,i}(0)) \bigg] 
\end{eqnarray}

Solving for $v_{\nu,i}(u)$:
\begin{eqnarray} \nonumber
 v_{\nu,i}(u)\ & = & \sqrt{a_{\nu,i}(0) b_{\nu,i}(0)} \\ \nonumber
 & = & \sqrt{[a_{\nu,i}(0) + v^F_{i}(0) f(u)][b_{\nu,i}(0) + v^F_{i}(0) f(u)]} \\
 & = & v_{\nu,i}(0) \bigg(1 + \frac{v^F_{i}(0) (a_{\nu,i}(0) + b_{\nu,i}(0)) }{2 a_{\nu,i}(0) b_{\nu,i}(0)} f(u) \bigg) .
\end{eqnarray}

\vspace{0.5cm}

$K_{\nu,i}(0)$ and $v_{\nu,i}(0)$ are the unboosted value of the initial points from where the flow starts. $K_{\nu,i}(0)$ is then expanded about its critical point $K^c_\nu$ : $ K_{\nu,i} (0)\ =\ K^c_\nu \big(1 + \frac{h_{\nu,i}(0)}{2} \big) $ . 
\vspace{0.5 cm}

Thus,
\begin{eqnarray} \nonumber
 K_{\nu,i}(u)\ &=& K^c_\nu \big(1 + \frac{h_{\nu,i}(0)}{2} \big) \bigg[1 + \frac{v^F_{i}(0) f(u)}{2 a_{\nu,i}(0) b_{\nu,i}(0)} (b_{\nu,i}(0) - a_{\nu,i}(0)) \bigg] \\ \nonumber
 &=& K^c_\nu \bigg[ 1 + \frac{h_{\nu,i}(0)}{2} \bigg(1 + \frac{v^F_{i}(0) f(u)}{2 a_{\nu,i}(0) b_{\nu,i}(0)} (b_{\nu,i}(0) - a_{\nu,i}(0)) \bigg) \\ \nonumber
 & + &  \frac{v^F_{i}(0) f(u)}{2 a_{\nu,i}(0) b_{\nu,i}(0)} (b_{\nu,i}(0) - a_{\nu,i}(0)) \bigg]  \\
 & = & K^c_\nu \left(1+\frac{h_{\nu,i}(u)}{2} \right)
\end{eqnarray}

where \begin{eqnarray} \nonumber
 h_{\nu,i}(u)\ &=& h_{\nu,i}(0) \bigg(1 + \frac{v^F_{i}(0) f(u)}{2 a_{\nu,i}(0) b_{\nu,i}(0)} (b_{\nu,i}(0) - a_{\nu,i}(0)) \bigg) \\
& & +  \frac{v^F_{i}(0) f(u)}{ a_{\nu,i}(0) b_{\nu,i}(0)} (b_{\nu,i}(0) - a_{\nu,i}(0)) \\
 &=& 2 \bigg( \frac{K_{\nu,i}(0)}{K^c_\nu}-1 \bigg)-\frac{v^F_{i}(0) f(u)}{v_{\nu,i}(0) K^c_\nu} \bigg(K_{\nu,i}(0)^2-1 \bigg) .
\end{eqnarray}
\vspace{0.5cm}

Since the interaction term in the microscopic hamiltonian remains unchanged, $ g^{\perp}_{\nu,i}(0) v_{\nu,i}(0) = g^{\perp}_{\nu,i}(u) v_{\nu,i}(u)$ \hphantom{} \\
which gives, \begin{eqnarray} \nonumber
      g^{\perp}_{\nu,i} (u)\ &=& g^{\perp}_{\nu,i}(0) \bigg( \frac{v_{\nu,i}(0)}{v_{\nu,i}(u)} \bigg) \\ \nonumber
      &=& g^{\perp}_{\nu,i}(0) \bigg(1 - \frac{v^F_{i}(0) (a_{\nu,i}(0) + b_{\nu,i}(0)) }{2 a_{\nu,i}(0) b_{\nu,i}(0)} f(u) \bigg) \\
      &=& g^{\perp}_{\nu,i}(0) \bigg[1- \frac{v^F_{i}(0) f(u)}{2 v_{\nu,i}(0)}\bigg(K_{\nu,i}(0)+\frac{1}{K_{\nu,i}(0)}\bigg) \bigg] . 
     \end{eqnarray}
\vspace{0.5cm}

Using the expressions for $h_{\nu,i}(u)$ and $g^{\perp}_{\nu,i}(u)$ in the flow equation we get the expression for the new fixed point,

\begin{eqnarray} \nonumber
 h_{\nu,i}(u)^2  - g^{\perp}_{\nu,i}(u)^2 \ &=& \bigg[h_{\nu,i}(0) \bigg(1 + \frac{v^F_{i}(0) f(u)}{2 a_{\nu,i}(0) b_{\nu,i}(0)} (b_{\nu,i}(0) - a_{\nu,i}(0)) \bigg) \nonumber \\
& + & \frac{v^F_{i}(0) f(u)}{ a_{\nu,i}(0) b_{\nu,i}(0)} (b_{\nu,i}(0) - a_{\nu,i}(0)) \bigg]^2 \\ \nonumber
  & - & g^{\perp}_{\nu,i}(0)^2 \bigg(1 - \frac{v^F_{i}(0) (a_{\nu,i}(0) + b_{\nu,i}(0)) }{2 a_{\nu,i}(0) b_{\nu,i}(0)} f(u) \bigg)^2 \\ \nonumber
&=& \left[h_\nu^*(0)\right]^2 + \frac{v^F_{i}(0) f(u) }{a_{\nu,i}(0) b_{\nu,i}(0)} \bigg[h_{\nu,i}(0) \bigg(h_{\nu,i}(0) + 2 \bigg)\bigg(b_{\nu,i}(0)-a_{\nu,i}(0)\bigg) \\
& + & g^{\perp}_{\nu,i}(0)^2 \bigg(a_{\nu,i}(0) + b_{\nu,i}(0)\bigg) \bigg] ,
\end{eqnarray}
\vspace{0.5cm}

which is, \begin{equation}
\begin{split}
\left[h_\nu^*(u)\right]^2\ &=\ \left[h_\nu^*(0)\right]^2 + \frac{v^F_{i}(0) f(u) }{a_{\nu,i}(0) b_{\nu,i}(0)} \bigg[h_{\nu,i}(0)  \bigg(h_{\nu,i}(0) + 2 \bigg)\bigg(b_{\nu,i}(0)-a_{\nu,i}(0)\bigg) \\
& + g^{\perp}_{\nu,i}(0)^2 \bigg(a_{\nu,i}(0) + b_{\nu,i}(0)\bigg) \bigg] \\
&=\ \bigg[4 \bigg( \frac{K_{\nu,i}(0)}{K^c_\nu}-1 \bigg)^2 - g^{\perp}_{\nu,i}(0)^2 \bigg] + \frac{v^F_{i}(0) f(u)}{v_{\nu,i}(0) K^c_\nu} \bigg[ 4 \bigg(K_{\nu,i}(0)^2-1 \bigg) \bigg(1-\frac{K_{\nu,i}(0)}{K^c_\nu} \bigg) \\
& + \frac{ g^{\perp}_{\nu,i}(0)^2 K^c_\nu}{K_{\nu,i}(0)} \bigg(K_{\nu,i}(0)^2+1 \bigg) \bigg] ,
\end{split}
\end{equation}
where $h_\nu^*(u)$ gives the new fixed point, as a function of the initial unboosted starting point $K_{\nu,i}(0)$ and $g^{\perp}_{\nu,i}(0)$, along with $v_{\nu,i}(0)$, $v^F_{i}(0)$ and the boost $u$ .

Thus, summarizing, the only effect of the boost is to change the values of the initial parameters in the following way: 
\begin{eqnarray} \nonumber
h_{\nu,i}(u) & = & 2 \bigg(\frac{K_{\nu,i}(0)}{K^c_\nu}-1 \bigg) - \frac{v^F_{i}(0)  f(u)}{v_{\nu}(0) K^c_\nu}  \left[\left(K_{\nu,i}(0) \right)^2 - 1\right]  \\ 
g^\perp_{\nu,i}(u)& = & g^\perp_{\nu,i}(0) \left[1 - \frac{v^F_{i}(0) f(u)}{2 v_{\nu}(0)} \bigg(K_{\nu,i}(0) + \frac{1}{K_{\nu,i}(0)} \bigg) \right].
\label{Eq:hg_boost}
\end{eqnarray}
where $f(u) = - \frac{v^F_{i}(0)-v^F_{i}(u)}{v^F_{i}(0)},$ is assumed to be small and $K_{\nu,i}(0) = K^c_\nu [1+h_{\nu,i}(0)/2]$. Note that the first of the above equations is the same as Eq. (\ref{Eq:Kwithu}) but with $K^c_\rho=1/2$ as is appropriate for spinless fermions. Even in this more general case with spin, the boost has the effect that $K_{\nu,i} (u) > K_{\nu,i}(0) [K_{\nu,i} (u) <  K_{\nu,i}(0)]$ if $K_{\nu,i}(0) > 1 [K_{\nu,i}(0) < 1]$.

The value $h_\nu^*(u)$ can be determined for a flow staring at $h_{\nu,i}(u)$ and $g^\perp_{\nu,i}(u)$ and is given by: 
\begin{equation}
\begin{split}
\left[h_\nu^*(u)\right]^2 &= 4 \left(\frac{K_{\nu,i}(0)}{K^c_\nu} - 1 \right)^2 -\left(g^\perp_{\nu,i}(0) \right)^2 \\
& + \frac{v^F_{i}(0) f(u) }{v_{\nu}(0) K^c_\nu} \bigg[4 (K_{\nu,i}(0)^2 - 1) \left(1 - \frac{K_{\nu,i}(0)}{K^c_{\nu,i}} \right) \\
& + \frac{g^\perp_{\nu,i}(0)^2 K^c_\nu}{K_{\nu,i}(0)} (K_{\nu,i}(0)^2 + 1) \bigg].
\end{split}
\end{equation}

It can be seen that in the presence of umklapp a sector is gapped (gapless) when $K_\nu^*(u) < K^c_\nu \left[ K_\nu^*(u) \geq K^c_\nu \right]$. For spinless fermions, $K^c_\rho=1/2 < 1$ ~\cite{giamarchi,gnt} and so when superconductivity dominates in the charge sector ($K_{\rho, i}(0)>1$), it is strengthened when the system is boosted just like in the absence of umklapp and the charge sector continues to be gapless. When $K_{\rho,i}(0) < 1$, gapless CDW order results down to a critical value of $g^\perp_{\rho,i}$, below which a gapped state is obtained, which can even result in a long range CDW case.

It can be seen from Fig.~\ref{fig:RGflow} that upon the application of a boost (which has the effect of reducing the value of $K_{\rho,i}$), a gapless CDW state can be transformed into a gapped one. {\em Thus a boost can convert quasi-long-ranged CDW order into true long-range order.} If the CDW state continues to remain gapless upon the application of a boost, the order is strengthened like in the case without umklapp.

For spinful fermions on a lattice, $K^c_\rho=K^c_\sigma=1$  ~\cite{giamarchi} .
Consequently, for $K_{\rho,i} < 1$, the system is always gapped and it is possible to have such a phase even when $K_{\rho,i} \geq 1$ depending on the value of $g^\perp_{\rho,i}$ as can be seen in Fig.~\ref{fig:RGflow}. A boost cannot open a charge gap in this case unlike for spinless fermions. However, it can close an existing gap for systems with a certain range of values of $K_{\rho, i}$ and $g^\perp_{\rho, i}$ as can be seen in Fig.~\ref{fig:PhaseDiagram}. This happens only for $K_{\rho,i}>1$. The boost has exactly the same effect in the spin sector as well (Note that $K_\sigma=1$ is a  rotationally invariant point and hence the boost has no effect on this point).
 
The above conclusions open up the possibility of transforming a system with a gap in the charge or spin sector or both into a different phase by closing one or both gaps upon the application of a boost. Of particular interest is a system with both a charge and spin gap. If $K_{\rho,i}(0)$ and $g^\perp_{\rho,i}(0)$ lie in the blue colored region between the dashed and the solid line in Fig.~\ref{fig:PhaseDiagram}, a boost can close a charge gap. If $K_{\sigma, i}(0)$ and $g^\perp_{\sigma,i}(0)$ lie in the red region (beyond the solid line), the boost cannot close the spin gap and the resultant state is a Luther-Emery fluid with gapped spin excitations and gapless charge excitations ~\cite{le1,le2}. {\em Thus, it is possible to obtain a Luther-Emery fluid from a fully gapped system by applying a boost which suggests a new way of obtaining such a fluid in experiments on trapped cold atoms} ~\cite{zwierlein,esslinger,wang}. 
It is also possible to destroy the spin gap of a Luther-Emery fluid by applying a boost if $K_{\sigma, i}(0)$ and $g^\perp_{\sigma,i}(0)$ for the systems lie in the blue colored region between the solid and the dashed lines, as shown in Fig.~\ref{fig:PhaseDiagram}.

\section{Conclusion} \label{conclude}
To conclude, we have shown that the application of a boost can strengthen the superconductivity of typical one dimensional systems with no Galilean invariance, in contrast to their higher dimensional counterparts. A similar effect exists for CDW order as well. At commensurate filling, the boost can open a charge gap for systems of spinless fermions. For spin $1/2$ fermions, a boost applied to a fully gapped system can produce a Luther-Emery fluid with gapped spin and gapless charge excitations.

\cleardoublepage

}


%% file: chapter4_2.tex
\chapter{\LARGE Symmetry classes and topological properties of 1D superconductors} \label{chapter3}
{\large

\section{Summary of main results}
In this chapter, we study the topological phases and the emergence of Majorana bound states (MBS) at the edges of 1D spinless p-wave superconductor (SC). We classify these systems according to the tenfold scheme on the basis of their topology, resulting from the presence or absence of discrete symmetries. The simplest system we study belongs to the BDI symmetry class, with $Z$ topological invariant and integer number of edge modes. This is reflected by the presence of MBS doublets at the edges. In our analysis, we have considered two types of spinless p-wave pairing with 
$\triangle_{\uparrow\uparrow}=\triangle_{\downarrow\downarrow}$ and $\triangle_{\uparrow\uparrow}=-\triangle_{\downarrow\downarrow}$. We study the type of edge states for these two cases for different values of the chemical potential (measured with respect to the SC gap). These edge states, which are MBS, have an oscillating part along with a decay for $\mu>1/2$ , similar to the form $\sim e^{-x} \sin(x)$. For $0<\mu<1/2$, the MBS are purely decaying. 
Adding perturbations like s-wave pairing and Zeeman fields induces transitions from one topological class to another. We show that there are 3 symmetry classes for the first type of system and 6 for the second, with these perturbations. In the succeeding sections, we analyse the possible perturbations and their combinations, which can place the system in the above number of classes, for each of the two cases.

\section{Introduction} \label{intro}
Majorana fermions (MF) appear in quantum field theory as real solutions of the Dirac equation, when all the non-zero entries of the $\Gamma$ matrices are imaginary~\cite{pal}. They are their own anti-particles. Topological superconductors are theoretically predicted to host Majoranas at their edges, especially in 1D~\cite{kitaev,stern,fu-kane,sau-sarma,oreg}, which appear as Bogulibov quasi-particles and can be expressed as equal superpositions of electron and hole states. MFs have also gathered massive interest due to their non-Abelian exchange statistics and have lead to the idea of low decoherence topological quantum computation~\cite{nayak}.

The 1D Kitaev model, a one-dimensional spinless p-wave superconductor (SC), has eigenstates which are spatially isolated Majoranas at the edges~\cite{kitaev}. Majoranas also appear at the points where transition occurs between a topological and a non-topological phase. For example, in the 1D Kitaev model, if $\mu$ varies as a function of $x$, then between the regions $|\mu|>2t$ and $-t<\mu<t$, two MFs appear at the transition point where the gap closes.  

Based on the discrete symmetries present in the system, anti-unitary  time reversal (TR) $\mathcal{T}$ and particle-hole (p-h) $\mathcal{C}$, and unitary chiral $\mathcal{S}$, all BdG (Bogulibov-de Gennes) Hamiltonians (quadratic Hamiltonian with a gapped spectrum) can be classified into different symmetry classes with a corresponding topological invariant~\cite{tenfold}.  The symmetry operations are defined by: $\mathcal{T}\mathcal{H}(p)\mathcal{T}^{-1}=\mathcal{H}(-p)$, $\mathcal{C}\mathcal{H}(p)\mathcal{C}^{-1}=-\mathcal{H}(-p)$ and $\mathcal{S}\mathcal{H}(p)\mathcal{S}^{-1}=-\mathcal{H}(p)$. If two such gapped quantum systems can be transformed onto one another through a adiabatic path, without closing the gap, they belong to the same topological class. States which can be continuously connected in such a way to an atomic insulator are topologically trivial and those which can not are topologically non-trivial.

\begin{figure}
 \centering
 \includegraphics[width=9cm,height=6cm]{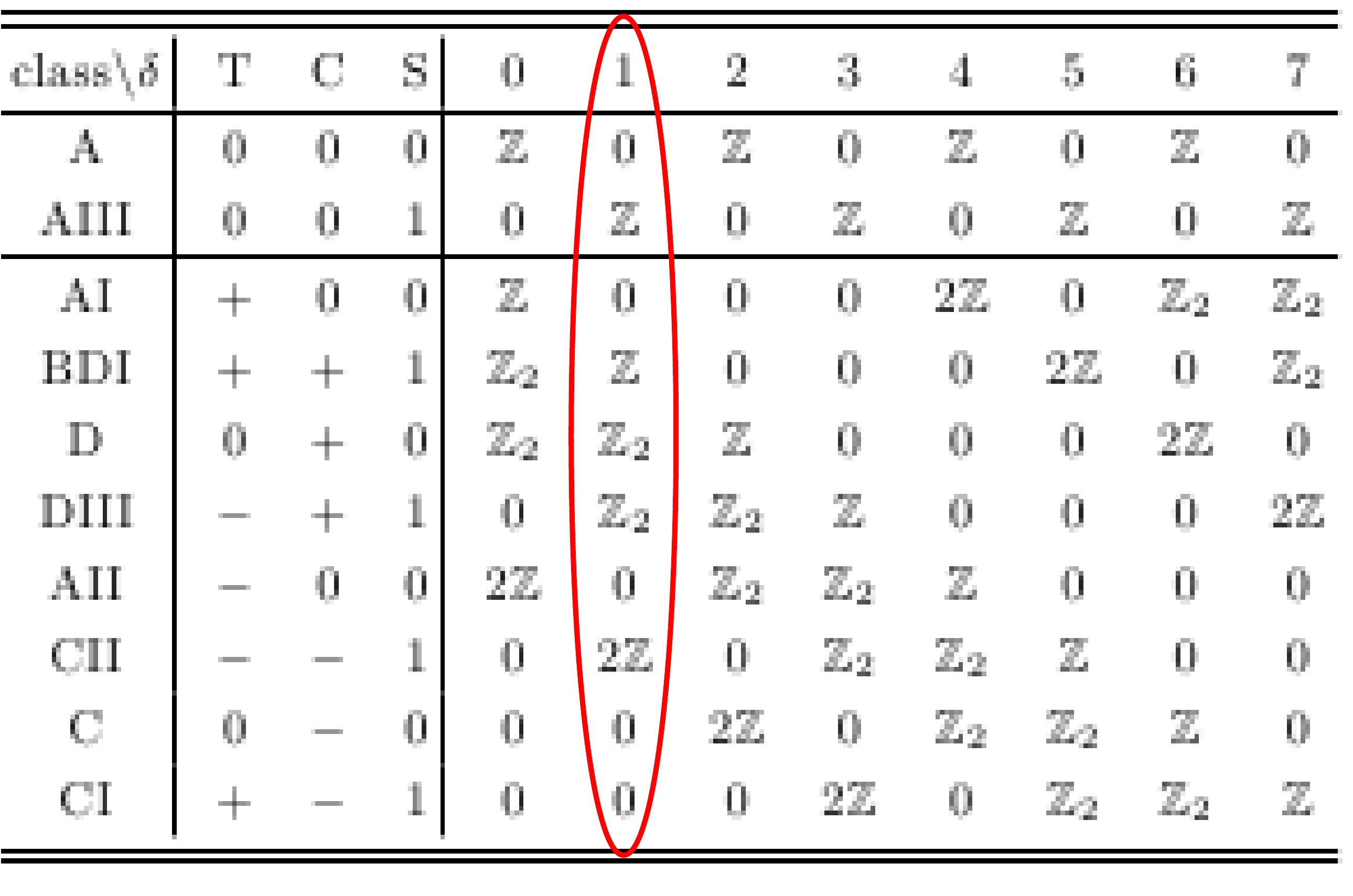}
\caption{Periodic table of Bloch-BdG Hamiltonian, focusing on spatial dimension $D=1$. The column with $\mathcal{T}$, $\mathcal{C}$ and $\mathcal{S}$ represent the values corresponding to $\mathcal{T}^2$, $\mathcal{C}^2$ and $\mathcal{S}^2$, respectively. When such an entry is $0$ it means there is no operator present which gives the corresponding symmetry in the Hamiltonian. The left most column denotes the symmetry class based on the above mentioned discrete symmetries present. The rightmost column gives the corresponding topological invariant for the symmetry class in each spatial dimension.  \small{Ryu, Schnyder {\it et.al}, RMP {\bf 88}, 035005 (2016)}}
\label{tenfold_table}
\end{figure}

For cases, when more than one anti-unitary operator is present for a particular symmetry operation, the Hamiltonian needs to be block diagonalized into irreducible diagonal blocks and an effective operator needs to be identified for each such block to get the correct symmetry operation~\cite{tenfold}. For example, let us consider a 1D spinless p-wave BdG Hamiltonian in the presence of a s-wave pairing $\triangle_1 \sigma_y\tau_y$,
\begin{equation} \label{p+s_1}
\mathcal{H}(p)=\left(\frac{p^2}{2m}-\mu\right)\sigma_0\tau_z+\triangle_0 p\sigma_0\tau_x+\triangle_1 \sigma_y\tau_y,
\end{equation}
where,  $\sigma$ and $\tau$ are Pauli matrices for the spin and the particle-hole sector, $\triangle_0$ and $\mu$ are effective parameters, depending on the underlying heterostructure. The Nambu spinor is $\begin{pmatrix}
c^{\dagger}_{k,\uparrow} & c_{-k,\uparrow} & c^{\dagger}_{k,\downarrow} & c_{-k,\downarrow}
\end{pmatrix}$, and $\triangle_0$ is the magnitude of the p-wave pairing $\triangle_{\uparrow\uparrow} (=\triangle_{\downarrow\downarrow})$. We define the time reversal (TR) operator as an anti-unitary operator which does not mix the particle-hole (p-h) sector. Whereas, the p-h operator is defined as an anti-unitary operator which rotates the particles into holes and holes into particle sector. It is not necessary for the TR operator to flip spin in all cases, which can be seen from the fact that the unperturbed 1D p-wave Kitaev model is TR invariant. TR operators, satisfying the condition  $\mathcal{T}\mathcal{H}(p)\mathcal{T}^{-1}=\mathcal{H}(-p)$, are $\mathcal{T}_1=\sigma_x\tau_z\mathcal{K}$ and $\mathcal{T}_2=\sigma_z\tau_z\mathcal{K}$. The Hamiltonian needs to be block diagonalized in the basis of $U_{\mathcal{T}_1}U^*_{\mathcal{T}_2}$, into $2\times2$ irreducible blocks, where $U_{\mathcal{T}}$ is the unitary operator for the corresponding TR operator $\mathcal{T}$. The above procedure allows us to diagonalize Eq.~\ref{p+s_1} into $\sigma_y=\pm1$ blocks:
\begin{equation} \label{p+s_block}
H_{1,2}=\left(\frac{p^2}{2m}-\mu\right)\sigma_0\tau_z+\triangle_0 p\sigma_0\tau_x\pm\triangle_1\tau_y
\end{equation}
The effective TR operator in each block is $\tau_z\mathcal{K}$. Similarly, there are two possible p-h operator for Eq.~\ref{p+s_1}, $\tau_x\mathcal{K}$ and $\sigma_y\tau_x\mathcal{K}$. However, in this case, due to the last term  $\triangle_1\tau_y$ in the block diagonalized form, neither of the two $\mathcal{C}$ operators is a symmetry. Thus, p-h symmetry is absent ($\mathcal{C}=0$) when a s-wave is present along with a p-wave of the given form. This class of Hamiltonian falls in the topologically trivial AI symmetry class (in $D=1$), with $\mathcal{T}^2=1$, $\mathcal{C}=0$ and $\mathcal{S}=0$, as shown in Fig.~\ref{tenfold_table}.
The possible physical realizations of these perturbations are given below:
\vspace{1cm}

\begin{minipage}{15cm}
\begin{tabular}{l | c }
Perturbation & Physical realization \\
\hline \hline
\begin{minipage}{2cm}
$\mathbf{B}_1\hat{\sigma}\tau_z$ 
$\mathbf{B}_2\hat{\sigma}\tau_0$ 
\end{minipage}
& 
Appropriate Zeeman fields \\
\\
\hline
$\triangle_1 \sigma_y\tau_y$ & 
\begin{minipage}{10cm}
s-wave pairing, which can be generated by proximity coupling to the bulk
of a weak s-wave SC.
\end{minipage}
\\ 
\\
\hline
\end{tabular} 
\end{minipage}
\vspace{1cm}

In this chapter, we study the topological phases and the emergence of Majorana bound states (MBS) at the edges of 1D spinless p-wave SC. It has been known that such systems lie in the DIII symmetry class, with the topological invariant being $Z_2$~\cite{flensberg,tewari1}. However, note that 1D spinless p-wave SC lies in the BDI class, and not in the DIII class, which can be seen by redefining the TR and p-h operators to a more general form as stated above (for standard spin-full systems, the TR operator is $ \sigma_y\mathcal{K}$).  

1D p-wave systems lying in the BDI symmetry class, have $Z$ topological invariant and an integer number of edge modes, which is reflected by the MBS doublets at the edges~\cite{flensberg,tewari1,tewari2}. Application of perturbations, like a s-wave pairing term or a Zeeman field, can cause a transition to a different symmetry class, like the topologically trivial AI class or the non-trivial AIII class, thus changing the nature of the edge state. In this work we study the type of edge states present in a pure p-wave 1D SC for different values of the chemical potential (measured with respect to the SC gap). These edge states, which are MBS, have an oscillating part along with its decaying nature for $\mu>1/2$ , similar to the form $\sim e^{-x} \sin(x)$. For $0<\mu<1/2$, the MBS are purely decaying. In the absence of a coupling term between the up-spin sector and the down spin sector, each MBS belong to one particular spin-sector.


The chapter is arranged as follows. In Sec.~\ref{Sec1}, we study the topological class and the MBS solutions for a spinless p-wave 1D SC, with $\triangle_{\uparrow\uparrow}=\triangle_{\downarrow\downarrow}$.
We have calculated the explicit form of the MBS as a function of $\mu$. Further, we study the three symmetry classes (AI, CI and BDI) that can be generated by perturbations like s-wave pairing and Zeeman term, in Sec.~\ref{1_a1}, Sec.~\ref{1_c1} and Sec.~\ref{1_bd1}. In Sec.~\ref{Sec2}, we study the MBS and the symmetry classes of the second type of spinless p-wave pairing, with $\triangle_{\uparrow\uparrow}=-\triangle_{\downarrow\downarrow}
$. In the consequent sections Sec.~\ref{2_a}, Sec.~\ref{2_c}, Sec.~\ref{2_a1}, Sec.~\ref{2_a3}, Sec.~\ref{2_bd1} and Sec.~\ref{2_d}, we present a detailed analysis of how the six symmetry classes, A, C, AI, AIII, BDI and D, can be generated by applying a s-wave term or Zeeman field on the p-wave SC of the second type. We have also explored the possibility when both the Zeeman fields ($B_1\hat{\sigma}\tau_z$ and $B_2\hat{\sigma}\tau_0$) are simultaneously present as perturbations~\cite{tewari1,tewari2,flensberg}. Without any loss of generality, we have fixed $B_1$ in the $x$-$z$ plane and have considered $B_2$ to be general.

\section{Case I: $\triangle_{\uparrow\uparrow}=\triangle_{\downarrow\downarrow}$
} \label{Sec1}
The first case we have studied is a BdG Hamiltonian (quadratic Hamiltonian describing gapped topological insulator and superconductor) with spinless p-wave superconductivity, such that the $ \triangle_{\uparrow\uparrow}=\triangle_{\downarrow\downarrow}$:
\begin{equation} \label{ham1}
\mathcal{H}_0(p) = \left(\frac{p^2}{2m}-\mu\right)\sigma_0\tau_z+\triangle_0 p\sigma_0\tau_x
\end{equation}
where, $\sigma$ and $\tau$ are Pauli matrices for the spin and the particle-hole sector, $\triangle_0$ and $\mu$ are effective parameters, depending on the underlying heterostructure. The Nambu spinor is $\begin{pmatrix}
c^{\dagger}_{k,\uparrow} & c_{-k,\uparrow} & c^{\dagger}_{k,\downarrow} & c_{-k,\downarrow}
\end{pmatrix}$, and $\triangle_0$ is the magnitude of the pairing $\triangle_{\uparrow\uparrow} (=\triangle_{\downarrow\downarrow})$. 
$\mathcal{T}$ and $\mathcal{C}$ are anti-unitary operators and are thus of the form $\mathcal{T}=U_T\mathcal{K}$ and $\mathcal{C}=U_C\mathcal{K}$, where $\mathcal{K}$ represents complex conjugation and $U_T$ and $U_C$ are unitary operators which can be represented as matrices. For the systems we consider, $U_C$ has to be of the form $\tau_x$, $\tau_y$ or linear combinations of the two, since it represents a particle-hole transformation. The above equation represents the unperturbed system with two decoupled spin sectors.

The symmetry class for Eq.~\ref{ham1} can be identified by studying the TR (time reversal), p-h (particle-hole) and chiral symmetries of the Hamiltonian. The symmetry operations are defined by: $\mathcal{T}\mathcal{H}(p)\mathcal{T}^{-1}=\mathcal{H}(-p)$, $\mathcal{C}\mathcal{H}(p)\mathcal{C}^{-1}=-\mathcal{H}(-p)$ and $\mathcal{S}\mathcal{H}(p)\mathcal{S}^{-1}=-\mathcal{H}(p)$. Using the TR operator $\mathcal{T}=\tau_z \mathcal{K}$, p-h operator $\mathcal{C}=\tau_x \mathcal{K}$ and chiral symmetry operator $\mathcal{S}=i\mathcal{T}.\mathcal{C}$, it can be seen that the Hamiltonian~\ref{ham1} lies in the BDI class, with $\mathcal{Z}$ topological invariant and integer number of edge states~\cite{tenfold}.

To get the explicit form of the edge states, Eq.~\ref{ham1} needs to be solved with the boundary condition $\psi(x=0)=0$, which gives four allowed values of $p$ (with $m,\triangle_0=1$):
\begin{equation}\label{pvalues}
p=\pm \sqrt{-1+\mu\pm\sqrt{1-2\mu}}
\end{equation}
Depending on whether $\mu>1/2$ or $\mu<1/2$, we get two different type of edge states. 
For $\mu>1/2$, the decaying modes are:
\begin{eqnarray} \label{p_mu1}
p_{1} &=& \sqrt{-1+\mu+\sqrt{1-2\mu}} \nonumber\\
p_2 &=& -\sqrt{-1+\mu-\sqrt{1-2\mu}}
\end{eqnarray}

Edge states with $E=0$:
\begin{align}\label{psi_mu1}
\psi_1 &= 2i \begin{pmatrix}
0\\
0\\
i\\
1\\
\end{pmatrix} e^{-\alpha x} \sin{\kappa x},&
\psi_2 &= 2i \begin{pmatrix}
i\\
1\\
0\\
0\\
\end{pmatrix} e^{-\alpha x} \sin{\kappa x},
\end{align}
with $\alpha=\rm{Im}(p_1)=\rm{Im}(p_2)$ and $\kappa=\rm{Re}(p_1)=-\rm{Re}(p_2)$. These edge states have both a decaying and oscillating nature, different from the usual purely decaying form.

For $0<\mu<1/2$, decaying modes: 
\begin{eqnarray} \label{p_mu2}
p_{1} &=& \sqrt{-1+\mu+\sqrt{1-2\mu}} \nonumber\\
p_3 &=& \sqrt{-1+\mu-\sqrt{1-2\mu}}
\end{eqnarray}
Purely decaying edge states with $E=0$:
\begin{align}\label{psi_mu2}
\psi_1 &= \begin{pmatrix}
0\\
0\\
i\\
1\\
\end{pmatrix} (e^{-\alpha_1 x}-e^{-\alpha_2 x}),&
\psi_2 &= \begin{pmatrix}
i\\
1\\
0\\
0\\
\end{pmatrix} (e^{-\alpha_1 x}-e^{-\alpha_2 x})
\end{align}
where, $\alpha_{1,2}=\rm{Im}(p_{1,2})$ and $\rm{Re}(p_1)=\rm{Re}(p_2)=0$. There are no zero energy edge states for $\mu<0$. In both the above cases, with $\mu>1/2$ and $0<\mu<1/2$, the edge states $\psi_1$ and $\psi_2$ are eigenstates of the p-h operator, and hence, are also Majorana Bound states (MBS).

In the following sections we see the accessible topological classes when perturbations like s-wave pairing and Zeeman terms are added. The different types of couplings can be understood from the schematic representation in Fig.~\ref{coupling_diag}.

\begin{figure}
 \centering
 \includegraphics[width=10cm,height=7cm]{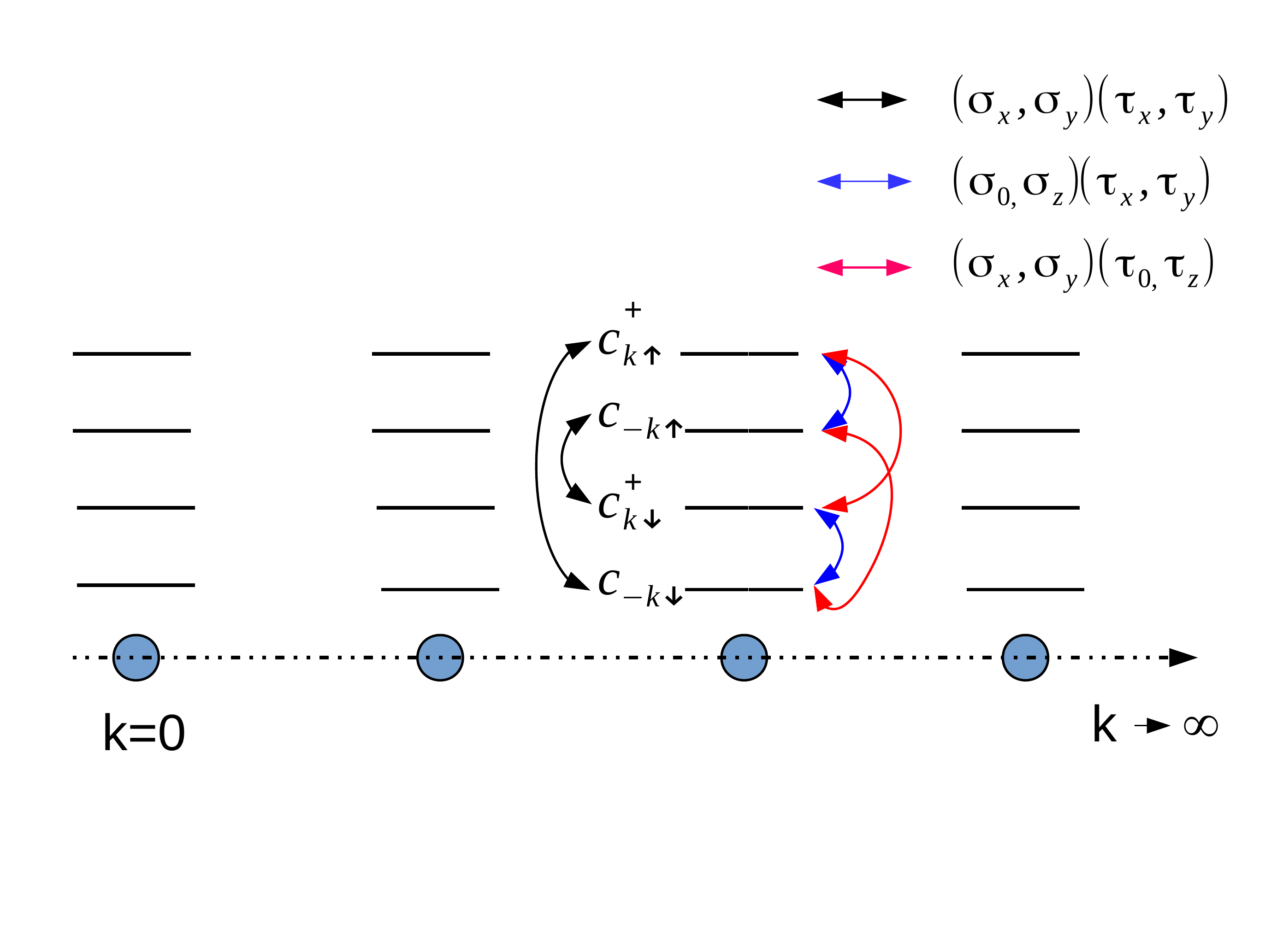}
\caption{Schematic diagram of the different types of coupling in a tight-binding type ladder in momentum space. At each momentum $k$, there is a Nambu spinor:
$\left(
c^{\dagger}_{k,\uparrow}, c_{-k,\uparrow}, c^{\dagger}_{k,\downarrow}, c_{-k,\downarrow}
\right)
$, the four components of which make up the rungs of the ladder. The Hamiltonians studied only contain on-site couplings among the rungs. The arrows stand for these different on-site couplings (including the superconducting pairing and the different Zeeman terms) between the rungs. $\sigma$ and $\tau$ act in spin and particle-hole space, respectively. The black arrows denote couplings of the form $\sigma_x\tau_x$, $\sigma_x\tau_y$, $\sigma_y\tau_x$ and $\sigma_y\tau_y$; the blue arrows for $\sigma_0\tau_x$, $\sigma_0\tau_y$, $\sigma_z\tau_x$ and $\sigma_z\tau_y$; and the red arrows for $\sigma_x\tau_0$, $\sigma_y\tau_0$, $\sigma_x\tau_z$ and $\sigma_y\tau_z$. The $\sigma_y\tau_y$ type of term denotes s-wave pairing. Any coupling of the form $\hat{\sigma}\tau_z$ is a conventional Zeeman term due to an applied magnetic field.}
\label{coupling_diag}
\end{figure}

\subsection{AI class} \label{1_a1}
The AI symmetry class is a topologically trivial class in 1D, with $\mathcal{T}^2=1$, $\mathcal{C}=0$ and $\mathcal{S}=0$. The presence of the Zeeman term, s-wave term or combinations of both, can cause a transition from the BDI class of the unperturbed Hamiltonian in Eq.~\ref{ham1} to the AI class. We elaborate on the effect of each type of perturbation below.

\subsection*{\underline{$\mathbf{B_2.\hat{\sigma}\tau_0}$}}
Since the Hamiltonian in Eq.~\ref{ham1} does not have a spin operator, any magnetic field of the form ${\bf B}_2.\hat{\sigma}\tau_0$ can always be aligned along the z-direction, without any loss of generality. Thus, we can still break our Hamiltonian into two $2\times2$ irreducible blocks. This adds a constant term $B_2\sigma_z\tau_0$ in each $\sigma_z$ block, thus breaking the chiral symmetry. Thus, the MBS are no longer at $E=0$ and a transition occurs from the BDI class to the AI class.

\subsection*{\underline{$\mathbf{\triangle_1 \sigma_y\tau_y}$}}
With the first type of p-wave pairing, the presence of a s-wave term, like $\triangle_1 \sigma_y\tau_y$, is necessary for transition into the AI class, as already discussed in Sec.~\ref{intro}.

\subsection*{\underline{$\triangle_1 \sigma_y\tau_y + B_2\sigma_y\tau_0$,  and,  $\triangle_1 \sigma_y\tau_y + B_1\sigma_y\tau_z$}}
Along with the spinless p-wave term and the s-wave term in Eq.~\ref{p+s_1}, we study the effect of two types of zeeman field $\mathbf{B_1\sigma.\tau_z}$ and $\mathbf{B_2\sigma.\tau_0}$:
\begin{eqnarray}\label{ham1_3}
\mathcal{H}^1_B(p) &=& \mathcal{H}_0(p)+\triangle_1 \sigma_y\tau_y+\mathbf{B}_1.\mathbf{\sigma} \mathbf{\tau_z} \nonumber\\
\mathcal{H}^2_B(p) &=& \mathcal{H}_0(p)+\triangle_1 \sigma_y\tau_y+\mathbf{B}_2.\mathbf{\sigma} \mathbf{\tau_0}
\end{eqnarray}
In the presence of $\triangle_1$, magnetic field along any $\hat{\sigma}$ cannot be aligned along $\sigma_z$ always. Thus, now the entire $4\times4$ Hamiltonian needs to be considered for symmetry analysis. However, when the zeeman field is along $\sigma_y$, the Hamiltonian can still be block diagonalized into $\sigma_y=\pm1$ blocks, and the symmetry analysis can be done for each block. This places $\mathcal{H}^1_B(p)$ and $\mathcal{H}^2_B(p)$  in the same class as $\mathcal{H}(p)$ in Eq.~\ref{p+s_1}. In each case, the effective TR operator is again $\tau_z\mathcal{K}$, p-h operator $\mathcal{C}=0$ and chiral operator $\mathcal{S}=0$.

\subsection*{\underline{$B_1\sigma_z\tau_z + B_2\sigma_z\tau_0$}}

For Hamiltonians with only spinless p-wave SC, $\triangle_{\uparrow\uparrow}=\triangle_{\downarrow\downarrow}$, in the simultaneous presence of both type of Zeeman fields:
\begin{equation} \label{ham_b1b2}
H({\bf p})=\left(\frac{p^2}{2m}-\mu\right)\tau_z +\triangle_0 p\tau_x + \mathbf{B}_1.\hat{\sigma}\tau_z +\mathbf{B}_2.\hat{\sigma}\tau_0
\end{equation}
(where, $\mathbf{B}_1$ and $\mathbf{B}_1$ are the two different Zeeman field), the AI class is again possible. There are two possible TR operators $\tau_z\mathcal{K}$ and $\sigma_z\tau_z\mathcal{K}$, with effective operator in each irreducible block being $\tau_z\mathcal{K}$. However, the p-h $\mathcal{C}$ and chiral operator $\mathcal{S}$ are still absent.

\begin{table}[H]
\centering
\caption{Summary table for perturbations giving AI class}
\setlength{\tabcolsep}{13pt}
\begin{tabular}{l | c | c | c }
Perturbation & $\mathcal{T}$ & $\mathcal{C}$ & $\mathcal{S}$ \\
\hline \hline
$\mathbf{B}_2\hat{\sigma}\tau_0$ & 
\begin{minipage}{2.5cm}
$\tau_z\mathcal{K}$\\
$\mathcal{T}^2=1$
\end{minipage}
 & 
 \begin{minipage}{2.5cm}
$\sigma_y\tau_x\mathcal{K},\
\tau_x \mathcal{K}
$ \\
$\mathcal{C}_{eff}=0$
\end{minipage}
& 0 \\ 
\\
\hline
$\triangle_1 \sigma_y\tau_y$ & 
\begin{minipage}{2cm}$
\tau_z\mathcal{K}
$\\
$\mathcal{T}^2=1$
\end{minipage}
 & 
 \begin{minipage}{2cm}
$ \tau_x\mathcal{K},\
\sigma_y\tau_x\mathcal{K}
$ \\
$\mathcal{C}_{eff}=0$
\end{minipage}
& 
0
\\ 
\\
\hline
$\triangle_1 \sigma_y\tau_y+B_2\sigma_y\tau_0$,\\
$\triangle_1 \sigma_y\tau_y+B_1\sigma_y\tau_z$ & 
\begin{minipage}{1.5cm}$
\tau_z \mathcal{K},
$
$\mathcal{T}^2=1$
\end{minipage}
 & 
 0
& 
0
\\ 
\\
\hline
$B_1\sigma_z\tau_z + B_2\sigma_z\tau_0$& 
\begin{minipage}{2cm}
$\tau_z \mathcal{K},\
\sigma_z\tau_z \mathcal{K}
$
$\mathcal{T}_{eff}=\tau_z\mathcal{K}$\\
$\mathcal{T}^2=1$
\end{minipage}
 & 
 0
& 
0
\\ 
\\
\hline
\end{tabular}

\label{table1}
\end{table}

\subsection{CI class} \label{1_c1}
The CI class is another topologically trivial class in 1D with $\mathcal{T}^2=1$, $\mathcal{C}^2=-1$ and $\mathcal{S}=1$. The p-wave Hamiltonian in Eq.~\ref{ham1} can be transferred to the CI class in the presence of perturbations like the s-wave pairing along with the Zeeman term $\mathbf{B_2}$.

\subsection*{\underline{$\triangle_1 \sigma_y\tau_y + B_2 \sigma_x\tau_0$}}
With the above perturbation over the p-wave case in Eq.~\ref{ham1}, the TR operator is $\sigma_x\tau_z\mathcal{K}$, p-h operator is $\sigma_y\tau_x\mathcal{K}$ and the chiral operator is $\sigma_z\tau_y$, rendering it to be in the CI class.

\subsection*{\underline{$\triangle_1 \sigma_y\tau_y + B_2 \sigma_z\tau_0$}}
Here again, the above perturbation places the Hamiltonian in Eq.~\ref{ham1} in the topologically trivial CI class, with $\mathcal{T}^2=1$, $\mathcal{C}^2=-1$ and $\mathcal{S}=1$. The TR operator is $\sigma_z\tau_z\mathcal{K}$, p-h operator is $\sigma_y\tau_x\mathcal{K}$ and the chiral operator is $\sigma_z\tau_y$. 

\begin{table}[H]
\centering
\caption{Summary table for perturbations giving CI class}
\setlength{\tabcolsep}{13pt}
\begin{tabular}{l | c | c | c }
Perturbation & $\mathcal{T}$ & $\mathcal{C}$ & $\mathcal{S}$ \\
\hline \hline
$\triangle_1 \sigma_y\tau_y+B_2\sigma_x\tau_0$ & 
\begin{minipage}{2.5cm}
$\sigma_x\tau_z\mathcal{K}$\\
$\mathcal{T}^2=1$
\end{minipage}
 & 
 \begin{minipage}{2.5cm}
$\sigma_y\tau_x\mathcal{K}
$ \\
$\mathcal{C}^2=-1$
\end{minipage}
& \begin{minipage}{2.5cm}
$\sigma_z\tau_y\mathcal{K}
$ \\
$\mathcal{S}=1$
\end{minipage} 
\\ 
\\
\hline
$\triangle_1 \sigma_y\tau_y+B_2\sigma_z\tau_0$ & 
\begin{minipage}{2cm}$
\sigma_z\tau_z\mathcal{K}
$\\
$\mathcal{T}^2=1$
\end{minipage}
 & 
 \begin{minipage}{2cm}
$
\sigma_y\tau_x\mathcal{K}
$ \\
$\mathcal{C}^2=-1$
\end{minipage}
& 
\begin{minipage}{2.5cm}
$\sigma_z\tau_y\mathcal{K}
$ \\
$\mathcal{S}=1$
\end{minipage} 
\\ 
\\
\hline
\end{tabular}

\label{table2}
\end{table}

\subsection{BDI class} \label{1_bd1}
This is a topologically non-trivial class (in $D=1$) in which the pure p-wave Hamiltonian in Eq.~\ref{ham1} belongs. Perturbations like the s-wave pairing $\triangle_1 \sigma_y\tau_y$ and the combinations of the two Zeeman fields ($\mathbf{B_1\hat{\sigma}\tau_z}$ and $\mathbf{B_2\hat{\sigma}\tau_0}$) can also generate the BDI class, with topological invariant $Z$.

\subsection*{\underline{$\mathbf{B_1.\hat{\sigma}\tau_z}$}}
In the absence of operator $\sigma$ in the p-wave Hamiltonian in Eq.~\ref{ham1}, any magnetic field can still be aligned along $\sigma_z$, and the Hamiltonian can be block diagonalized into two $2\times2$ irreducible blocks ($\sigma_z=\pm1$). However, due to the $\tau_z$ term, $\mathbf{B_1}$ does not get added as an overall constant in each block in this case. The chiral symmetry is still preserved in the blocks, and the system still remains in the same symmetry class BDI.
The allowed $p$-values will be of the same form as in Eq.~\ref{pvalues}, Eq.~\ref{p_mu1} and Eq.~\ref{p_mu2}, with different chemical potentials $\mu_{\uparrow,eff}=\mu_{\uparrow}-B_1$ for up-spin and $\mu_{\downarrow,eff}=\mu_{\uparrow}+B_1$ for down spin. 


\begin{figure}[h]
\centering
\begin{tabular}{c}
\includegraphics[width=9cm,height=7.5cm]{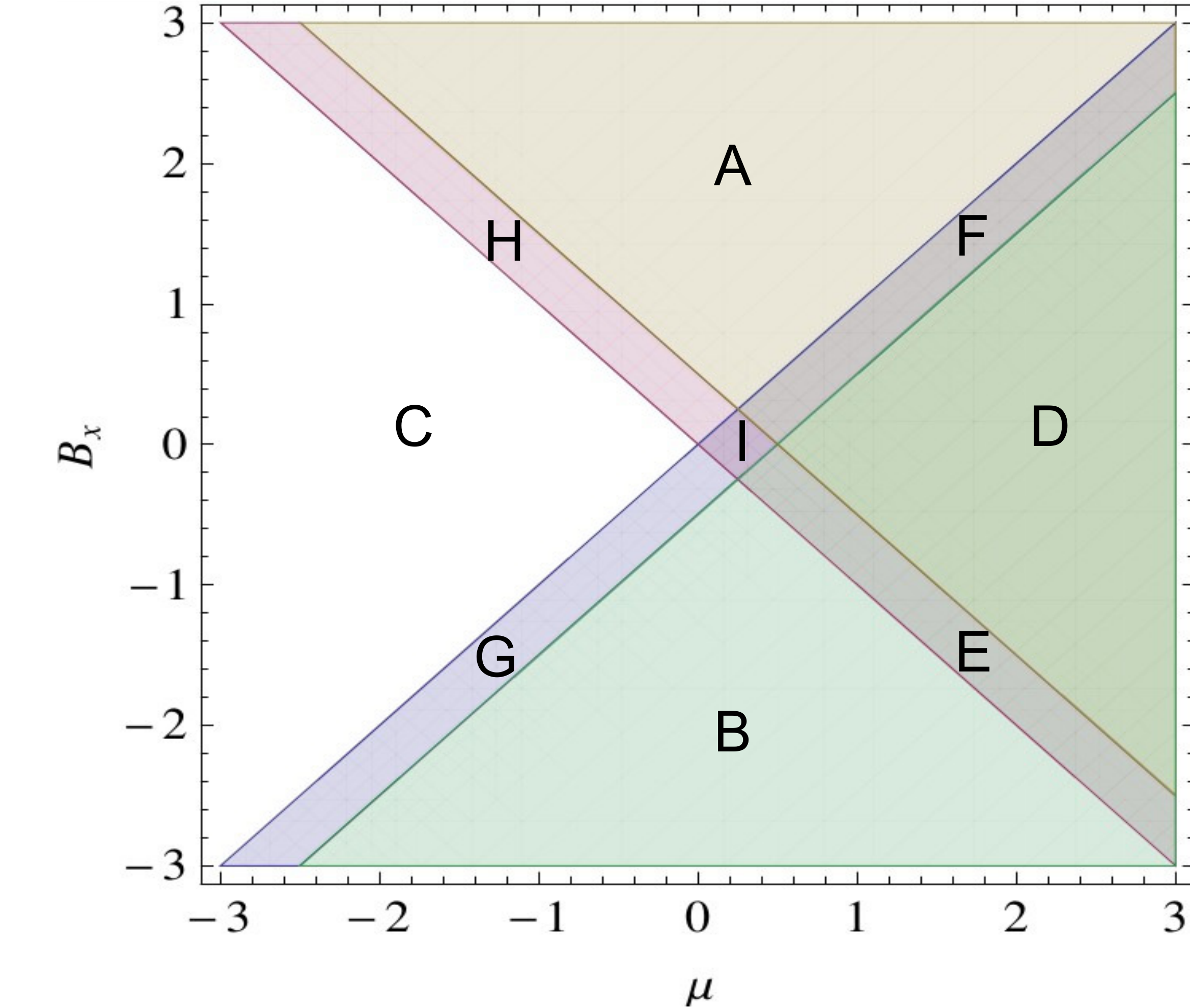}
\includegraphics[width=7cm,height=5.5cm]{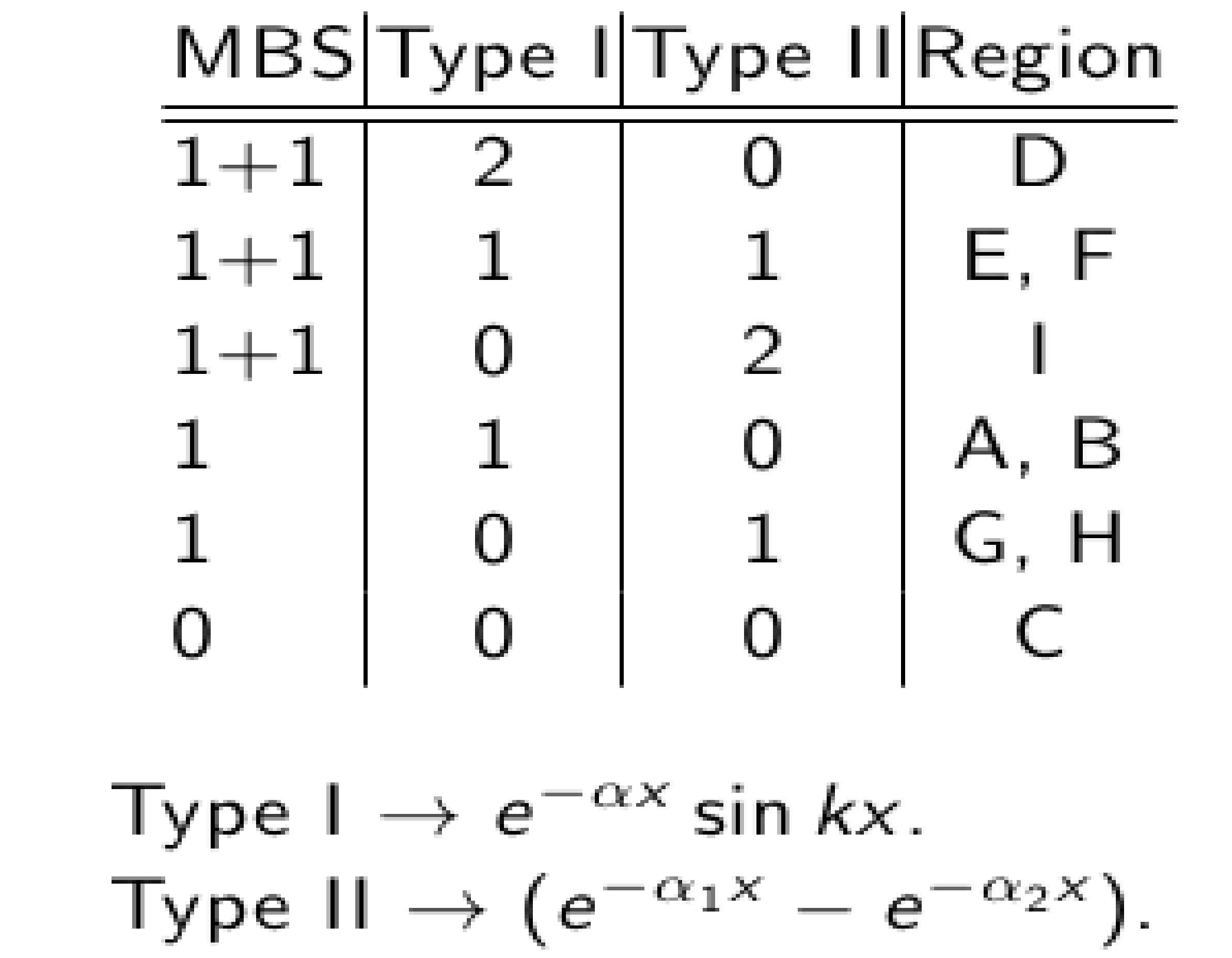}
\end{tabular}
\caption{Left panel: Phase diagram in the $\mu-B_1$ space, showing the number of MBS. For the plot we have considered $\mathbf{B_1}$ along $\sigma_x$. However, as discussed, $\mathbf{B_1}$ along any other direction would still give the same phase diagram. Region C does not have any Majorana modes. All the other coloured regions have a Majorana singlet or a doublet, as indicated in the table on the right. Each of the modes can be of two types, purely decaying or damped oscillating, depending on which region of $\mu$ and $\mathbf{B}_1$ it is in. With the p-wave pairing $\triangle_{\uparrow\uparrow}=\triangle_{\downarrow\downarrow}$, and perturbation of the form $\mathbf{B}_1.\hat{\sigma}.\tau_z$, $\mathbf{B}_1$ along any direction will always give this phase diagram and all phases belong to the BDI class. }
\label{phase1}
\end{figure}

The MBS have the same form as in Eq,~\ref{psi_mu1} and Eq,~\ref{psi_mu2}, now for particular ranges of values in the $\mu-B_1$ space, as shown in Fig.~\ref{phase1}.

\subsection*{\underline{$\triangle_1 \sigma_y\tau_y + B_1 \sigma_x\tau_z$ and $\triangle_1 \sigma_y\tau_y + B_1 \sigma_z\tau_z$}}
Here again the entire $4\times 4$ Hamiltonian, Eq.~\ref{ham1} along with the above perturbations, becomes irreducible and need to be considered in its entirety for symmetry classification. With $\triangle_1 \sigma_y\tau_y + B_1 \sigma_x\tau_z$, the TR operator is $\sigma_x\tau_z\mathcal{K}$, p-h operator is $\sigma_0\tau_x\mathcal{K}$ and chiral operator is $\sigma_x\tau_y$.  For $\triangle_1 \sigma_y\tau_y 
+ B_1 \sigma_z\tau_z$, the TR operator is $\sigma_z\tau_z\mathcal{K}$, p-h operator remains $\sigma_0\tau_x\mathcal{K}$ and chiral operator is $\sigma_z\tau_y$, giving $\mathcal{T}^2=1$, $\mathcal{C}^2=1$ and $\mathcal{S}=1$.

\subsection*{\underline{$B_1 \sigma_z\tau_z+B_2 \sigma_x\tau_0$ and $B_1 \sigma_z\tau_z+B_2 \sigma_y\tau_0$}}
When the two Zeeman fields $\mathbf{B}_1$ and $\mathbf{B}_2$ are simultaneously present in Eq.~\ref{ham1}, and $\mathbf{B}_1$ is fixed along $\hat{z}$, both $B_2 \sigma_x \tau_0$ and $B_2 \sigma_y \tau_0$ give the BDI class separately. For the first case the TR operator is $\tau_z\mathcal{K}$, and the p-h operator is
$\sigma_z\tau_x\mathcal{K}$.
For the second case the TR operator is $\sigma_z\tau_z\mathcal{K}$, and the p-h operator is
$\tau_x\mathcal{K}$. In both cases, the chiral operator is
$\tau_y$, giving $\mathcal{T}^2=1$, $\mathcal{C}^2=1$ and $\mathcal{S}=1$.

\begin{table}[H]
\centering
\caption{Summary table for perturbations giving BDI class}
\setlength{\tabcolsep}{13pt}
\begin{tabular}{l | c | c | c }
Perturbation & $\mathcal{T}$ & $\mathcal{C}$ & $\mathcal{S}$ \\
\hline \hline
$\mathbf{B}_1\hat{\sigma}\tau_z$ & 
\begin{minipage}{2.5cm}
$\tau_z\mathcal{K}$\\
$\mathcal{T}^2=1$
\end{minipage}
 & 
 \begin{minipage}{2.5cm}
$
\tau_x \mathcal{K}
$ \\
$\mathcal{C}^2=1$
\end{minipage}
& \begin{minipage}{2.5cm}
$
\tau_y
$ \\
$\mathcal{S}=1$
\end{minipage}
\\ 
\\
\hline
$\triangle_1 \sigma_y\tau_y+B_1\sigma_x\tau_z$  & 
\begin{minipage}{2cm}$
\sigma_x\tau_z\mathcal{K}
$\\
$\mathcal{T}^2=1$
\end{minipage}
 & 
 \begin{minipage}{2cm}
$ \tau_x\mathcal{K}
$ \\
$\mathcal{C}_{eff}=1$
\end{minipage}
& 
\begin{minipage}{2.5cm}
$
\sigma_x\tau_y
$ \\
$\mathcal{S}=1$
\end{minipage}
\\ 
\\
\hline
$\triangle_1 \sigma_y\tau_y+B_1\sigma_z\tau_z$ & 
\begin{minipage}{1.5cm}$
\sigma_z\tau_z \mathcal{K},
$
$\mathcal{T}^2=1$
\end{minipage}
 & 
 \begin{minipage}{1.5cm}$
\tau_x \mathcal{K},
$
$\mathcal{C}^2=1$
\end{minipage}
& 
\begin{minipage}{1.5cm}$
\sigma_z\tau_y,
$
$\mathcal{S}=1$
\end{minipage}
\\ 
\\
\hline
$B_1\sigma_z\tau_z + B_2\sigma_x\tau_0$& 
\begin{minipage}{2cm}
$\tau_z \mathcal{K}
$\\
$\mathcal{T}^2=1$
\end{minipage}
 & 
 \begin{minipage}{2cm}
$\sigma_z\tau_x \mathcal{K}
$\\
$\mathcal{C}^2=1$
\end{minipage}
& 
\begin{minipage}{2cm}
$\tau_y
$\\
$\mathcal{S}=1$
\end{minipage}
\\ 
\\
\hline
$B_1\sigma_z\tau_z + B_2\sigma_y\tau_0$& 
\begin{minipage}{2cm}
$\sigma_z\tau_z \mathcal{K}
$\\
$\mathcal{T}^2=1$
\end{minipage}
 & 
 \begin{minipage}{2cm}
$\tau_x \mathcal{K}
$\\
$\mathcal{C}^2=1$
\end{minipage}
& 
\begin{minipage}{2cm}
$\tau_y
$\\
$\mathcal{S}=1$
\end{minipage}
\\
\\
\hline
\end{tabular}

\label{table3}
\end{table}
 
\section{Case II: $\triangle_{\uparrow\uparrow}=-\triangle_{\downarrow\downarrow}$
} \label{Sec2}

The second case we have studied is a BdG Hamiltonian with spinless p-wave superconductivity, such that the $ \triangle_{\uparrow\uparrow}=-\triangle_{\downarrow\downarrow}$:
\begin{equation} \label{ham2}
\mathcal{H}_0(p) = \left(\frac{p^2}{2m}-\mu\right)\sigma_0\tau_z+\triangle_0 p\sigma_z\tau_x
\end{equation}
where, $\sigma$ and $\tau$ are Pauli matrices for the spin and the particle-hole sector, and $\triangle_0$ and $\mu$ are the effective parameters. The Nambu spinor is again, $\begin{pmatrix}
c^{\dagger}_{k,\uparrow} & c_{-k,\uparrow} & c^{\dagger}_{k,\downarrow} & c_{-k,\downarrow}
\end{pmatrix}$, and $\triangle_0$ is the magnitude of the pairing $\triangle_{\uparrow\uparrow} (=|\triangle_{\downarrow\downarrow}|)$.

The symmetry class can be identified by studying the TR (time reversal), p-h (particle-hole) and chiral symmetries of the Hamiltonian. The symmetry operations are defined by: $\mathcal{T}\mathcal{H}(p)\mathcal{T}^{-1}=\mathcal{H}(-p)$, $\mathcal{C}\mathcal{H}(p)\mathcal{C}^{-1}=-\mathcal{H}(-p)$ and $\mathcal{S}\mathcal{H}(p)\mathcal{S}^{-1}=-\mathcal{H}(p)$. However, in this case there is more than one possible TR and p-h operator each: $\mathcal{T}=\sigma_x\mathcal{K}$, $\sigma_y\mathcal{K}$ and $\tau_z \mathcal{K}$, and, $\mathcal{C}=\tau_x \mathcal{K}$, $\sigma_z\tau_x \mathcal{K}$, $\sigma_y\tau_y \mathcal{K}$ and $\sigma_x\tau_y \mathcal{K}$. By block diagonalizing the Hamiltonian in Eq.~\ref{ham2} into $\sigma_z=\pm1$ blocks, the effective TR operator is $\tau_z\mathcal{K}$ and p-h operator is $\tau_x\mathcal{K}$. It can be seen that the Hamiltonian lies in the BDI class ($\mathcal{T}^2=1$, $\mathcal{C}^2=1$ and $\mathcal{S}=1$) , with $\mathcal{Z}$ topological invariant and integer number of edge states~\cite{tenfold}.

The allowed $p$ values have the same form as Eq.~\ref{pvalues}. Here again, we have two MBS $\psi_1$ and $\psi_2$, Eq.~\ref{psi_mu1} (for $\mu>1/2$) and Eq.~\ref{psi_mu2} (for $0<\mu<1/2$). Only the structure of the eigenvectors of the MBS is different:
\begin{align} \label{MBS_2}
\begin{pmatrix}
0\\
0\\
i\\
1\\
\end{pmatrix}, &\ \text{and},\ \begin{pmatrix}
-i\\
1\\
0\\
0\\
\end{pmatrix} 
\end{align}
The phase diagram for the MBS in this type of p-wave SC is also the same as in Fig.~\ref{phase1}.

\subsection{A class} \label{2_a}
The A symmetry class is a topologically trivial class in 1D with $\mathcal{T}$, $\mathcal{C}$ and $\mathcal{S}$ all equal to zero. Combination of the s-wave pairing with the $\mathbf{B}_1$ Zeeman field can cause a transition from the BDI class to the A class in the unperturbed p-wave Hamiltonian in Eq.~\ref{ham2}.

\subsection*{\underline{$\triangle_1 \sigma_y\tau_y+B_1 \sigma_x \tau_z$}}
A perturbation of the above form on Eq.~\ref{ham2}, will generate the A symmetry class. Even though for the $4\times4$ Hamiltonian there are two possible p-h operators $\tau_x \mathcal{K}$ and $\sigma_x\tau_y\mathcal{K}$, after block diagonalization, each irreducible block becomes:
\begin{equation}
H_{1,2}= -\left(\frac{p^2}{2m}-\mu\right)\tau_z-\triangle_0 p \tau_x+\triangle_1\tau_x-B_1 \tau_0,
\end{equation}
and none of the above p-h operators is a symmetry. This is another example which shows that even though the entire $4\times4$ Hamiltonian has more than one possible symmetry operator, when block diagonalized into an irreducible form the symmetry might be absent in each block.

\subsection{C class} \label{2_c}
The C symmetry class is again a topologically trivial class in $D=1$ with $\mathcal{T}=0$, $\mathcal{C}^2=-1$ and $\mathcal{S}=0$. Here again, perturbations like the s-wave term and the Zeeman terms $\mathbf{B}_1$ and $\mathbf{B}_2$ on the unperturbed p-wave case, can cause a transition to the C class.

\subsection*{\underline{$\triangle_1 \sigma_y\tau_y+B_2 \sigma_z \tau_0$ and $\triangle_1 \sigma_y\tau_y+B_1 \sigma_y \tau_z$}}
For each of the above perturbations on Eq.~\ref{ham2}, the symmetry operations follow the above conditions, rendering the system topologically trivial. In both cases, the entire $4\times4$ Hamiltonian is irreducible, with the p-h operator being $\sigma_x\tau_y\mathcal{K}$ in each case.

\subsection{AI class} \label{2_a1}
Another topologically trivial class which can be accessed by the 1D spinless p-wave SC is the AI class, as seen in the previous section Sec.~\ref{1_a1}. The symmetry conditions are, $\mathcal{T}^2=1$, $\mathcal{C}=0$ and $\mathcal{S}=0$. Note that unlike in Sec.~\ref{1_a1}, s-wave perturbations will not cause a transition to the AI class. For this second type of p-wave, only combinations of the Zeeman fields induce a transition to this topologically trivial class.

\subsection*{\underline{$B_2 \sigma_z \tau_0$}}
With the unperturbed Hamiltonian in Eq.~\ref{ham2}, $B_2 \sigma_z \tau_0$ adds as a constant term in each diagonal block, thus breaking chiral symmetry in each. The TR operator is $\tau_z\mathcal{K}$ and the effective p-h operator in each block is 0. Thus, $\mathcal{T}^2=1$, $\mathcal{C}=0$ and $\mathcal{S}=0$ for this perturbation, placing it in the topologically trivial AI class.

\subsection*{\underline{$B_1 \sigma_x \tau_z$}}
With the above perturbation, the possible TR operators are $\sigma_x\mathcal{K}$ and $\tau_z\mathcal{K}$, with the effective $\mathcal{T}$ for each block being $\tau_z\mathcal{K}$. Similarly, possible p-h operators are $\tau_x\mathcal{K}$ and $\sigma_x\tau_y\mathcal{K}$, but the effective $\mathcal{C}$ is 0. Chiral symmetry $\mathcal{S}$ is also zero for the above perturbation.

\subsection*{\underline{$B_1 \sigma_y \tau_z$}}
Even though $\sigma_x\mathcal{K}$ is the TR operator for the entire $4\times4$ Hamiltonian with $B_1 \sigma_y \tau_z$, the effective $\mathcal{T}$ in each irreducible block is $\tau_z\mathcal{K}$. Similarly, possible p-h operators are $\sigma_x\tau_y\mathcal{K}$ and $\sigma_z\tau_x\mathcal{K}$, but the effective $\mathcal{C}$ is 0. Chiral symmetry, again, is zero for the above perturbation.

\subsection*{\underline{Combination of $\mathbf{B_1 \sigma \tau_z}$ and $\mathbf{B_2 \sigma \tau_0}$}}
Certain combinations of the two Zeeman fields in Eq.~\ref{ham2} (present simultaneously on the p-wave SC) also give the AI class. 
\begin{itemize}
\item $B_1 \sigma_z \tau_z+B_2 \sigma_z \tau_0$: The possible TR operators are $\tau_z\mathcal{K}$ and $\sigma_z\tau_z\mathcal{K}$, with $\mathcal{T}_{eff}=\tau_z\mathcal{K}$ in each block. $\mathcal{C}$ and $\mathcal{S}$ are 0.

\item $B_1 (\sigma_x+\sigma_z)\tau_z+B_2 \sigma_z \tau_0$ and $B_1 (\sigma_x+\sigma_z)\tau_z+B_2 \sigma_x \tau_0$: For each combination, $\mathcal{T}$ is $\tau_z\mathcal{K}$ for the entire $4\times4$ Hamiltonian. $\mathcal{C}$ and $\mathcal{S}$ are again 0.

\end{itemize}

\begin{table}[H]
\centering
\caption{Summary table for perturbations giving AI class}
\setlength{\tabcolsep}{13pt}
\begin{tabular}{l | c | c | c }
Perturbation & $\mathcal{T}$ & $\mathcal{C}$ & $\mathcal{S}$ \\
\hline \hline
$B_2\sigma_z\tau_0$ & 
\begin{minipage}{2.5cm}
$\tau_z\mathcal{K}$\\
$\mathcal{T}^2=1$
\end{minipage}
 & 
 0
&
0
\\ 
\\
\hline
$B_1\sigma_x\tau_z$  & 
\begin{minipage}{2cm}$
\sigma_x\mathcal{K},\
\tau_z\mathcal{K}
$\\
$\mathcal{T}_{eff}=\tau_z\mathcal{K}$\\
$\mathcal{T}^2=1$
\end{minipage}
 & 
 \begin{minipage}{2cm}
$ \tau_x\mathcal{K},\
\sigma_x\tau_y\mathcal{K}
$ \\
$\mathcal{C}_{eff}=0$
\end{minipage}
& 
0
\\ 
\\
\hline
$B_1\sigma_y\tau_z$ & 
\begin{minipage}{1.5cm}$
\sigma_x\mathcal{K},
$\\
$\mathcal{T}_{eff}=\tau_z \mathcal{K}$\\
$\mathcal{T}^2=1$
\end{minipage}
 & 
 \begin{minipage}{2cm}
$ \sigma_z\tau_x\mathcal{K},\
\sigma_x\tau_y\mathcal{K}
$ \\
$\mathcal{C}_{eff}=0$
\end{minipage}
& 
0
\\ 
\\
\hline
$B_1\sigma_z\tau_z + B_2\sigma_z\tau_0$& 
\begin{minipage}{2cm}
$\tau_z \mathcal{K},\
\sigma_z\tau_z \mathcal{K}
$\\
$\mathcal{T}_{eff}=\tau_z \mathcal{K}$\\
$\mathcal{T}^2=1$
\end{minipage}
 & 
0
& 
0
\\ 
\\
\hline
$B_1(\sigma_x+\sigma_z)\tau_z + B_2\sigma_z\tau_0$\\ 
$B_1(\sigma_x+\sigma_z)\tau_z + B_2\sigma_x\tau_0$&
\begin{minipage}{2cm}
$\tau_z \mathcal{K}
$\\
$\mathcal{T}^2=1$
\end{minipage}
 & 
 0
& 
0
\\
\\
\hline
\end{tabular}
\label{table4}
\end{table}

\subsection{AIII class} \label{2_a3}
AIII symmetry class is a topologically non-trivial class, with $Z$ invariant. The symmetry conditions are : $\mathcal{T}=0$, $\mathcal{C}=0$ and $\mathcal{S}=1$. Spinless p-wave SC of the particular type in Eq.~\ref{ham1} can access this symmetry class in the presence of certain perturbations like the s-wave $\mathbf{\triangle_1 \sigma_y\tau_y}$, and combinations of the s-wave and Zeeman term along $\tau_0$.

\subsection*{\underline{$\mathbf{\triangle_1 \sigma_y\tau_y}$}}
In the presence of a s-wave term in Eq.~\ref{ham2}:
\begin{equation} \label{ham2_2}
\mathcal{H}(p)=\mathcal{H}_0(p)+\triangle_1 \sigma_y\tau_y
\end{equation}
TR operator, satisfying the condition  $\mathcal{T}\mathcal{H}(p)\mathcal{T}^{-1}=\mathcal{H}(-p)$, is $\sigma_y\mathcal{K}$. 
There are two possible p-h operator for Eq.~\ref{ham2_2}, $\tau_x\mathcal{K}$ and $\sigma_x\tau_y\mathcal{K}$. On block diagonalization into $\sigma_z=\pm1$ irreducible blocks, we get,
\begin{equation} \label{ham2_3}
H_{1,2}=-\left(\frac{p^2}{2m}-\mu\right)\tau_z-\triangle_0 p\tau_x\pm\triangle_1\tau_x
\end{equation}
However, each block in Eq.~\ref{ham2_3}, no longer has TR or p-h symmetry, since neither the two p-h operators nor the TR operator is a symmetry. But a chiral operator still exists, $\mathcal{S}=\tau_y$, which gives, $\mathcal{S}^2=1$. This class of Hamiltonian falls in the topologically non-trivial AIII symmetry class (in $d=1$), with $Z$ invariant.

\subsection*{\underline{$\triangle_1 \sigma_y\tau_y + B_2\sigma_x\tau_0$}}
With the above perturbation, when both the s-wave term and the zeeman term  $B_2\sigma_x\tau_0$ are present in Eq.~\ref{ham2}, we again get back the AIII class. $\mathcal{T}$ and $\mathcal{C}$ is zero, but $\mathcal{S}$ in each block is $\sigma_y\tau_x$, giving $\mathcal{S}=1$.

\begin{table}[H]
\centering
\caption{Summary table for perturbations giving AIII class}
\setlength{\tabcolsep}{13pt}
\begin{tabular}{l | c | c | c }
Perturbation & $\mathcal{T}$ & $\mathcal{C}$ & $\mathcal{S}$ \\
\hline \hline
$\triangle_1 \sigma_y\tau_y$ & 
\begin{minipage}{2.5cm}
$\sigma_y\mathcal{K}$\\
$\mathcal{T}_{eff}=0$
\end{minipage}
 & 
 \begin{minipage}{2.5cm}
$\tau_x\mathcal{K},\
\sigma_x\tau_y\mathcal{K}$\\
$\mathcal{C}_{eff}=0$
\end{minipage}
&
\begin{minipage}{2.5cm}
$\tau_y$\\
$\mathcal{S}=1$
\end{minipage}
\\ 
\\
\hline
$\triangle_1\sigma_y\tau_y +B_2\sigma_x\tau_0$  & 
0
 & 
0
& 
\begin{minipage}{2.5cm}
$\sigma_y\tau_x$\\
$\mathcal{S}=1$
\end{minipage}
\\ 
\\
\hline
\end{tabular}
\label{table5}
\end{table}

\subsection{BDI class} \label{2_bd1}
As discussed earlier in Sec.~\ref{1_bd1}, the topologically non-trivial BDI class can appear in 1D p-wave SC. However, with the particular type in Eq.~\ref{ham2}, an s-wave pairing term cannot induce such a transition. It is necessary for the Zeeman terms, along $\tau_0$ and $\tau_z$, to be present. The symmetry conditions are $\mathcal{T}^2=1$, $\mathcal{C}^2=1$ and $\mathcal{S}=1$.

\subsection*{\underline{$\mathbf{B_2.\hat{\sigma}\tau_0}$}}
\begin{itemize}
\item $B_2\sigma_x\tau_0$: Possible TR operators are $\sigma_x\mathcal{K}$ and $\tau_z\mathcal{K}$, with the effective $\mathcal{T}$ in each block being $\tau_z\mathcal{K}$. Similarly, possible p-h operators are $\sigma_z\tau_x\mathcal{K}$ and $\sigma_y\tau_y\mathcal{K}$, with the effective $\mathcal{C}$ being $\tau_x\mathcal{K}$. Chiral operator $\mathcal{S}$ is $\tau_y$.

\item $B_2\sigma_y\tau_0$: Here again the effective $\mathcal{T}$ is $\tau_z\mathcal{K}$. The possible p-h operators are $\tau_x\mathcal{K}$ and $\sigma_y\tau_y\mathcal{K}$, with the effective $\mathcal{C}$ being $\tau_x\mathcal{K}$. Chiral operator $\mathcal{S}$ is $\tau_y$.
\end{itemize}

\subsection*{\underline{$B_1\sigma_z\tau_z$}}
The TR operator $\mathcal{T}$ is $\tau_z\mathcal{K}$. The possible p-h operators are $\tau_x\mathcal{K}$ and $\sigma_z\tau_x\mathcal{K}$, with the effective $\mathcal{C}$ being $\tau_x\mathcal{K}$. Chiral operator $\mathcal{S}$ is $\tau_y$.

\begin{table}[H]
\centering
\caption{Summary table for perturbations giving BDI class}
\setlength{\tabcolsep}{13pt}
\begin{tabular}{l | c | c | c }
Perturbation & $\mathcal{T}$ & $\mathcal{C}$ & $\mathcal{S}$ \\
\hline \hline
$B_2\sigma_x\tau_0$ & 
\begin{minipage}{2.5cm}
$\sigma_x\mathcal{K},\
\tau_z\mathcal{K}
$\\
$\mathcal{T}_{eff}=\tau_z\mathcal{K}$\\
$\mathcal{T}^2=1$
\end{minipage}
 & 
 \begin{minipage}{2.5cm}
$\sigma_y\tau_y\mathcal{K},\
\sigma_z\tau_x\mathcal{K}$\\
$\mathcal{C}_{eff}=\tau_x\mathcal{K}$\\
$\mathcal{C}^2=1$
\end{minipage}
&
\begin{minipage}{2.5cm}
$\tau_y$\\
$\mathcal{S}=1$
\end{minipage}
\\ 
\\
\hline
$B_2\sigma_y\tau_0$  & 
\begin{minipage}{2cm}
$\mathcal{T}_{eff}=\tau_z\mathcal{K}$\\
$\mathcal{T}^2=1$
\end{minipage}
 & 
\begin{minipage}{2.5cm}
$\sigma_y\tau_y\mathcal{K},\
\tau_x\mathcal{K}$\\
$\mathcal{C}_{eff}=\tau_x\mathcal{K}$\\
$\mathcal{C}^2=1$
\end{minipage}
& 
\begin{minipage}{2.5cm}
$\sigma_y\tau_x$\\
$\mathcal{S}=1$
\end{minipage}
\\ 
\\
\hline
$B_1\sigma_z\tau_z$  & 
\begin{minipage}{2cm}
$\tau_z\mathcal{K}$\\
$\mathcal{T}^2=1$
\end{minipage}
 & 
\begin{minipage}{2.5cm}
$\sigma_z\tau_x\mathcal{K},\
\tau_x\mathcal{K}$\\
$\mathcal{C}_{eff}=\tau_x\mathcal{K}$\\
$\mathcal{C}^2=1$
\end{minipage}
& 
\begin{minipage}{2.5cm}
$\sigma_y\tau_x$\\
$\mathcal{S}=1$
\end{minipage}
\\ 
\\
\hline
\end{tabular}
\label{table6}
\end{table}

\subsection{D class} \label{2_d}
The D symmetry class is another topologically non-trivial class in 1D, with $\mathcal{T}=0$, $\mathcal{C}^2=1$ and $\mathcal{S}=0$, with the invariant being $Z_2$. The unperturbed p-wave Hamiltonian in Eq.~\ref{ham2} can access this class in the presence of s-wave and Zeeman terms $B_1\sigma_z\tau_z$ and  $B_2\sigma_y\tau_0$ , and also with combinations of both the Zeeman term $\mathbf{B}_1$ and $\mathbf{B}_2$.

\subsection*{\underline{$\triangle_1 \sigma_y\tau_y + B_2\sigma_y\tau_0$ and $\triangle_1 \sigma_y\tau_y + B_1\sigma_z\tau_z$}}
With each of the above perturbations, the p-h operator $\mathcal{C}$ is $\tau_x\mathcal{K}$. However, since the chiral symmetry is absent the edge states do not appear at $E=0$ and is not captured in the present MBS calculation.

\subsection*{\underline{$B_1(\sigma_x +\sigma_z) \tau_z+B_2\sigma_y\tau_0$}}
Here again we consider the combination of two Zeeman fields in Eq.~\ref{ham2}, $\mathbf{B}_1 \tau_z$ in the $x$-$z$ plane and $\mathbf{B}_2\tau_0$ along $\sigma_y$. Here again, the p-h operator $\mathcal{C}$ is $\tau_x\mathcal{K}$.

\begin{table}[H]
\centering
\caption{Summary table for perturbations giving D class}
\setlength{\tabcolsep}{13pt}
\begin{tabular}{l | c | c | c }
Perturbation & $\mathcal{T}$ & $\mathcal{C}$ & $\mathcal{S}$ \\
\hline \hline
$\triangle_1\sigma_y\tau_y+B_2\sigma_y\tau_0$\\
$\triangle_1\sigma_y\tau_y+B_1\sigma_z\tau_z$ & 
0
 & 
 \begin{minipage}{1.5cm}
$\tau_x\mathcal{K}$\\
$\mathcal{C}^2=1$
\end{minipage}
&
0
\\ 
\\
\hline
$B_1(\sigma_x+\sigma_z)\tau_z+B_2\sigma_y\tau_0$  & 
0
 & 
\begin{minipage}{1.5cm}
$
\tau_x\mathcal{K}$\\
$\mathcal{C}^2=1$
\end{minipage}
& 
0
\\ 
\\
\hline
\end{tabular}
\label{table7}
\end{table}

\section{Conclusion}
The two cases of the p-wave pairing in Sec.~\ref{Sec1} and Sec.~\ref{Sec2} have very distinct features in the presence of perturbations, even though both the unperturbed Hamiltonians start from the topologically non-trivial BDI class. Firstly, in the presence of the s-wave pairing term alone, the first type of p-wave Hamiltonian ($\triangle_{\uparrow\uparrow}=\triangle_{\downarrow\downarrow}$) becomes topologically trivial (AI) whereas the second type ($\triangle_{\uparrow\uparrow}=-\triangle_{\downarrow\downarrow}$) shifts to another non-trivial class, AIII. Secondly, the first type of p-wave can still remain in the BDI class in the presence of certain combinations of the Zeeman fields and the s-wave term. Whereas, the second type of p-wave cannot remain in the BDI class when the s-wave perturbation is present. This is true for also the AI class, which the second type of p-wave cannot access in the presence of s-wave pairing, but the first type can.

We have also shown that the MBS in each of the two unperturbed p-wave cases have the same spatial dependence, with the emergence of two types of Majorana modes depending on the value of $\mu$. For $\mu>1/2$ we get damped oscillating modes and for $0<\mu<1/2$ the MBS are purely decaying. The only difference between them is that the eigenvectors have a different structure for the two cases. This nature is also reflected in the presence of a Zeeman field, where both the p-wave cases have the same phase diagram, Fig.~\ref{phase1}.

The presence of an s-wave term in each of the cases changes this spatial dependence of the MBS. For cases with s-wave which lie in the non-trivial topological classes with chiral symmetry (like the AIII class), zero energy MBS still appear for $\mu>0$, but they are only of the damped oscillating form, shown in Fig.~\ref{phase_swave}. But the eigenvector structure of the Majoranas still remain the same, as in the p-wave case. This is another unique result in our study of Majorana fermions and topological properties of 1D p-wave SC.

\begin{figure}
\begin{tabular}{c}
\includegraphics[width=8.8cm, height=6.8cm]{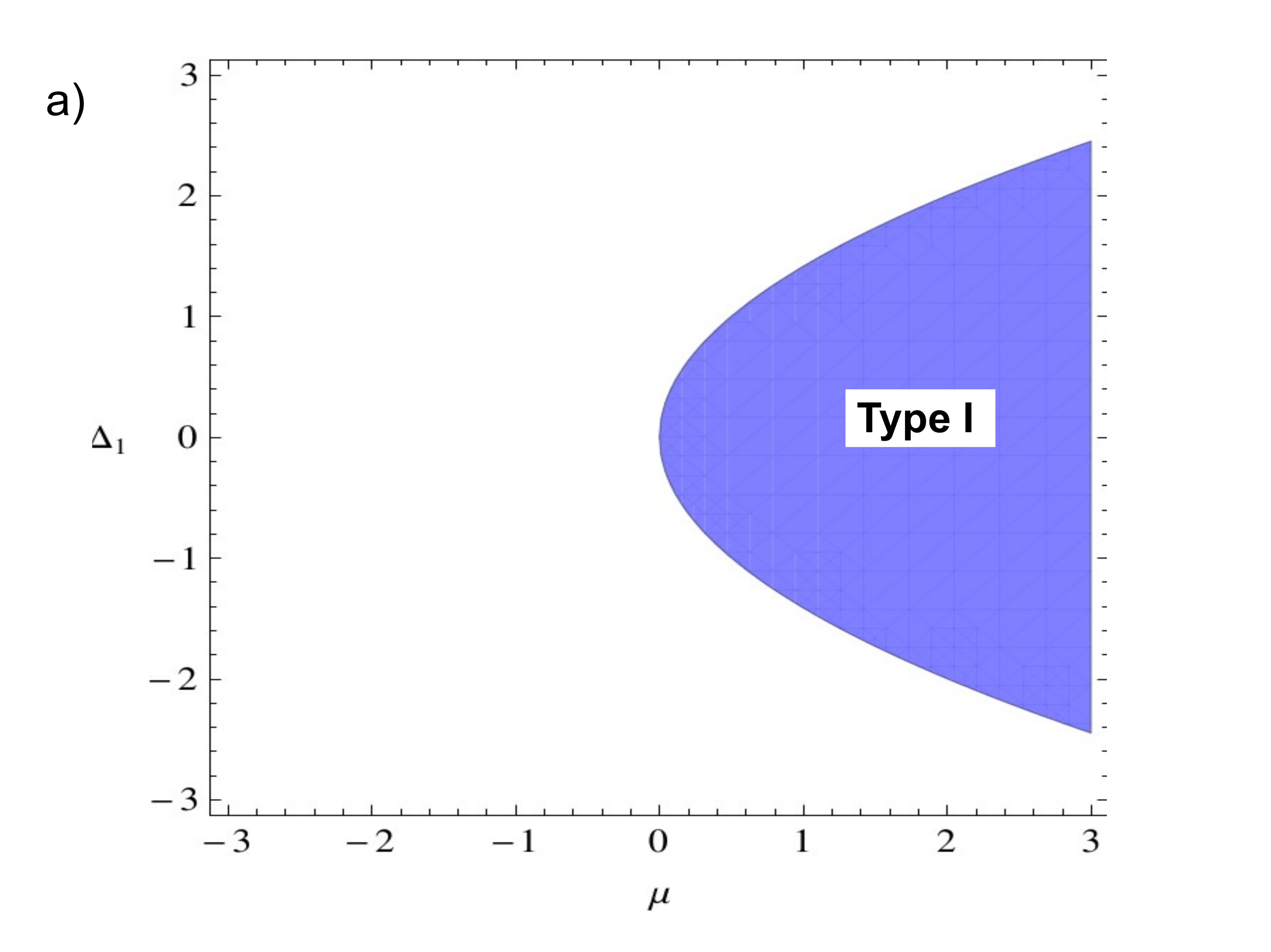}
\includegraphics[width=8.5cm, height=6.5cm]{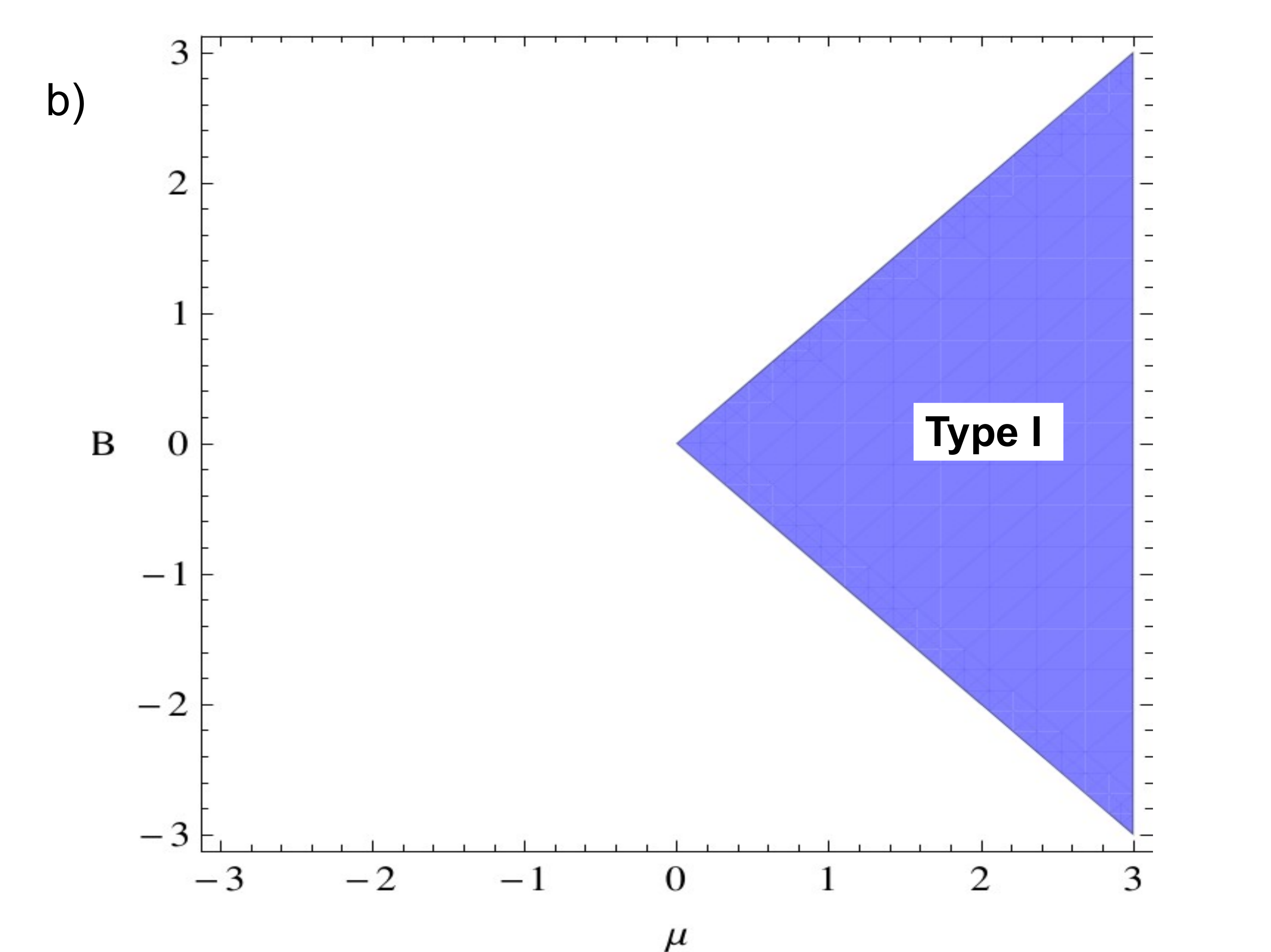} \\
\includegraphics[width=10cm, height=8.5cm]{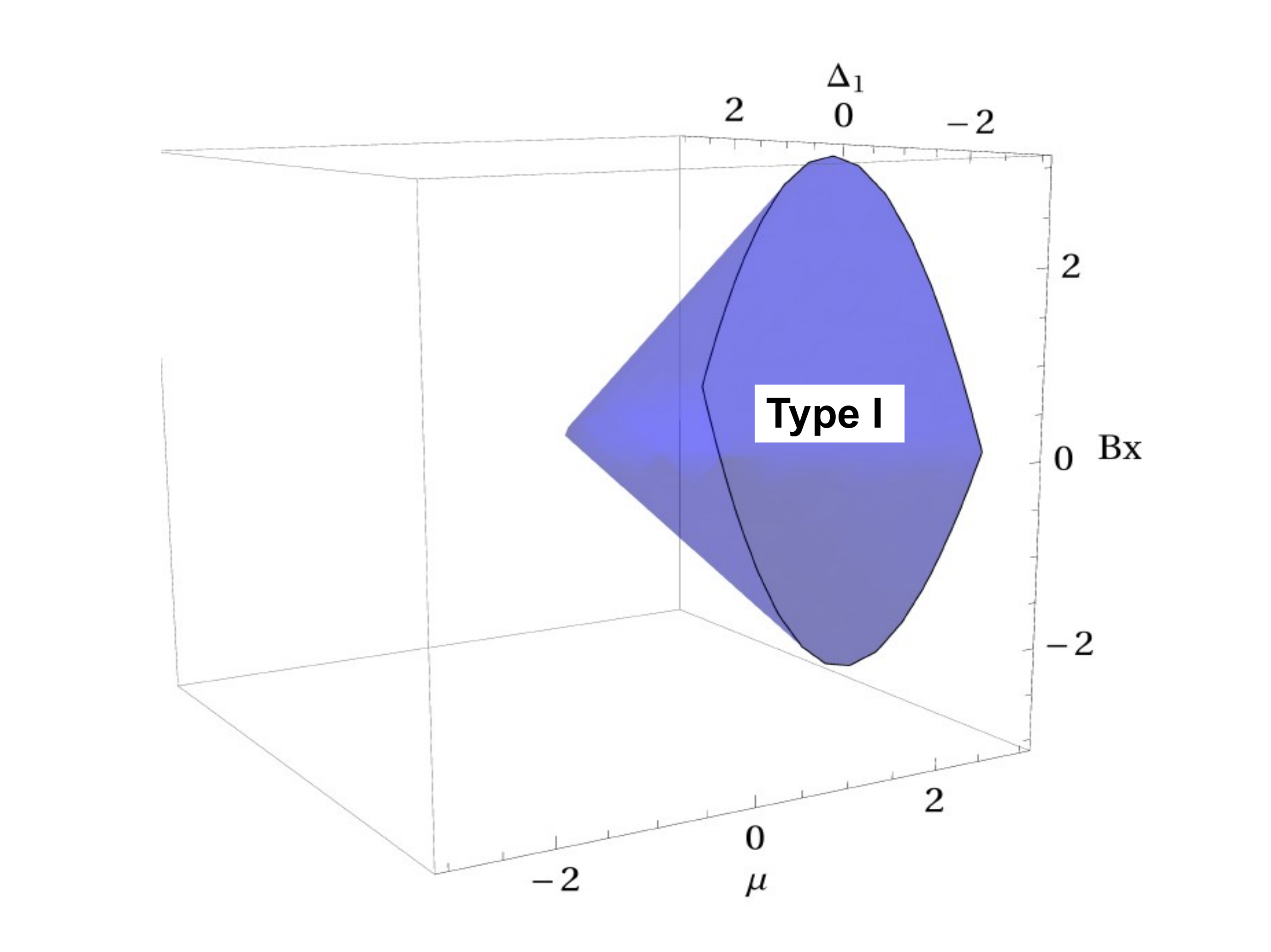}
\end{tabular}
\caption{Phase diagram for MBS doublets in the presence of s-wave perturbations in Eq.~\ref{ham2}. (Top Left) Region plot showing the regions in $\mu-\triangle_1$ space over which MBS doublets of Type I exist in the presence of s-wave. Top Right: Region plot in $\mu-B$ space over which again MBS doublets of Type I exist. Bottom: Three dimensional region plot in $\mu-\triangle_1-B$ space, showing MBS exists only for regions above $\mu>0$, all belonging to the Type I category.}
\label{phase_swave}
\end{figure}

}


%% file: mychapter3.tex
\chapter{{\LARGE Studies of SU(3) topological phases in two-dimensional systems} \\
\vspace{0.5cm}
\small{Ref: S. Ray, A. Ghatak and T. Das, Phys. Rev. B. {\bf 95}, 165425 (2017)}} \label{chapter4}
{\large

\section{Introduction}

The Hoftstadter model, with spinless electrons trapped in a square lattice in the presence of a magnetic field, has non-trivial topology. This results in the appearence of chiral edge modes and also a non-zero Chern number for each band.
Consider $N$ decoupled copies of the Hofstadter model with $\frac{2\pi}{N}$ Abelian flux per plaquette. The Hamiltonian for this system is given by~\cite{galitski}:
\begin{equation}
{\it H} = -t\sum_i(\Psi^{\dagger}_i\Psi_{i+\hat{x}}+ \Psi^{\dagger}_i e^{-i2\pi\alpha(x_i + \hat{S}_z)}\Psi_{i+\hat{y}} + h.c), 
\end{equation}
where, $\Psi_i$ is a N-component spinor, $\hat{S}_z = \rm{diag}(s,s-1,...-s)$ with $2 s+1=N$, and $\alpha=\frac{1}{N}$. Each component has topologically non-trivial band structure with non-zero Chern number, and thus, so does the entire system (except for $N \le 2$, for $N=2$ the flux per each plaquette is $\pi$, which itself is time reversal invariant).
Under a gauge transformation, $\Psi_i \rightarrow \mathcal{U}^{x_i}\Psi_i$ for an appropriate $\mathcal{U}$, the Hamiltonian can be written in terms of a uniform non-Abelian gauge field which couples the components.  The lowest value of $N$ where a non-zero Chern number appears, in the presence of a uniform non-Abelian gauge field, is $N=3$. The resulting Hamiltonian for $N=3$ is~\cite{galitski}:
\begin{equation}
{\it H} = -t\sum_i(\Psi^{\dagger}_ie^{-i \hat{A}_x}\Psi_{i+\hat{x}}+ \Psi^{\dagger}_i e^{-i \hat{A}_y}\Psi_{i+\hat{y}} + h.c),
\end{equation}
with the non-Abelian gauge fields being, $\hat{A}_x= \frac{2\pi}{3\sqrt{3}}(\hat{\lambda}_2-\hat{\lambda}_5+\hat{\lambda}_7)$ and $\hat{A}_y= \frac{\pi}{3}(\hat{\lambda}_3+\sqrt{3}\hat{\lambda}_8)$, where $\hat{\lambda}$ are the Gell-mann matrices. Such a system will be referred to as an SU(3) system in our further analysis.
Note that the SU(3) Hamiltonians being  referred to in our work are $3\times3$ Hamiltonians which can be written as linear combinations of the Gell-mann matrices and the identity. They are not invariant under SU(3) transformations in general.

In this work, we aim to build a lattice model, using three component atoms (three different sublattices or orbitals) trapped in a non-Abelian SU(3) gauge field, using nearest neighbour (NN) hopping to  mimic the gauge field. We uncover two salient ingredients required to express a general three-component lattice Hamiltonian in terms of the Gell-mann matrices, with non-trivial topological invariant. We find that all three components must be coupled via a gauge field, with opposite Bloch phase (in momentum space if the NN hopping between two components is $\sim -t e^{i k}$, then for the other two components, this should be $\sim -t e^{-i k}$) between any two components, and there must be band inversion between all {\it three} components in a given eigenstate. For spinless particles, we show that such states can be obtained in a tripartite lattice with three inequivalent lattice sites, in which the Bloch phase associated with the NN hopping acts as $k$-space gauge field. The second criterion is the hopping amplitude $t$ should have an opposite sign in the diagonal element for one of the two components, which can be introduced via a constant phase $e^{i\pi}$ along the direction of hopping. The third and a more crucial criterion is that there must also be an odd-parity Zeeman-like term (as $k\rightarrow-k$, the term  changes sign), i.e. $\sin(k)\sigma_z$ term, where $\sigma_z$ is the third Pauli matrix defined with any two components of the three component basis. In the presence of a constant vector potential, the kinetic energy of the electron gets modified when the vector potential causes a flux to be enclosed. This can generate the desired odd parity Zeeman term, via a site-selective polarization of the vector potential. This can be achieved in principle by suitable modifications of techniques used in Sisyphus cooling, and with a suitable arrangement of polarizer plates, etc. 
The presence of the topological phase is affirmed by edge state calculation, obeying the bulk-boundary correspondence.

The rest of the chapter is arranged as follows. In Sec.~\ref{Sec:II}, we describe the general characteristics that the SU(3) Hamiltonian should possess to obtain non-zero Chern number. In Sec.~\ref{model_build} we discuss the setup. Derivation of the odd-parity Zeeman term in the presence of a constant vector potential is given in Sec.~\ref{Sec:EPcoupling}. The tight-binding model for a tripartite lattice is discussed in Sec.~\ref{Sec:lattice}. A simplistic approach towards designing such a lattice with spatially dependent polarization is illustrated in Sec.~\ref{Sec:design}. In Sec.~\ref{cal_chern} we elaborate on the geometrical method for calculating the Berry curvature and Chern number. We detail the calculation of edge states using the strip geometry approach in Sec.~\ref{cal_edge}. In Sec.~\ref{spin} we discuss the robustness of the spinless SU(3) topological phases to spinful perturbations. We end the chapter with discussions and conclusions in Sec.~\ref{conclusion}. 

\section{Model}\label{Sec:II}

\subsection{General characteristics of SU(3) topological Hamiltonians}\label{gen_char}
We design a SU(3) Hamiltonian with on eye on finite Chern number in a bottom-up approach. A generic Hamiltonian, obeying the SU(3) decomposition, can be written as~\cite{galitski,gellmann},
\be \label{su3_gellmann}
\hat{H}({\bf k})=a({\bf k})\hat{{\mathbb{I}}}_3+{\bf b}({\bf k})\cdot {\bf \hat{\l}}\,,
\ee
where $\hat{{\mathbb{I}}}_3$ is a 3$\times$3 identity matrix, $\hat{\l}_i$ are the SU(3) generators (Gell-Mann matrices), and $a({\bf k})$, $b_i({\bf k})$ are the corresponding coefficients. The explicit matrix form of the Hamiltonian (the $k$- dependencies in $a$ and $b_i$ are implied) is,
\be\label{gensu3}
\hat{H}({\bf k})=\begin{pmatrix}
a+b_3+\frac{b_8}{\sqrt{3}}& b_1-ib_2& b_4-i b_5\\
b_1+i b_2& a-b_3+\frac{b_8}{\sqrt{3}}& b_6-i b_7\\
b_4+i b_5& b_6+i b_7& a-\frac{2b_8}{\sqrt{3}}\\
\end{pmatrix}.
\ee
We work with a three component spinor $\Psi_k^{\dag}=\left( {\psi}_1^{\dag}(k), {\psi}^{\dag}_2(k),{\psi}^{\dag}_3(k)\right)^{T}$, where $\psi_i(k)$ are the basis  representing different orbitals, or sublattices, and so on (but we do not consider spin here). Each $b_i$ term requires special treatment such that opposite Bloch phase, and odd-parity Zeeman term can be simultaneously achieved in such a way that Berry curvature singularities at discrete $k$-points can be attained.  

\subsubsection{Diagonal terms} \label{diag_terms}
We start with the diagonal terms of Eq.~\eqref{gensu3}. We denote the three onsite, inter-basis (hopping between the same sub-lattice of NN unit cell), dispersions as $\xi_{i}({\bf k})$ where $i=1,~2,~3$. In general tight-binding Hamiltonians, diagonal terms comprise of cosine functions of momentum, and chemical potential. A sine function of momentum arise only in the presence of a flux or a spin-orbit coupling (SOC). We have taken the NN hopping along only the $y$ direction to contribute in the diagonal terms, following the SU(3) Hamiltonian in ~\cite{galitski}. With the analysis of Berry curvature, we recognize that the essential requirements for non-zero Chern number in this case are: the diagonal terms $\xi_1$ and $\xi_3$ must contain $\pm \sin(k_y)$ terms, which is equivalent to having odd-parity Zeeman term, and the hopping amplitude along $y$-direction in $\xi_2$ should be opposite to that of $\xi_1$ and $\xi_3$. Without specifying the origin at this point, we start with a combination of three diagonal terms in a 1D lattice:
\beq\label{diagonal}
\xi_i({\bf k}) &=& t_i\cos{(k_y \alpha)}+ m_i\sin{(k_y\beta)}-\mu, 
\eeq
%
%
%
where $t_i$, $m_i$ are the expansion parameters, and $\mu$ is the chemical potential. $\alpha$ and $\beta$ are arbitrary parameters depending on the crystal structure and lattice constants. Finite Chern number arises for a set of parameters as $t_1=-2t_y$, $t_2=\frac{3}{2}t_y$, $t_3=-t_y$, and $m_1=-m_3$, and $m_2=0$. The cosine terms arise from the nearest neighbor hopping along the $y$-direction. In Sec.~\ref{Sec:EPcoupling} below, we discuss how to obtain $m_i\sin{k_y}$ term with the help of an interaction between an electron and a constant vector potential $\mathbf{A}$ (which traps a flux), in which we find that $m_i$ depends on both $t_i$ as well as the vector potential $\mathbf{A}$. Thus its sign can be simultaneously reversed by using antiparallel polarization of $\mathbf{A}$. We notice that all three diagonal terms are taken to depend only on $k_y$ which is consistent with the setup drawn in Fig.~\ref{model_grid}.  By comparing Eqs.~\eqref{diagonal} and \eqref{gensu3}, we obtain
\beq\label{ab3b8}
a({\bf k}) &=& \frac{\xi_1+\xi_2+\xi_3}{3} = \frac{1}{6} \big(-3 \cos{\alpha k_y} + 2 (m_1 + m_3) \sin{\beta k_y} \big), \nonumber\\
b_3({\bf k}) &=& \frac{\xi_1-\xi_2}{2} = \frac{1}{4}(-7 \cos \alpha k_y  +2 m_1 \sin\beta k_y), \nonumber\\
b_8({\bf k}) &=& \frac{2\xi_3-\xi_1-\xi_2}{2\sqrt{3}} \nonumber\\
&=& \frac{1}{4\sqrt{3}}(3 \cos\alpha k_y + 2(m_1 - 2 m_3) \sin\beta k_y), 
\eeq
Looking at the Hamiltonian in Eq.~\eqref{gensu3}, we notice that $a({\bf k})$ gives  a overall shift to all the bands and thus does not play any specific role on the topology. $b_3({\bf k})$ gives an anisotropic Zeeman splitting between 1st and 2nd basis in the Hamiltonian, while $b_8({\bf k})$ gives a similar splitting of the 3rd basis from the other two ones. It is easy to see that the band inversion along the $k_y$ direction is driven by $b_3$ and $b_8$ terms. And also, since the eigenvalues are proportional to $b_3$ and $b_8$, we see that the bands become anisotropic between $\pm k_y$. On the other hand, along the $k_x$ direction they are symmetric, since the eigenvalues depend on the absolute value of the other $b_m$ terms. This asymmetry also reflects in the Berry curvature maps shown in Fig.~\ref{3Dbren}.

\subsubsection{Off-diagonal terms} \label{off_diag}
Next we consider the three off-diagonal terms which follow a general form $b_{\nu}({\bf k}) \pm i b_{\sigma}({\bf k})$, where $(\nu,\sigma)=(1,2), (4,5), (6,7)$. In spin-1/2 systems, such a complex term usually has two origins: (1) Rashba- or Dresselhaus-type spin-orbit coupling (SOC), (2) Bloch phase ($e^{i k}$ due to NN hopping in the momentum space) from nearest neighbor electron's hopping. (1) Rashba and Dresselhaus SOC yields $b_{\nu}({\bf k})=\alpha_R \sin{k_x}$, and $b_{\sigma}({\bf k})=\alpha_R\sin{k_y}$ (where $\alpha_R$ is the SOC strength). SOC is however difficult to achieve for all three spins in the basis in both condensed matter and optical lattice setups. More importantly, we find that the computation of Chern number with SOC in the off-diagonal terms often gives zero Chern number. Therefore, we focus on the possibility (2). Assigning $b_{\nu}({\bf k})=t_x\cos{k_x}$, and $b_{\sigma}({\bf k})=t_x\sin{k_x}$ (where $t_x$ is a parameter which can be different for different $\nu$ and $\sigma$), we see that this term simplifies to $\sim t_x\exp(ik_x)$. This is just a Bloch phase associated with the hopping between different sublattices of the nearest neighbour unit cell (it appears in the off-diagonal term in the Hamiltonian). We thus find that to write the tight-binding hopping terms in terms of linear combinations of the Gell-mann matrices as given in \cite{galitski}, {\it the Bloch phase must be reversed in at least one of the off-diagonal terms, compared to the other two.} 

\subsubsection{Full Hamiltonian}
Based on the constraints for both diagonal and off-diagonal terms, we now seek a minimal model for the realization of SU(3) topological insulator in the spinless basis:
\begin{eqnarray}
H(k)=\left(
 \begin{array}{ c c c c }
\xi_1(k_y) & -t_x e^{ik_x} &  -t_x e^{-ik_x}    \\
-t_x e^{-ik_x}  & \xi_2(k_y)  & - \frac{1}{2}t_x\ e^{i k_x}    \\
-t_x e^{ik_x} & -\frac{1}{2}t_x\ e^{-i k_x} &  \xi_3( k_y)\\
\end{array} \right),
\label{Ham1}
\end{eqnarray}
where $t_x$ is the nearest neighbour tight-binding hopping parameter between different basis. Without losing generality, we set $t_x=t_y$=1. This gives all eight components of the ${\bf b}$ vector to be:
\begin{align} \label{b_vector}
\begin{split} 
{\bf b}(k) &= \big[-\cos k_x, -\sin k_x, 
 \frac{1}{4}(-7 \cos \alpha k_y  +2 m_1 \sin\beta k_y),\\ 
&-\cos k_x, \sin k_x, -\frac{1}{2}\cos k_x, -\frac{1}{2}\sin k_x,\\
& \frac{1}{4\sqrt{3}}(3 \cos\alpha k_y + 2(m_1 + 2 m_3) \sin\beta k_y)\big].
\end{split}
\end{align}

\subsection{Setup} \label{model_build}
Next we discuss how to obtain such a Hamiltonian using orbitals or different sub-lattices. The opposite sign of hopping $t$ for the second component, and the Zeeman term $\sin(\beta k_y)\sigma_z$ term can be simultaneously obtained in a tripartite lattice by applying linearly polarized vector potential on each sub-lattice, such that it encloses a flux. At the end of this section, we discuss how to simulate such a lattice.

\subsubsection{Tight-binding (TB) model for electron in a vector potential (origin of $\mathbf{\sin{(\beta k_y)}}$)}\label{Sec:EPcoupling}
The motivation for the origin of $\sin{\beta k_y}$ term can be drawn from the fact that the interaction between an electron with momentum ${\bf p}=\hbar{\bf k}$ and a vector potential ${\bf A}=A{\hat \epsilon}$ (${\hat \epsilon}$ is the light polarization) is $H_{\rm int}=-\frac{e}{m}{\bf p}\cdot{\bf A}=-\frac{e\hbar A}{m}{\bf k}\cdot{\hat \epsilon}$. We choose the polarization oriented along the $y$-direction. We take a single electron Hamiltonian under the periodic potential $U({\bf r})$ of the lattice as $H =\frac{ p^2}{2m^*} + U({\bf r})$. The corresponding Bloch wavefunction is $\eta_{\bf k}=\frac{1}{\sqrt{N}}\sum_{n}e^{i{\bf k}\cdot{\bf R}_n}u_n({\bf r})$, where $N$ is the total number of unit cells, $u_n({\bf r)}$ is the Wannier state at the $n^{\rm th}$ site located at ${\bf R}_n$. In the presence of vector potential ${\bf A}$, the Hamiltonian becomes $H^{\prime} =\frac{ ({\bf p}-e{\bf A})^2}{2m^*} + U({\bf r})$. 
The new Bloch wavecfunction simply changes to $\eta^{\prime}_{\bf k}=\frac{1}{\sqrt{N}}\sum_{n}e^{i{\bf k}\cdot{\bf R}_n}u^{\prime}_n({\bf r})$, where $u^{\prime}_n({\bf r)}=u_n({\bf r})e^{i\frac{e}{\hbar}\int_{{\bf R}_n}^{\bf r}{\bf A}\cdot d{\bf l}}=u_n({\bf r})e^{i\phi_n({\bf r})}$. $\phi_n({\bf r})$ is called the Peierls phase at ${\bf r}$ acquired by the charged particle in traversing from the $n^{\rm th}$ lattice site. It can be shown that $H^{\prime}|u^{\prime}_n({\bf r)}\rangle =e^{i\phi_n({\bf r})}H|u_n({\bf r)}\rangle$. Using these ingredients, we can now derive the tight-binding dispersion as 
\beq\label{TBdispersion}
\xi({\bf k}) &=& \langle \eta^{\prime}_{\bf k}| H^{\prime} | \eta^{\prime}_{\bf k}\rangle\nonumber\\
&=& \frac{1}{N}\sum_{n,n^{\prime}} e^{i{\bf k}\cdot({\bf R}_n-{\bf R}_{n^{\prime}})}  \int d{\bf r} \langle {u^{\prime}_{n^{\prime}}}| H^{\prime}| u^{\prime}_n \rangle\nonumber\\
&=& \frac{1}{N}\sum_{n,n^{\prime}} e^{i{\bf k}\cdot({\bf R}_n-{\bf R}_{n^{\prime}})}  e^{i (\phi_n-\phi_{n^{\prime}})}  \int d{\bf r} \langle {u_{n^{\prime}}}| H| u_n \rangle\nonumber\\
&=& \sum_{n,n^{\prime}} t_{nn^{\prime}}e^{i{\bf k}\cdot({\bf R}_n-{\bf R}_{n^{\prime}})}  e^{i (\phi_n-\phi_{n^{\prime}})}.
\eeq
Here $t_{nn^{\prime}}=\frac{1}{N}\int d{\bf r} \langle {u_{n^{\prime}}}| H| u_n \rangle$ is the TB hopping amplitude between $n$ and $n^{\prime}$ sites {\it without} the vector potential. Note that we consider $\mathbf{A}$ as a constant vector along $\hat{y}$ with periodic boundary conditions along $\hat{y}$. This ensures that a constant flux is enclosed and hence cannot be gauged away. We here restrict ourselves to the nearest neighbor hopping, i.e., $n^{\prime}=n\pm 1$. Let the lattice constant along the $y$-direction be $b$. By setting $t_{n(n\pm1)}=t_y$, and $\pm\phi=\phi_n-\phi_{n\pm1} =\pm\frac{e}{\hbar}Ab= \frac{e}{\hbar}Ab{\hat y}\cdot{\hat \epsilon}$, we obtain,
\beq\label{TBdispersion2}
\varepsilon({\bf k}) &=& t_y\left[e^{i(k_yb + \phi)}  + e^{-i(k_yb + \phi)}\right].\nonumber\\
&=& 2t_y\left[\cos{(k_yb)}\cos{\phi} -\sin{(k_yb)}\sin{\phi}\right].
\eeq
We absorb $\cos{\phi}$ in to the TB term as $t_y(\phi)=2t_y(0)\cos{\phi}$, and define $m(\phi)=-2t_y\sin\phi$. Then we see that Eq.~\eqref{TBdispersion2} is the same as Eq.~\eqref{diagonal}. From this definition, it is easy to see that as the direction of polarization is reversed, $\phi\rightarrow -\phi$, $m\rightarrow -m$ and $t_y\rightarrow t_y$, while the perpendicular polarization yields $m (\phi=0) = 0$, and $t_y$ remains the same.  

\subsubsection{Tripartite lattice}\label{Sec:lattice}

\begin{figure}[H]
\centering
\includegraphics[width=0.7\columnwidth]{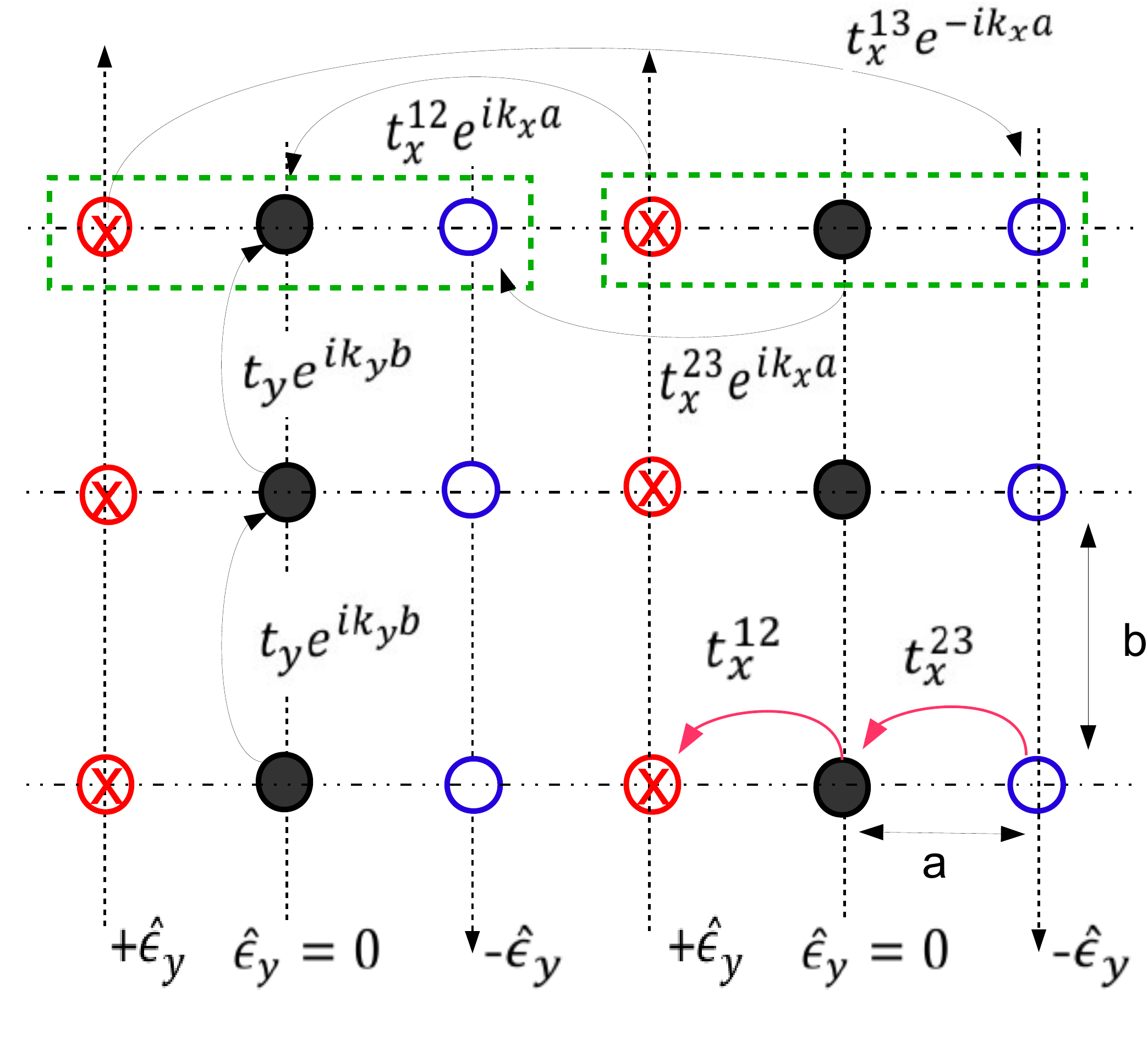}
\caption{(Color online) Schematic drawing of the proposed setup. The different coloured spheres denote three different basis of the Hamiltonian, which can be three different atomic species or orbitals, $\psi_1$, $\psi_2$ and $\psi_3$. $\mathbf{\hat{\epsilon}}$ denotes the polarization direction of the constant vector potential, which causes the different Pierl's phase coupling for the different orbitals. For our choice of the model, we have taken red sphere to denote the $s$-orbital, black sphere for the $p_x$-orbital and blue sphere for the $p_y$-orbital. The hopping amplitudes are multiples of $t_x$ and $t_y$ , and are different for the different components (orbitals). The black arrows show hopping between orbitals of different unit cells and red arrows represent hopping between NN within the same unit-cell. Also, in the present model, $b=2 a$ and $a=1$.
}
\label{model_grid}
\end{figure}

For the SU(2) case, an uncompensated Bloch phase in the hopping term $\sim t\exp(ik_x)$ is required for generating a Chern number,  and can be obtained in bipartite lattice (c.f. Su-Schrieffer-Heeger model in 1D~\cite{ssh_rmp}, or honeycomb lattice~\cite{castroneto} in 2D) or for hopping between even and odd-parity orbitals~\cite{WeylTD}. Similarly, for the SU(3) case, we need the same term for all three off-diagonal terms. Therefore, we propose a tripartite lattice as depicted in Fig.~\ref{model_grid}. Also note that, the phase of the hopping term only includes $k_x$ terms, implying that different basis elements (inequivalent sub-lattice sites) should be aligned along the $x$-direction only (along $y$ direction in each wire, the same type of sub-lattice should be placed). Therefore, we consider three chains of different species which are connected via quantum tunneling in both directions. We assume periodic boundary conditions along both directions. 
As for the opposite $t$ for the middle component, a constant gauge field with phase $e^{i\pi}$ can be applied along the $y$-direction, which does not affect the Zeeman term ($\sim \sin{\phi}$) but reverses the hopping amplitude.
Note that in the off-diagonal elements, hopping between different sub-lattices within the same unit cell will introduce an additional constant term $\sim t$, along with the hopping between NN unit cells. Such a term still gives a finite Chern number, and combinations of the two hoppings with different magnitudes can be used to generate different realizable lattice models.

\subsubsection{Scheme for site-selective polarization}\label{Sec:design}


\begin{figure}
\centering
\includegraphics[width=0.75\textwidth]{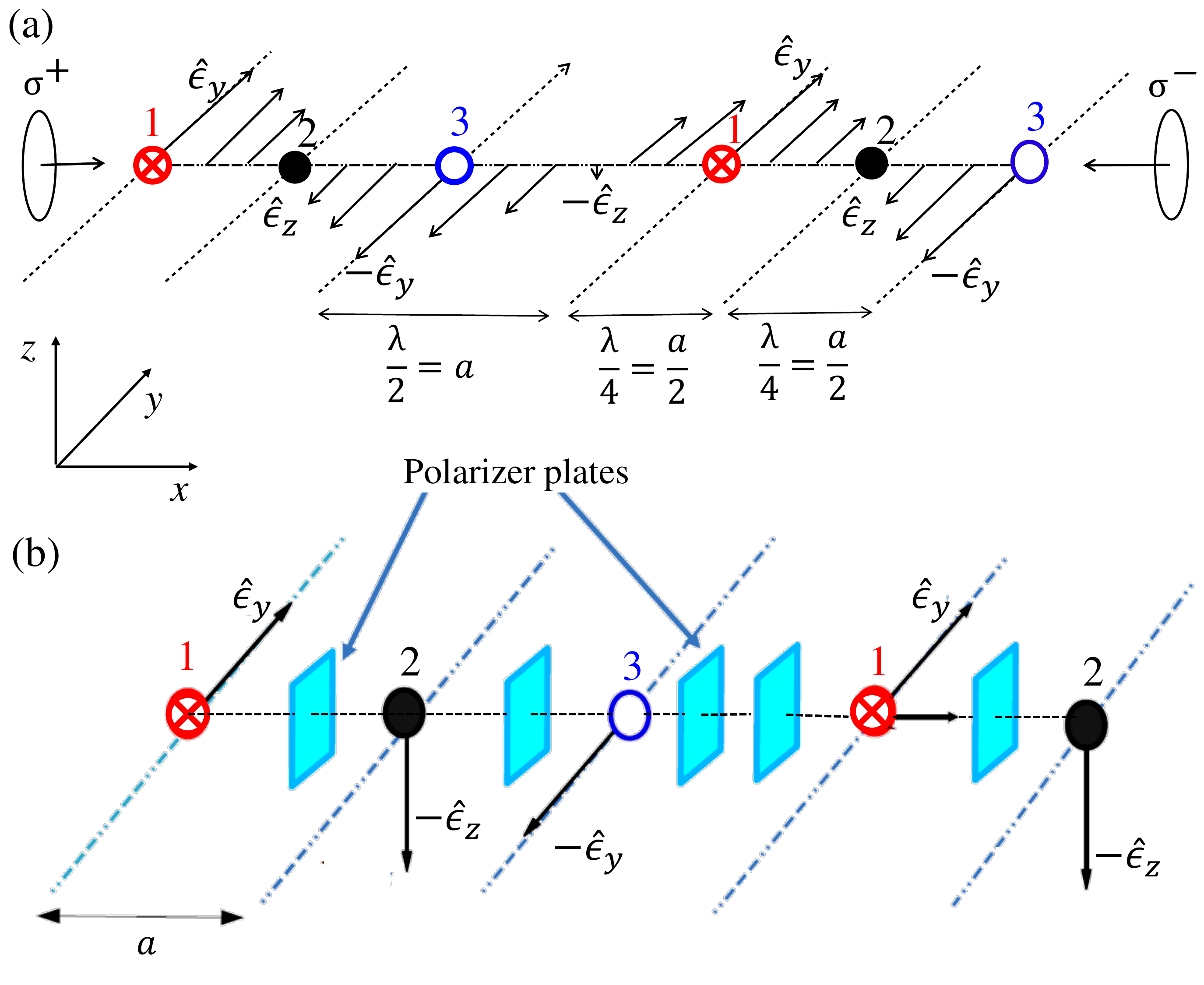}
\caption{(Color online) Schematic diagram for site-selective polarization. (a) Sisyphus cooling technique gives polarization gradient in a 1D lattice, when counter-propagating circularly polarized waves $\sigma^{\pm}$ are used. This creates a linear polarization that rotates in space (at $x=0$ polarization is along $\hat{\epsilon}_y$, at $x=\lambda/4$ and $\lambda/2$ polarization along $\hat{\epsilon}_z$ and $-\hat{\epsilon}_y$, respectively). Atoms trapped at these $x$-values can create the necessary odd parity Zeeman term.
b) A linearly polarized vector potential along $\hat{\epsilon}_y$ is incident on the atom trapped in the 1st atomic chain. A polarizer plate, placed between the 1st and 2nd wires, can rotate the polarization vector by $\pi/2$, i.e, perpendicular to $\hat{\epsilon}_y$. Another polarizer plate, between the wires with the 2nd and the 3rd atom, can rotate the incident vector $\hat{\epsilon}_z$ by $\pi/2$ again, thus aligning the polarization along $-\hat{\epsilon}_y$. These techniques are in essence similar to our proposal.}
\label{plate_diag}
\end{figure}

For all the above terms, no constraint arose about the specific parity (or orbital nature) of each basis. Therefore, coupled chain structure can be engineered with ultracold fermionic or bosonic atoms in optical lattice setup, or with quasi-1D quantum wires of electrons with lithography or pulse laser deposition method. An alternative approach towards visualizing the site-selective $\sin{\beta k_y}$ term is via coupling to a linearly polarized vector potential. Techniques used in Sisyphus cooling can generate polarization along a particular direction~\cite{sisyphus1,sisyphus2}. A polarizer plate set-up can also accomplish the purpose of spatially dependent polarization. Ours is a similar proposal with a constant vector potential having a site-selective polarization, as shown in the schematic set-up in Fig.~\ref{plate_diag}. If it were possible to use a very low frequency electromagnetic wave, such a proposal could be experimentally realizable (for example, in a medium with very high refractive index).


\section{Band topology and Berry curvature}\label{cal_chern}

We start with a generalized form of the SU(3) Hamiltonian Eq.~\ref{su3_gellmann}, where ${\bf b}({\bf k})$ is the eight component vector made of electron hopping in a lattice, as given in Eq.~\ref{b_vector}. 
The product of two Gell-mann matrices can be written as $\hat{\lambda}_a\hat{\lambda}_b=\frac{2}{3} \delta_{ab}+d_{abc}\hat{\lambda}_c+if_{abc}\hat{\lambda}_c$, where $d_{abc}$ and $f_{abc}$ are symmetric and anti-symmetric structure constants of SU(3) algebra,~\cite{galitski,mukunda}. These structure constants define three bilinear operations for the eight-component vectors, which are, $ \mathbf{u\cdot v}=u_av_a$, $(\mathbf{u}\times\mathbf{v})_a=f_{abc}u_bv_c$ and the star product $(\mathbf{u}*\mathbf{v})_a=\sqrt{3}d_{abc}u_bv_c$, where $\mathbf{u}$ and $\mathbf{v}$ are two arbitrary vectors. 

The eigenstate projection operator for the SU(3) case can be written in terms of the Gell-mann matrices:
\begin{equation} \label{projsu3}
\hat{P}_{{\bf k},n}=\frac{1}{3}\left(1+\sqrt{3}\mathbf{n}_{{\bf k},n}.\mathbf{\hat{\lambda}} \right),
\end{equation}
where, $\mathbf{n}_{{\bf k},n}$ lies on the surface of the eight-dimensional sphere of the $\mathbf{\hat{\lambda}}$ vectors, and $Tr\hat{P}_{{\bf k},n}=1$. The condition $(\hat{P}_{{\bf k}n})^2=\hat{P}_{{\bf k}n}$ gives two constraints on $\mathbf{n}_{{\bf k},n}$. They are $\mathbf{n}_{{\bf k},n}\cdot\mathbf{n}_{{\bf k},n}=1$ and $\mathbf{n}_{{\bf k},n}*\mathbf{n}_{{\bf k},n}=\mathbf{n}_{{\bf k},n}$.
To express $\mathbf{n}_{{\bf k},n}$ in terms of the $\mathbf{b}({\bf k})$'s, the relation $\left[\hat{P}_{{\bf k},n},\mathcal{\hat{H}}({\bf k})\right]=0$ can be used, (true for projection operators for the eigenstates) which leads to the condition, $\mathbf{b}({\bf k})\times \mathbf{n}_{{\bf k},n}=0$. This relation along with the above conditions on $\mathbf{n}_{{\bf k},n}$, give the resulting expression for the Berry curvature which can be obtained from the projection operator Eq.~\ref{projsu3},
\begin{align} \label{su3berry}
\Omega_n({\bf k}) &= -\frac{4}{3^{3/2}}f^3_{1{\bf k}n} \bigg(f_{2{\bf k}n}^2 \partial_{k_x} {\bf b} \times \partial_{k_y} {\bf b} + f_{2{\bf k}n} \partial_{k_x} {\bf b} \times \partial_{k_y} ({\bf b}*{\bf b}) 
+ f_{2{\bf k}n}  \partial_{k_x} ({\bf b}*{\bf b}) \times \partial_{k_y}{\bf b} \nonumber \\
&+ \partial_{k_x} ({\bf b}*{\bf b}) \times \partial_{k_y}({\bf b}*{\bf b}) \bigg). \bigg(f_{2{\bf k}n} {\bf b}+({\bf b}*{\bf b})\bigg)
\end{align}
where, $f_{1{\bf k}n}=\frac{1}{|b(k)|^2\left(4 \cos^2{(\theta_k+\frac{2\pi}{3} n)}-1\right)}$, $f_{2{\bf k}n}=2|b(k)|\cos{\big(\theta_k+\frac{2\pi}{3}n\big)}$, and, $\theta_k=\frac{1}{3}\arccos{\left[\frac{b(k).b(k)*b(k)}{|b(k)|^3} \right]}$ ($n$ runs from one to three). Using this expression and the $\mathbf{b({\bf k})}$'s in Eq.~\ref{b_vector}, we arrive at the Chern number by integrating over the Brillouin zone.

\begin{figure}[h]
\centering
\includegraphics[width=0.7\columnwidth]{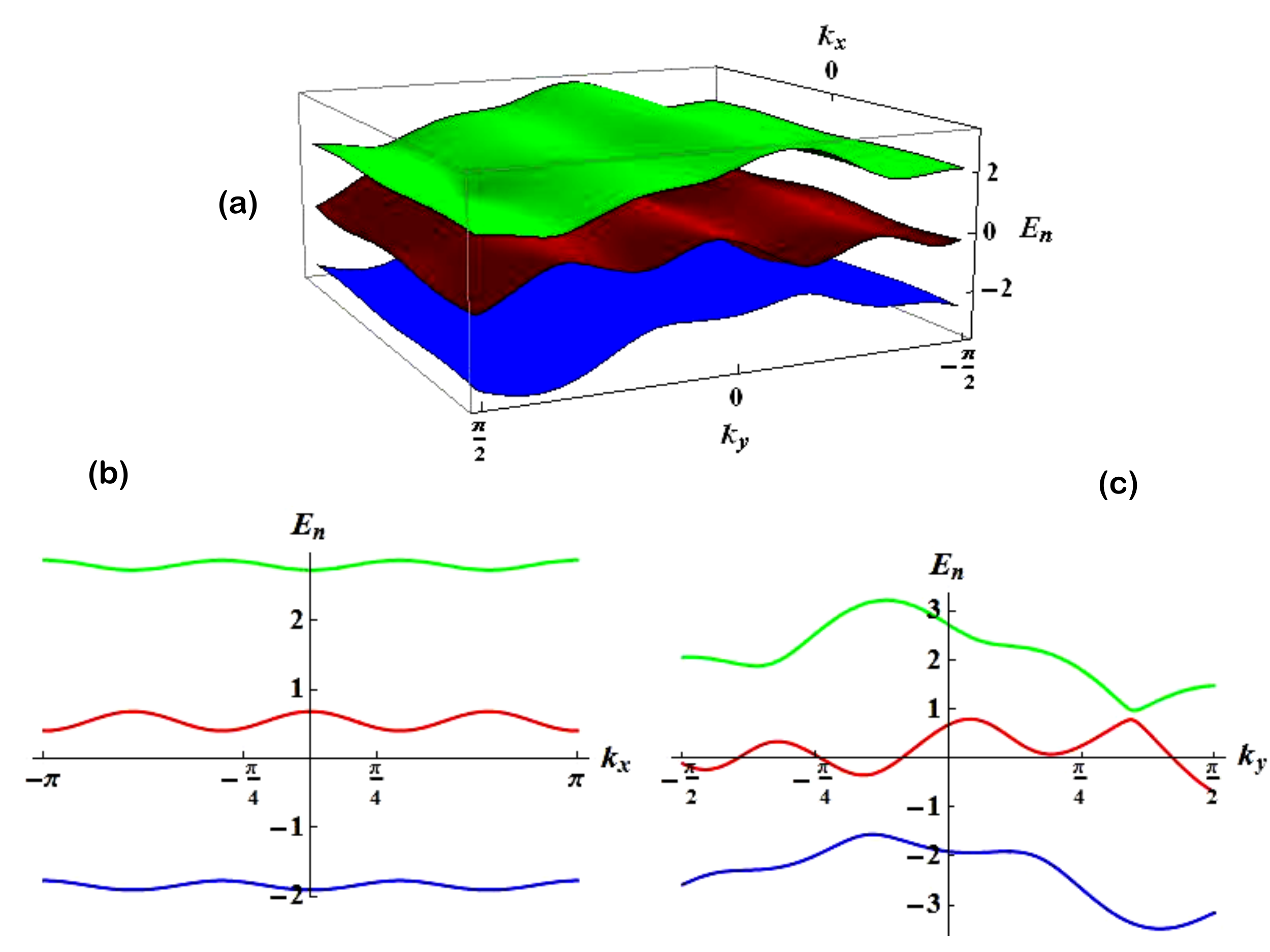}
\caption{(Color online) Bands demonstrating the energy dispersion for the proposed model are shown. (a) Surface plot with both $k_x$ ($-\pi$ to $\pi$) and $k_y$($-\frac{\pi}{2}$ to $\frac{\pi}{2}$), (b) Keeping $k_y=0$, dispersion plot along $k_x$, from $-\pi$ to $\pi$,  and (c)  Keeping $k_x=0$, dispersion plot along $k_y$, from $-\frac{\pi}{2}$ to $\frac{\pi}{2}$. All quantities are measured in units $t_x=t_y=1$, and $|m_1|=|m_3|=\sqrt{3} t_y$. 
 }
\label{3Dbren}
\end{figure}

In Fig.~\ref{3Dbren}, we plot the band structure in the momentum-space for the parameter values of $m_1=-m_3=\sqrt{3}t$ and $\alpha=\beta=2$, and $t_x=t_y$ (with this set of parameter values the Chern numbers are $\{-3,6,-3\}$, as will be discussed in details in the following sections). The electronic structure consists of three well separated bands, with Dirac-like nodes at various discrete non-high symmetric $k$-points (see Fig.~\ref{3Dbren}(a)). Therefore, a topological invariant can be separately assigned for each band. However, projecting the orbital character onto each band, we observe that substantial exchange of orbital character occurs in each band. We visualize the three orbital characters (in different columns) for three different bands (in different rows) in the entire 2D $k$-space in Fig.~\ref{supp_band}. 

\begin{figure*}
\includegraphics[width=\columnwidth]{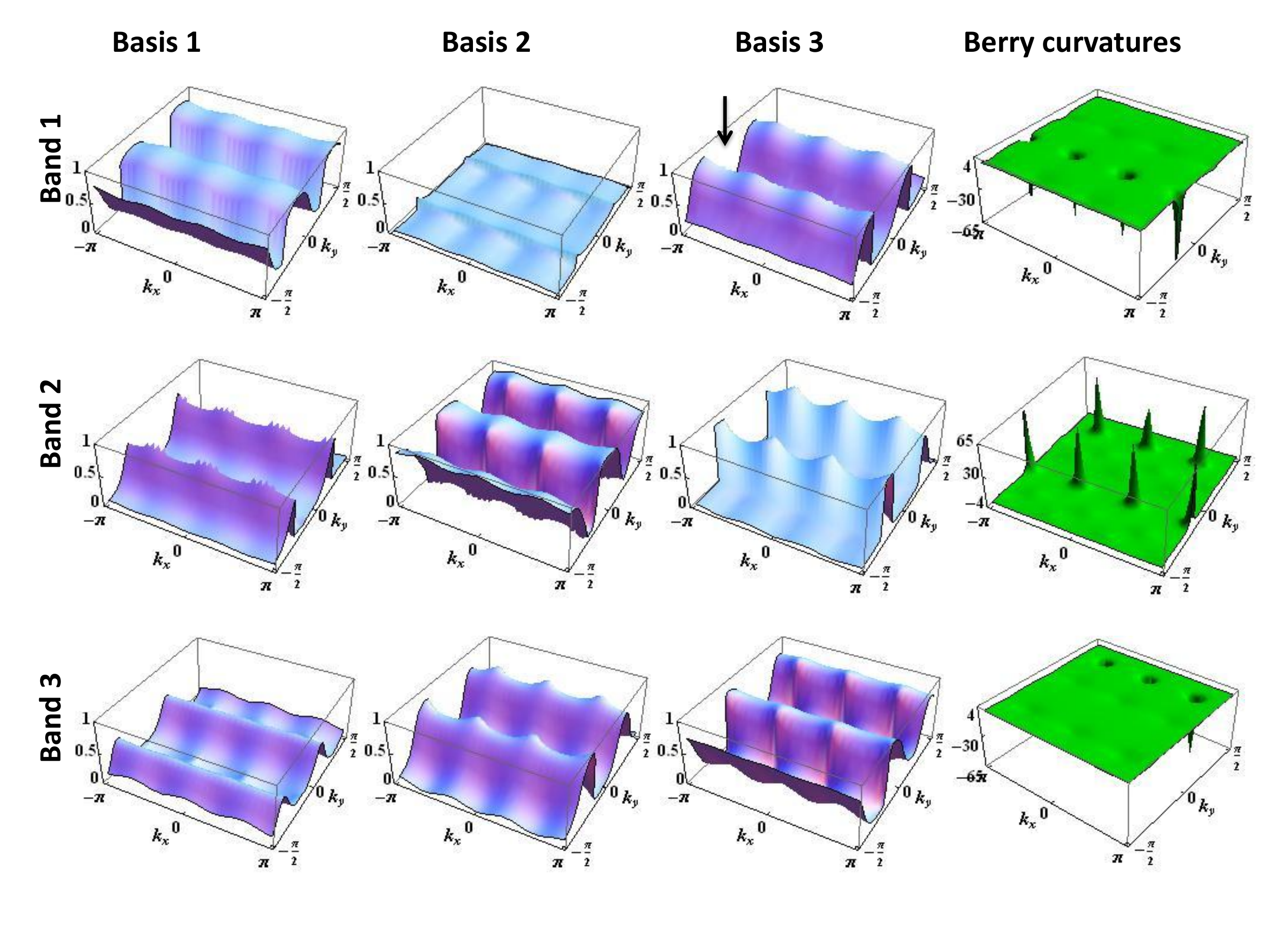}
\caption{(Color online) First three columns: The orbital weights of the basis states $\psi_j$ ($j=1,2,3$) are shown for each energy bands $E_1$, $E_2$ and $E_3$ respectively. The arrow in the figure corresponding to the third column shows the position of one minima which is corresponding to the gapless point of one of the edge states in the system. Fourth column: The berry curvature $\Omega_i (i=1,2,3)$ corresponding to each energy band is shown in the fourth column. The berry curvatures show a sharp peak at the corresponding gapless points of the edge states, indicative of a band inversion at the respective $k$-points. }
\label{supp_band}
\end{figure*}

As discussed in the introduction section, band inversion is an important criterion for both the SU(2) and SU(3) topological classes. In time-reversal invariant SU(2) topological classes, bands are only required to be inverted at the time-reversal invariant $k$-points. This makes it easier to define the band inversion strength simply by defining the band gap between the two bands at the time-reversal invariant $k$-points~\cite{AdiabaticTI,TIreviewTD}. Such simple definition becomes difficult to implement for SU(3) systems. On the other hand, we recognize that the Berry curvature acquires spike at the discrete band degenerate $k$-points across which bands are inverted in two orthogonal directions, $k_x$- and $k_y$.
In this spirit we can define a band inversion strength via occupation number or the `orbital weight' (here orbital refers to the three component basis) of the band at each $k$-point. Another interesting feature of these discrete $k$-points is that it represents a saddle point in the orbital weight, as seen in Fig.~\ref{supp_band}.
The rightmost column of Fig.~\ref{supp_band} refers to the Berry curvature as a function of $k$. We see that at all the ${\bf k}^*$-points where $\Omega({\bf k}^*)$ diverges in a given band, the corresponding orbital weight profile exhibits a saddle point. The system is in a topological phase whenever there are peaks in the Berry curvature plots. This is because the peaks always have the same sign for a given band, and so their existence implies a non-zero Chern number. Note that the bulk is always gapped in our system and there is no topological phase transition. The peaks in the Berry curvature occur at the $k$-points where the edge states become gapless.

If the orbital weight of the $\nu$th band and the $i$th orbital is given by $\gamma^\nu_i=|\psi^\nu_i(k_x,k_y)|^2$, then, at the discrete $k_x$, $k_y$ points for which  $\left(\frac{\partial^2\gamma^\nu_i}{\partial k^2_l} \right)\left(\frac{\partial^2\gamma^\nu_j}{\partial k^2_l} \right)<0$ for atleast two orbitals, the Berry curvature peaks. For example, in Fig.~\ref{supp_band}, it can be seen for band 2, the second orbital has a minimum at $k_x=\pi$ and $k_y=k^*_y$, whereas, the third orbital attains a maximum at these $k_x$, $k_y$ values. Similarly, in band 1, the first orbital and third orbital shows this behaviour, so does second and third orbital in band 3.


For the higher energy band, among the six visible saddle-points, three of them reside in the $-k_y$ region. They give three negative spikes in the Berry curvatures, as shown in the corresponding rightmost column of Fig.~\ref{supp_band}. (The two extreme peaks occurring at the zone boundary are related by reciprocal lattice vectors). This gives the corresponding Chern number $C_n=\sum_{\bf k} \Omega_n({\bf k})$ (where $n$ stands for bands) to be -3. In this band, inversion occurs between orbitals 1 and 3  across the three saddle points. In the lowest band, the Berry curvature peaks occur in the corresponding $+k_y$ side due to the band inversions between orbitals 2 and 3. The Chern number of this band comes out to be the same as -3. The middle band shows band inversions between all three orbitals at the same locations in both $\pm k_y$ sides, with positive Berry curvatures and thus obtains a Chern number of $+6$. 

It can also be shown that the Chern number does not depend if the strength of the Pierels' phase ($m_i$) is different in the different orbitals. As long as the $\sin{\beta k_y}$ terms have opposite signs in the diagonal terms (i.e. resembling a Zeeman-like term), the model is topologically non-trivial. Changing the off-diagonal elements, say from $e^{i k_x}$ to $e^{2 i k_x}$, we find higher  Chern number $(-4,8,-4)$. In the Appendix Sec.~\ref{otherhamiltonians}, we discuss several other parameter sets where Chern number can be tuned by different values of $\alpha$, $\beta$, and other terms in the Hamiltonian.

\section{\label{cal_edge} Calculation of edge states} 

\begin{figure}
\centering
\includegraphics[width=0.65\columnwidth]{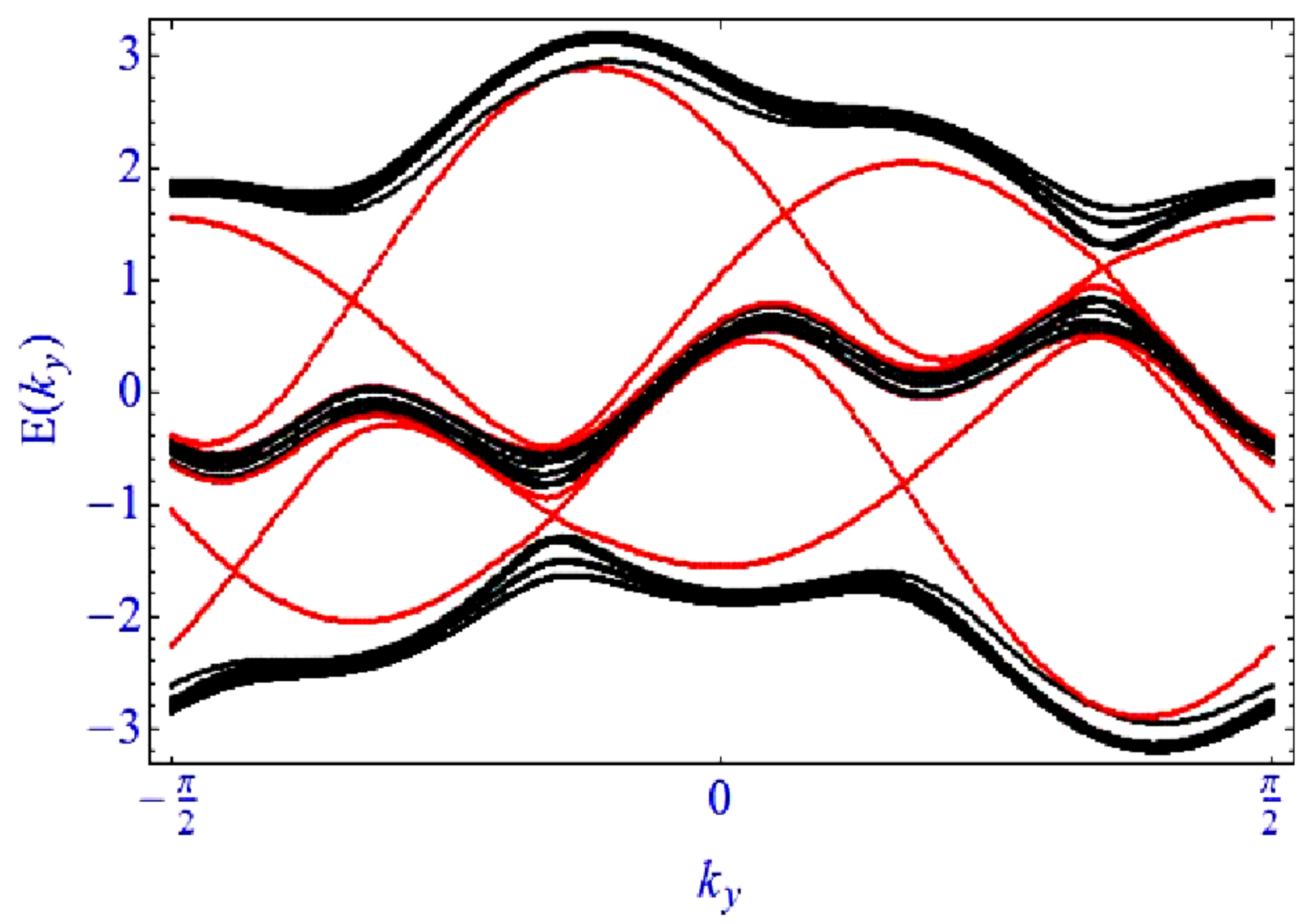}
\caption{(Color online) Edge state structure for $N=30$, with $\alpha=\beta=2$ and $|m_1|=|m_3|=\sqrt{3} t_y$. Also, $t_x=t_y=1$. There are six gapless points within the first Brillouin zone, as indicated by the chern numbers being $\{-3,6,-3 \}$.
}
\label{edgestate}
\end{figure} 

Non-trivial topological character can be observed from the edge state dispersion~\cite{TIreviewTD,TIreviewCK,TIreviewSCZ}. We study the characteristics of the edge state parallel to the $y$-direction for the above parameter set which gives Chern numbers $(-3,6,-3)$. We solve the Hamiltonian in Eq. ~\ref{Ham1} with periodic boundary conditions in the $y$-direction, and open boundary condition in the $x$-direction with a finite size lattice of $N$ atoms. Considering $\Phi_i({\bf k}_y)$ as the Wannier state localized on the $i^{\rm th}$ atom we can now expand the Hamiltonian in a $N\times N$ matrix as
\begin{eqnarray}\label{edge}
H(k)&=& -\sum_i\Big[\Phi_i^\dagger(k_y) {\mathcal{B}_{i,i}}(k_y)\Phi_i(k_y)\nonumber\\
&&\quad\quad+ \Phi_i^\dagger(k_y) {\mathcal{A}_{i,i+1}}\Phi_{i+1}(k_y)\Big] + {\rm h. c.}, 
\end{eqnarray}
\begin{align} \label{bmatrix}
\begin{split}
{\rm where,}~~
{\mathcal{B}}(k_y)=&\cos {\alpha k_y} \begin{pmatrix} 
-2t_y&0&0\\
0&\frac{3}{2}t_y&0\\
0&0&-t_y\\
\end{pmatrix}\\
+
\sin {\beta k_y}
&\begin{pmatrix}
-\frac{2m_1-m_3}{3}&0&0\\
0&\frac{m_1+m_3}{3}&0\\
0&0&\frac{m_1-2m_3}{3}\\
\end{pmatrix}\,,
\end{split}
\end{align}
and,
\begin{eqnarray}\label{amatrix}
{\mathcal{A}}= t_x\begin{pmatrix}
0&0&-1\\
-1&0&0\\
0&-1/2&0\\
\end{pmatrix}\,.
\end{eqnarray}
Eigenvalues of the Eq.~\eqref{edge} are plotted in Fig.~\ref{edgestate} with the same parameter set. The essence of the topological edge state is that it must adiabatically connect to the bulk states which is clearly observed in the present case. If the chemical potential is placed between the lowermost and the middle bulk band, the Chern number is -3, corresponding to three gapless points in the edge states (in red). Similarly, if the chemical potential is placed between the topmost band and the middle band, the Chern number is 3, corresponding to three gapless points in the edge states, which in this case have an opposite dispersion to that of the former.
To summarize, edge states with opposite dispersion, connecting two different bulk bands with opposite sign of the Chern number, meet at discrete ${\bf k}^*$-points where the Berry curvature obtained singularities in Fig.~\ref{supp_band}. Consistently, there are total of 6 such band touching points for the edge states.

We see that $\mathcal{B}$ term is diagonal in this basis which gives the dispersion along the $k_y$ direction. These states become gapped by $\mathcal{A}$. However as the number of lattice site is increased, the gap at the edge state vanishes at the $k_y$-points where Berry phase acquires divergence. Therefore, expanding the edge state near these points, we find that three eigenstates up to linear-in-$k_y$ as (substituting $|m_1|=|m_3|=m$)
\begin{eqnarray} \nonumber
E_1(k_y) &=& -t_y  \bigg(2 + \frac{m}{t_y} k_y \bigg) \nonumber \\
E_2(k_y) &=& \frac{3}{2}t_y \nonumber \\
E_3(k_y) &=& -t_y \bigg(1 - \frac{m}{t_y} k_y \bigg).
\end{eqnarray}
We notice that the second term represents a localized bound state, while the other two bands are linear with $k_y$ in the low-energy region. As $\mathcal{A}$ term is turned on, these three states split into six states, in accordance with the higher Chern number in the bulk state.

\section{\label{spin} Extension to spinful case}

In our model, SU(3) topological insulator is obtained for spinless fermions, in which spin of the particles is a dummy variable. This Hamiltonian respects the spin-rotational symmetry. Therefore, as long as this symmetry is held (in the absence of spin orbit coupling), the topological invariant remains the same for all values of spin in a given system. We now discuss how the result changes when the spin rotational symmetry is broken. For generality, we assume the atoms/ electrons have a spin value $S$, which splits into 2$S$+1 multiplets once the rotational symmetry is lifted. In this case, our starting SU(3) spinor has the dimension of 3(2$S$+1). For illustration we take $S=1/2$, while the obtained conclusions below remain the same for any other values of $S$. In this case, the spinor is, $$\Psi_k^{\dag} = \left( {\psi}_{1\uparrow}^{\dag}(k), {\psi}^{\dag}_{2\uparrow}(k),{\psi}^{\dag}_{3\uparrow}(k), {\psi}_{1\downarrow}^{\dag}(k), {\psi}^{\dag}_{2\downarrow}(k),{\psi}^{\dag}_{3\downarrow}(k)\right)^{T},$$ in which the Hamiltonian becomes a $6\times 6$ matrix. The Hamiltonian can be split into two $3\times 3$ diagonal blocks as $H_{\uparrow\uparrow}$, and $H_{\downarrow\downarrow}$, and off diagonal blocks $H_{\uparrow\downarrow}$, and $H_{\uparrow\downarrow}^{\dagger}$. In the absence of the spin-flip terms, i.e., when $H_{\uparrow\downarrow}=0$ at all momentum, the Hamiltonian breaks into a block diagonal one, and the bands remain doubly degenerate for the two spin species. We get two copies of the same SU(3) topological insulator, with the same set of Chern numbers for each spin structure.
If now a perturbation like the Zeeman term in $\hat{\sigma_z}$ direction is added, it will appear as a constant term in each spin block, thus splitting the band structure by a constant amount for the two spins. However, the Chern number will still be the same in each spin block.
 

Thus, as long as a simple exchange field is present to split the spin states, such as a Zeeman field, the system maintains its topological property. 
When the spin-flip term $H_{\uparrow\downarrow}$ is introduced for the case of spin-orbit coupling, spin is no longer a good quantum number. In this case, Chern number cannot be defined for each spin state or band. Therefore, there will be a topological phase transition into either a trivial case, or to another topological class, such as $Z_2$ family for fermions. 


\section{ \label{conclusion} Discussion and Conclusion}
This chapter delivers the following messages. 
We engineered the complex phase dependent off-diagonal terms in a tripartite lattice through uncompensated Bloch phase (from hopping between NN unit cells). The sign reversal of the hopping amplitude for the middle component of the unit cell can be obtained by applying a constant flux $e^{i \pi}$ along the $y$-direction.
The multi-channel set-up of one-dimensional atomic chains suggested in our model can be visualized as an array of quantum wires, hosting different types of orbitals, being subjected to a linearly polarized constant vector potential. Quantum wires are being studied extensively to describe varied topological phenomena theoretically~\cite{haim,seroussi,sela,alicea}. Fractional topological phases are being studied in weakly coupled quantum wires, in both two and three dimensions~\cite{lubensky,mudry,teo,yoreg}. Periodically driven systems can also show interesting non-trivial topological effects in spinless systems ~\cite{yan,fulga,ezhao,diehl,goldman} and can also be extended to SU(3) systems. 

The second criterion for the SU(3) topological Hamiltonian is that two of the diagonal terms must contain an odd parity term, such as a sine function of momentum. This poses an important bottleneck to engineer SU(3) topological phase in a condensed matter setup. Here we suggest a simplistic scheme for such term by generalizing the tight-binding Hamiltonian in the presence of a constant vector potential. This gives a spinless SU(3) topological material. We have illustrated two schematics, using either the Sisyphus cooling technique or polarizer plates, to visualize the desired site-selective polarization. 

Robustness of the spinless SU(3) topological phase when spin is introduced is also discussed. We showed that as long as there is only a Zeeman-kind of term present  without any spin-flip term, the topological invariant is robust upto a new band inversion. When a spin-flip term is introduced (such as spin orbit coupling term), spin is no longer a good quantum number, and Chern number can no longer be defined for each spin or band. So our formalism does not hold any more. 

There are several methods for detection of edge states, tailored specifically for cold atom systems~\cite{mancini,Stuhl}. Direct imaging of edge states after a sudden quench in cold atoms is one such technique ~\cite{pnas}. Another relevant method of detection of chiral edge states in cold atoms is by the shelving method, demonstrated in ~\cite{shelving1,shelving2}.


\begin{subappendices}

\section*{Appendix}

\section{\label{otherhamiltonians} Other forms of Hamiltonians}

So far, we have considered a specific form of the most general Hamiltonian given in Eq.~\eqref{gensu3}. This model Eq.~\eqref{Ham1} is realized in a tripartite lattice with site-selective polarization of the vector potential. With respect to the structure of our model, the Hamiltonian can be rewritten in a more generalized form by expressing the off-diagonal terms as,
\beq
H_{12}=  t_x^{12}\ e^{-ik_x} \ ;  H_{13}=   t_x^{13} \ e^{-ik_x} \ ; H_{23}=   t_x^{23} \ e^{-ik_x} .
\label{Hamgen}
\eeq
Where the $t_{ij}$ provide the inter-component hopping strengths between nearest neighbour unit cells. The diagonal terms are kept in the same form as in Eq. (\ref{diagonal}) with various choice of $t_i$ and $m_i$. Here we discuss various other combinations of the diagonal and off-diagonal terms which give finite Chern number, some of which may require different lattice structure than the tripartite lattice discussed in the main text. It should be noted that the following list is not necessarily exhaustive, and more combinations can be derived based on the basic principles deduced in the main text.  In all combinations, the Hamiltonian is represented by the eight Gell-Mann matrices. \hphantom{} \\

{\bf \underline{Case I: $\mathbf{\alpha=\beta=1}$}}
\begin{enumerate}
\item{ With $t_1 = t_3= -t_y$, $ t_2 = 2t_y, m_1 = -m_3 = -\sqrt{3}t_y$, $m_2=0$, $t_x^{12}=t_x^{13}=t_x^{23}=-t_x$, this Hamiltonian provides integer Chern number set $(-3,6,-3)$.
}
\item{Same as (1) but with $t_x^{12}=-t_x\cos k_y$. This Hamiltonian provides Chern numbers $(-1,2,-1)$.
}
\item{ With $t_2=m_2=0$ i.e. $\xi _2({\bf k})=0$, and $t_3=t_y/2 $ with rest of the coefficients as in 2. This Hamiltonian again, gives Chern numbers $(-3,6,-3)$.
}
\end{enumerate}

{\bf \underline{Case II: $\mathbf{\alpha=\beta=2}$}}\\
\begin{enumerate}
\item{With $t_2=m_2=0$ and $t_3=t_y/2 $, as in point 3 of Case I, replace $t_x^{12}=-t_x$, $t_x^{13}=2t_x \cos 2k_x$ and $t_x^{23}=-t_x \sin (k_y-\sqrt{3})$, keeping the rest of the coefficients same as in point 1 in Case I. This Hamiltonian gives the Chern numbers $(4,0,-4)$.
}
\item{The almost similar configuration as in point 1 in Case II, only changing $t_x^{12}$ as $t_x^{12}=\pm t_x(\cos k_y+\sqrt{3}\sin k_y)$ and using the same sign (either $+$ or $-$) for $t_x^{ij}$ i.e with $t_x^{13}=\pm t_x\cos k_x$ and $t_x^{23}=\pm t_x\sin(k_y-\sqrt{3})$ the Chern number for this hamiltonian is $(3,0,-3)$.  
}
\end{enumerate}

In all the above calculations, $t_y=t_x=1$.

\end{subappendices}

}
